\documentclass[11pt,a4paper]{article}
\pdfoutput=1
\usepackage{jheppub}

\usepackage[utf8]{inputenc}

\usepackage{graphicx}
\usepackage{amsmath,amsfonts,amssymb}
\usepackage{epsfig,color}
\usepackage[outdir=./]{epstopdf}
\usepackage[T1]{fontenc}
\usepackage{slashed}
\usepackage{soul}
\preprint{\begin{tabular}{r}JLAB-THY-19-3024\end{tabular}}
\begin{document}

\newcommand{\beq}{\begin{equation}}
\newcommand{\eeq}{\end{equation}}
\newcommand{\bea}{\begin{eqnarray}}
\newcommand{\eea}{\end{eqnarray}}
\newcommand{\msb}{\overline{\rm MS}}
\newcommand{\be}{\begin{eqnarray}}
\newcommand{\ee}{\end{eqnarray}}
\newcommand\del{\partial}
\newcommand\nn{\nonumber}
\newcommand{\Tr}{{\rm Tr}}
\newcommand{\Str}{{\rm Str\,}}
\newcommand{\Sdet}{{\rm Sdet\,}}
\newcommand{\Pf}{{\rm Pf\,}}
\newcommand{\U}{{\rm U\,}}
\newcommand{\mat}{\left ( \begin{array}{cc}}
\newcommand{\emat}{\end{array} \right )}
\newcommand{\vect}{\left ( \begin{array}{c}}
\newcommand{\evect}{\end{array} \right )}
\newcommand{\tr}{{\rm Tr}}
\newcommand{\hm}{\hat m}
\newcommand{\ha}{\hat a}
\newcommand{\hz}{\hat z}
\newcommand{\hx}{\hat x}
\newcommand{\tm}{\tilde{m}}
\newcommand{\ta}{\tilde{a}}
\newcommand{\tz}{\tilde{z}}
\newcommand{\tx}{\tilde{x}}

\definecolor{red}{rgb}{1.00, 0.00, 0.00}
\newcommand{\rd}{\color{red}}
\definecolor{blue}{rgb}{0.00, 0.00, 1.00}
\definecolor{green}{rgb}{0.10, 1.00, .10}
\newcommand{\blu}{\color{blue}}
\newcommand{\green}{\color{green}}

\def\emg{\color{green}}
\def\emb{\color{blue}}
\def\emr{\color{red}}
\def\embr{\color{brown}}
\def\emm{\color{magenta}}
\def\emo{\color{orange}}



\title{Parton Distribution Functions from Ioffe time pseudo-distributions}

\author[a]{B\'alint Jo\'o}
\author[b,a,c]{, Joseph Karpie}
\author[b,a]{, Kostas Orginos}
\author[d,a]{, Anatoly Radyushkin}
\author[a]{, David Richards}
\author[e,f]{and Savvas Zafeiropoulos}
\affiliation[a]{Thomas Jefferson National Accelerator Facility, \\ 12000 Jefferson ave., Newport News, VA 23606, U.S.A.}
\affiliation[b]{Department of Physics, William \& Mary, \\ 300 Ukrop Way Williamsburg, VA 23187, U.S.A.}
\affiliation[c]{Department of Physics, Columbia University \\ 538 West 120th Street, New York, NY 10027,U.S.A.}
\affiliation[d]{Physics Department, Old Dominion University, \\ 4600 Elkhorn Ave., Norfolk, VA 23529, U.S.A.}
\affiliation[e]{Institute for Theoretical Physics, Heidelberg University, \\ Philosophenweg 12, 69120 Heidelberg, Germany}
\affiliation[f]{Aix Marseille Univ, Universit\'e de Toulon, CNRS, CPT, \\ UMR 7332, F-13288 Marseille, France}
\emailAdd{bjoo@jlab.org}
\emailAdd{jmk2289@columbia.edu}
\emailAdd{kostas@wm.edu}
\emailAdd{aradyush@odu.edu}
\emailAdd{dgr@jlab.org}
\emailAdd{savvas.zafeiropoulos@cpt.univ-mrs.fr}

%
\abstract{
In this paper, we present a detailed study of the unpolarized nucleon parton distribution function (PDF) employing the approach of parton pseudo-distribution functions. We perform a systematic analysis using three lattice ensembles at two volumes,  with lattice spacings $a=$ 0.127 fm and  $a=$ 0.094 fm, for a pion mass of roughly 400 MeV. With two lattice spacings and two volumes, both continuum limit and infinite volume extrapolation systematic errors of the PDF are considered. In addition to  the $x$ dependence of the PDF, we compute their  first two moments  and compare them with the pertinent phenomenological determinations. }

\maketitle
\date{\today}

\flushbottom

\section{Introduction}
Parton distribution functions (PDF)~\cite{Feynman:1973xc} describe the structure of hadrons in terms of the momentum and spins
of the quarks and gluons. Deep inelastic scattering (DIS) experiments have allowed for a phenomenological determination of the PDFs, but a direct calculation using Quantum Chromodynamics (QCD) remains out of reach. The theoretical definition of PDFs requires calculation of hadronic matrix elements with separations on the light cone. A calculation on a
Euclidean lattice is therefore not possible. Previously, the Mellin moments of PDFs and Distribution Amplitudes (DA) of baryons and mesons have been calculated with Lattice QCD~\cite{Bali:2013gya,Abdel-Rehim:2015owa,Alexandrou:2017oeh,Oehm:2018jvm,Bali:2019ecy,Bali:2019dqc}, but the reduced rotational symmetry of the lattice only allowed access to the lowest few moments. Unfortunately more moments than are available are required for an accurate reconstruction of the $x$ dependence of the nucleon PDF~\cite{Detmold:2001dv}. The signal-to-noise ratio and the power divergent mixings are two pressing bottlenecks in this approach.

To avoid the difficulties stemming from the light cone coordinates, it has been proposed to calculate the nucleon matrix elements with purely space-like separations. A momentum space formulation, proposed by X. Ji~\cite{Ji:2013dva}, calculates the parton quasi-distribution function (quasi-PDF), $\tilde{q}(y,p_3^2)$, which describes the  parton distribution of the third component of the hadron momentum $p_3$ 
 rather than that of  the  ``plus'' light-cone component $p_+$. 
  In the limit 
 $p_3 \to \infty$,  
the quasi-PDF can be factorized into the light-cone PDF, $f(x,\mu^2)$. This technique
has been explored extensively in numerical lattice calculations,
for several different quasi-PDFs, as well as for the pion
quasi-distribution amplitude
(DA)~\cite{Lin:2014zya,Chen:2016utp,Alexandrou:2015rja,Alexandrou:2016jqi,
  Monahan:2016bvm, Zhang:2017bzy, Alexandrou:2017huk, Green:2017xeu,
  Stewart:2017tvs,Monahan:2017hpu,
  Broniowski:2017gfp,Alexandrou:2018pbm,Alexandrou:2018eet,Alexandrou:2019lfo,Izubuchi:2019lyk}. Other
approaches to obtaining PDFs and meson Distribution Amplitudes from the lattice include those of
references~\cite{Detmold:2005gg,Braun:2007wv,Chambers:2017dov, Liang:2019frk}.

Some of the main difficulties with the quasi-PDF method arise from the high
momenta necessary for the calculation. One issue is that the
signal-to-noise ratio of correlation functions decreases
exponentially with the momentum which requires increasing
computational costs to achieve a precise matrix element
extraction. Another issue is that the momenta must be large enough for
the perturbative matching formulae to apply and still must be small
enough to be free of lattice artifacts. Recent work suggests
that  non-perturbative effects may dominate the
evolution of the quasi-PDF up to rather large momenta. 

A series of papers by one of the authors (AR) discusses the
nonperturbative $p_3^2$-evolution of quasi-PDFs and
quasi-DAs~\cite{Radyushkin:2016hsy,Radyushkin:2017gjd} based on the
formalism of virtuality distribution functions. Using the approach
in~\cite{Radyushkin:2016hsy,Radyushkin:2017gjd}, a connection was
established between the quasi-PDF and ``straight-link'' transverse
momentum dependent distributions (TMDs) ${\cal F} (x, k_T^2)$, whose
Fourier transform has been calculated
on the lattice  
 in~\cite{Musch:2010ka}.
Using simple assumptions about  
TMDs, models were built for the
non-perturbative evolution of quasi-PDFs. It was made clear that the
convolution nature of the quasi-PDFs leads to a complicated pattern of
\mbox{$p_3^2$-evolution,} which consequently enforces the use of large
values of momenta, namely, $p_3 \gtrsim 3$ GeV to ensure a controlled
approach to the PDF limit. The derived curves agree qualitatively with
the patterns of \mbox{$p_3$-evolution} produced by lattice
calculations.

The structure  of quasi-PDFs was further studied in Ref. ~\cite{Radyushkin:2017cyf}. It was  shown  that, 
when a  hadron is moving, the parton $k_3$ momentum
may be treated  as 
coming  from two sources. The hadron's motion as a whole yields 
the  $xp_3$  part,  which is governed by the dependence of
 the TMD  ${\cal F} (x, \kappa^2)$ on its first argument, $x$.
  The residual part $k_3-xp_3$  is controlled by the  way that the TMD depends on its second argument,
$\kappa^2$, which dictates the shape of the primordial rest-frame momentum distribution.   

A position space formulation was proposed by one of the authors
in~\cite{Radyushkin:2017cyf}. The suggestion was to perform the
calculation of the Ioffe time pseudo-distribution function
(pseudo-ITD), $\mathcal{M}(\nu,z^2)$, where the Ioffe time, $\nu$, is
dimensionless and describes the amount of time the DIS probe interacts
with the nucleon, in units of the inverse hadron mass.  The related
pseudo-PDF (or parton pseudo-distribution function) 
$\mathcal{P}(x,z^2)$
can be determined from its Fourier transform. The pseudo-PDF and the
pseudo-ITD are Lorentz invariant generalizations of the PDF and of the
Ioffe time distribution function (ITD) to space-like field
separations. Unlike the quasi-PDF, the pseudo-PDF has canonical
support in $-1 \leq x \leq 1$ for all values of $z^2$ even when the PDF
limit has not yet been reached. In a super renormalizable theory, the
pseudo-PDF will approach the PDF in the 
 $z^2 \to 0$ limit. 
 In renormalizable theories, the pseudo-PDF will have a logarithmic
divergence at small $z^2$ which
corresponds to 
the DGLAP evolution of
the PDF. The pseudo-PDF and the pseudo-ITD can be factorized into the
PDF and perturbatively calculable kernels, similar to experimental
cross sections. This fact means that the pseudo-PDF and pseudo-ITD
fall into the category of ``Good Lattice Cross Sections'' as described
in~\cite{Ma:2017pxb}.  The first lattice implementation of this
technique was performed in ~\cite{Orginos:2017wcl, Karpie:2017bzm} to
compute the iso-vector quark pseudo-PDF in the quenched
approximation. Other Good Lattice Cross Sections have been calculated
to extract the pion DA~\cite{Bali:2017gfr,Bali:2018spj} and the pion
PDF~\cite{Sufian:2019bol}. We refer the reader to~\cite{Lin:2017snn,
  Cichy:2018mum, Monahan:2018euv, Qiu:2019kyy} for detailed reviews of
these topics.

Possible difficulties with these approaches were raised
in~\cite{Rossi:2017muf, Rossi:2018zkn}. In~\cite{Rossi:2017muf}, the
authors observed that the power divergent mixing of moments of the PDF
calculated in Euclidean space would cause a divergence of the moments
of the quasi-PDF. Due to this issue, they argued the PDF could not be
extracted from the quasi-PDF. This claim was refuted
in~\cite{Ji:2017rah}, where the authors showed that the non-local
operator can be matched to the PDF without the presence of power
divergent mixings. In~\cite{Rossi:2017muf, Rossi:2018zkn}, the authors
noted that the Fourier transformation of the logarithmic $z^2$
dependence, generated by the DGLAP evolution of the PDF, will create
contributions to the qPDF in the region
of $|y|>1$ which do not vanish in the limit $p_3 \to \infty$. This
effect is unavoidable in the quasi-PDF formalism since the Fourier
transform must be performed before matching to the $\overline{\rm MS}$
PDF.  It is this contribution which generates the divergent moments of
the quasi-PDF.  In~\cite{Radyushkin:2018nbf}, the origin of this
contribution was described in terms of the ``primordial transverse
momentum distribution''. 
It was argued that the non-perturbative part of the $|y|>1$ contributions 
vanishes  in the $p_3 \to \infty$  limit, while the non-vanishing 
perturbatively calculable contributions are canceled after implementation of the 
matching procedure. As a result, the moments of extracted PDFs are finite. 

It should be noted that in the pseudo-PDF formalism 
the $z^2$-dependence of  $\mathcal{M}(\nu,z^2)$ is not subject to a Fourier transform, 
and the issue is completely avoided.  As
was shown in~\cite{Radyushkin:2017cyf}, pseudo-PDFs
 $\mathcal{P}(x,z^2)$ 
  have the canonical
support $[-1,1]$ for the momentum fraction $x$.  The unphysical region
of $|x|>1$ is avoided and the moments of the pseudo-PDF are finite.
Finally in~\cite{Karpie:2018zaz}, it was demonstrated using lattice
data that the finite moments of the PDF can be extracted from the
non-local matrix element for the reduced pseudo-ITD, refuting the
claim in~\cite{Rossi:2018zkn} that the pseudo-PDF moments would be
power divergent.  In the OPE, the power divergent mixings of the
moments are explicitly canceled by the corresponding Wilson
coefficients.  This feature of the OPE has been known for some
time~\cite{Dawson:1997ic},  and this method of extracting moments from
non-local operators is referred to as ``OPE without
OPE''~\cite{Martinelli:1998hz}. This cancellation of divergences is
unsurprising. The reduced pseudo-ITD is by design a renormalization
group invariant quantity. There can be no difference between this
object calculated with lattice regularization or dimensional regularization in
the continuum. Since all of the moments are finite, a matching
relationship between the pseudo-ITD and the $\overline{\rm MS}$ ITD can be
derived from these Wilson coefficients.

In this paper, we show the first calculation of the Ioffe time
pseudo-distribution function with dynamical fermion ensembles. The
other aspect new to pseudo-ITD analysis is that we have applied the
method of momentum smearing~\cite{Bali:2016lva} to the pertinent
matrix element which substantially improves our results at high
momenta when compared to Gaussian smearing. The remainder of the paper is as follows.
In Section~\ref{sec:itpd}, the Ioffe time distribution is outlined, and in Section~\ref{dlc}
the details of its lattice implementation are described. In Section~\ref{pdfe},
the results of the calculation are presented, and finally in Section~\ref{sec:con}, we
summarize our findings and propose future research directions.

\section{Ioffe time pseudo distributions}\label{sec:itpd}
The unpolarized ITD is
described in terms of a special case of the helicity-averaged,
forward, non-local matrix element,
\begin{equation}
  \label{eq:a_matrix_element} M^\alpha (p,z) = \langle p |
  \bar\psi(z) \gamma^\alpha U(z;0) \psi(0) | p \rangle \,,
\end{equation}
for $p = (p^+, \frac{m^2}{2p^+},0_T)$, $z = (0,z_- , 0_T)$, and
$\alpha = +$, where we use light-cone coordinates, i.e.\ $a_\mu=(a_+,a_-,a_T)$ with $a_\pm = (a_t \pm
a_z)/\sqrt 2$ and $a_T = (a_x,a_y)$.
Given arbitrary choices of $p$,
$z$, and $\alpha$, the Lorentz decomposition of the matrix element in
Eq.~\eqref{eq:a_matrix_element} is
\begin{equation}
  \label{eq:pseudo_lor_decomp}
M^\alpha(z,p) = 2 p^\alpha \mathcal{M}(\nu,z^2) + 2 z^\alpha
\mathcal{N}(\nu,z^2) \,,
\end{equation}
where the Lorentz invariant $\nu=p\cdot
z$ is called the Ioffe time. For the choice of parameters which
corresponds to the ITD, only $\mathcal{M}$ contributes to the matrix
element. For arbitrary $z^2$, the $\mathcal{M}$ function, called the
Ioffe time pseudo-distribution function, can be thought as a
generalization of the ITD to separations other than light-like. The
pseudo-ITD contains the leading twist contributions, but also contains
higher twist contributions at $O(z^2 \Lambda^2_{\rm QCD})$. The
removal of these contributions, through cancellation or small $z^2$,
is necessary for accurately determining the ITD from the pseudo-ITD.

From the relevant handbag diagrams, it has been shown
\cite{Radyushkin:2017cyf} that the Fourier conjugate of $\nu$, denoted
by $x$, has a restricted range of $-1\le x \le 1$. The Fourier
transform of the pseudo-ITD, called the pseudo-PDF $\mathcal{P}(x,z^2)$ is
given by
 \be\label{eq:ppdf} \mathcal{M}(\nu,z^2) = \int_{-1}^1 dx
e^{i\nu x} \mathcal{P}(x,z^2) \,.  \ee 
This definition of $x$ is completely
Lorentz covariant and there is no need to restrict displacements onto
the light cone or to require infinite momenta. This feature is
promising for a lattice calculation where only space-like
displacements are possible and large momenta are plagued by both large
statistical and systematic errors.  An extended discussion of this
approach is provided in~\cite{Radyushkin:2017cyf, Orginos:2017wcl,
  Radyushkin:2017lvu,
  Radyushkin:2017sfi,Karpie:2017bzm,Karpie:2018zaz,Karpie:2019eiq}.

 \subsection{PDFs, TMDs, and pseudo-PDFs}
Specific choices of $z$, $p$, and $\alpha$ can connect the Lorentz-invariant pseudo-PDF to the standard light cone PDF. Using light-cone coordinates, and assuming a light-like displacement $z_\mu = (0,z_-,0_T)$, longitudinal hadron momenta $p_\mu=(p_+,p_-,0_T)$, and the Dirac matrix for $\alpha=+$, the standard light-cone PDF can be determined as the Fourier transform of the ITD
\be
f(x) = \mathcal{P}(x,0) = \frac 1 {2\pi} \int_{-\infty}^{\infty} d(p_+z_-) e^{-i (p_+ z_-) x} \mathcal{M}(p_+ z_-,0)\,.
\ee
One can now introduce transverse degrees of freedom through the displacement $z = (0,z_-,z_T)$ in order to define a TMD, $\mathcal{F}(x,k_T^2)$, as
\be
\mathcal{P}(x,-z_T^2) = \int d^2 k_T e^{-i k_T \cdot z_T} \mathcal{F}(x,k^2_T) \,.
\ee 
This definition of the TMD uses a straight link operator for $U(z,0)$
of Eq.~\ref{eq:a_matrix_element} and describes the undisturbed or
``primordial'' distribution of the nucleon. Standard definitions of the
TMD rely on staple links. These staple links account for interactions
with either the initial or final state colored particles which exist in
scattering processes. The primordial TMD may not be capable of being
determined in a scattering experiment, but it can still be treated
properly within the realm of Quantum Field Theory.
 Unfortunately, light cone coordinates are not suitable for lattice
field theory applications. Instead by choosing the displacement and
momenta along a particular lattice axis, $z=(0,0,z_3,0)$ and $p
=(0,0,p_3,E)$ and the Dirac matrix for $\alpha = 4$, the pseudo-PDF
can be determined by \be \mathcal{P}(x,-z_3^2) = \frac 1 {2\pi}
\int_{-\infty}^{\infty} d\nu \,e^{-i\nu x}
\mathcal{M}(\nu,-z_3^2)\,,  \ee where $\nu=p\cdot z$. Ignoring the logarithmic
divergences in a renormalizable theory, the pseudo-PDF will converge
to the PDF in the limit of $z_3^2 \to 0$. Due to Lorentz invariance,
the pseudo-PDF calculated in either light cone or Cartesian
coordinates will produce the same function. 
The difference between
the pseudo-PDF and the PDF can then be described by using the $k_T$
dependence of the primordial TMD, $\mathcal{F}(x,k^2_T)$.

This limit can be shown to be the same convergence limit of the quasi-PDF
after one recognizes $z_3^2 = \frac{\nu^2}{p_3^2}$. The complicated
evolution of the quasi-PDF can be explained by the fact that, in the
space of the Lorentz invariants $\nu$ and $z^2$, it is an integral
with respect to $\nu$ of $\mathcal{M}(\nu,\nu^2/p^2)$ along the curve
$z^2 = \frac{\nu^2}{p^2}$. This feature makes the quasi-PDF at a given
value of the momentum $p$ a mixture of the Ioffe time distribution at
different scales some of which may not be in the perturbative
regime. In order to ensure the applicability of perturbative matching
formulae, the momenta used in the quasi-PDF determination need  to be very
large in order to neglect the $\nu$ dependence in the second
argument. In contrast, the pseudo-PDF is the integral along the
line $z^2 = \rm{const}$. The single scale makes the validity of
perturbation theory, or lack there of, more transparent. A
verification of the validity of the perturbative formula will be necessary
for any lattice-calculated PDF to be believable.

\subsection{Reduced distribution}
In order to improve the calculation of the ITD from the pseudo-ITD, it has been suggested~\cite{Orginos:2017kos} to remove the $O(z^2 \Lambda^2_{\rm QCD})$ contributions by considering the reduced pseudo-ITD
\beq\label{eq:redu}
\mathfrak{M}(\nu,z^2) = \frac{\mathcal{M}(\nu,z^2)}{D(z^2)} \,.
\eeq
Ideally, $D(z^2)$ will contain all the non-trivial $z^2$ dependence from higher twist effects. The chief requirement is that in the $z^2\to 0$ limit, $D(z^2)$ approaches a non-zero finite constant. This feature ensures that the OPE for this reduced pseudo-ITD will be the same as the original pseudo-ITD,  and factorization will lead to the same PDF moments. For future purposes in a lattice calculation, the choice of $D(z^2) = \mathcal{M}(0,z^2)$ is particularly useful and in the following this choice will be assumed. An alternative choice has been proposed in~\cite{Braun:2018brg}. The authors claim that if one used a vacuum matrix element of the same operator, instead of the rest frame hadron matrix element, then the pseudo-PDF would have smaller higher twist corrections in the limit of $x\to1$.

In order to take a continuum limit, the operator $O_{\rm WL}= \bar\psi(z) \gamma^\alpha U(z;0) \psi(0)$ must first be renormalized. In QCD, the renormalization constant of a Wilson line is proportional to $e^{-gm\frac{z}a + c(z)}$ which when expanded in $g$ gives a power divergence in perturbation theory and $c(z)$ is a term which depends on the number of cusps in the Wilson line and the specific angles formed at the cusps~\cite{Dotsenko:1979wb,Brandt:1981kf}. The full Wilson line operator $O_{\rm WL}$ is multiplicatively renormalizable~\cite{Ishikawa:2017faj}. This convenient feature allows the ratio in Eq.~\eqref{eq:redu}, with the choice $D(z^2) = \mathcal{M}(0,z^2)$, to have a finite continuum limit. The renormalization constants in the numerator and denominator, which only depend on the length and shape of the Wilson line, not the external momentum, cancel exactly. Even more importantly, this ratio is Renormalization Group Invariant (RGI). Therefore, it can be factorized using the $\overline{\rm MS}$ scheme into the PDF and an appropriate Wilson coefficient contrary to what was claimed by~\cite{Rossi:2017muf, Rossi:2018zkn}.

\subsection{$z^2$ evolution and $\overline{\rm MS}$ matching}\label{sec:MS_pseudo}
\begin{figure}[ht]
\includegraphics[width=3in]{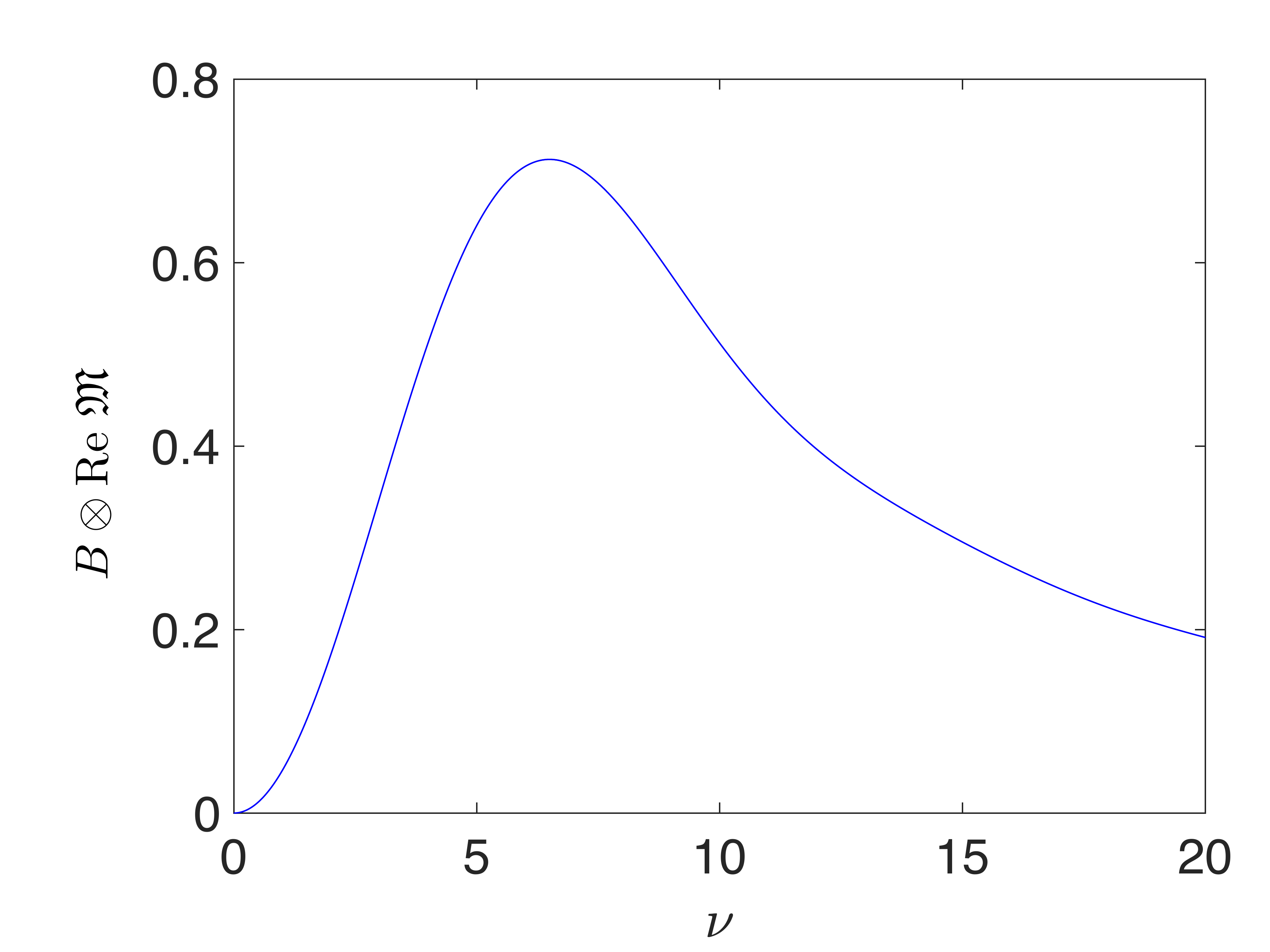}
\includegraphics[width=3in]{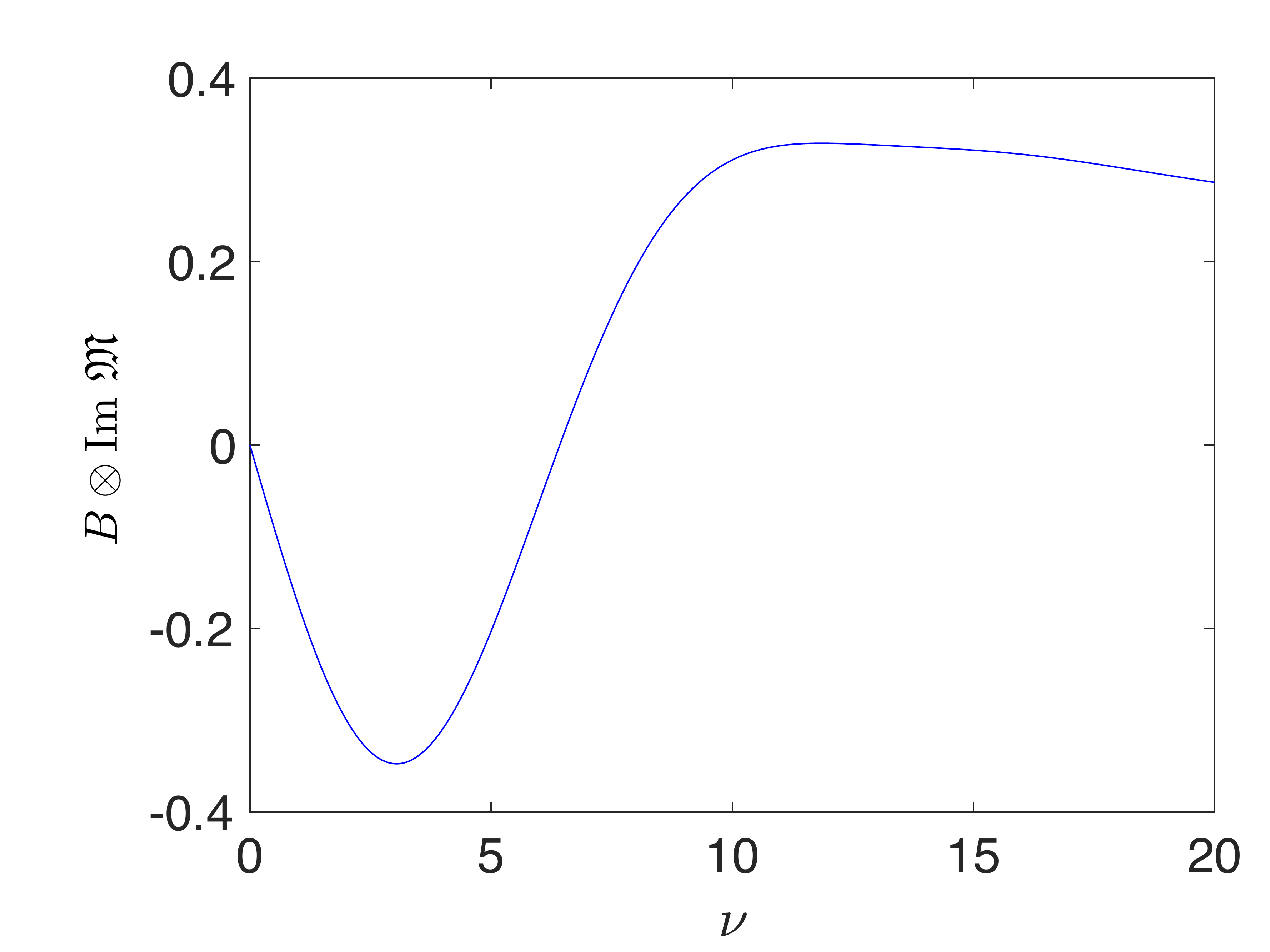}
\includegraphics[width=3in]{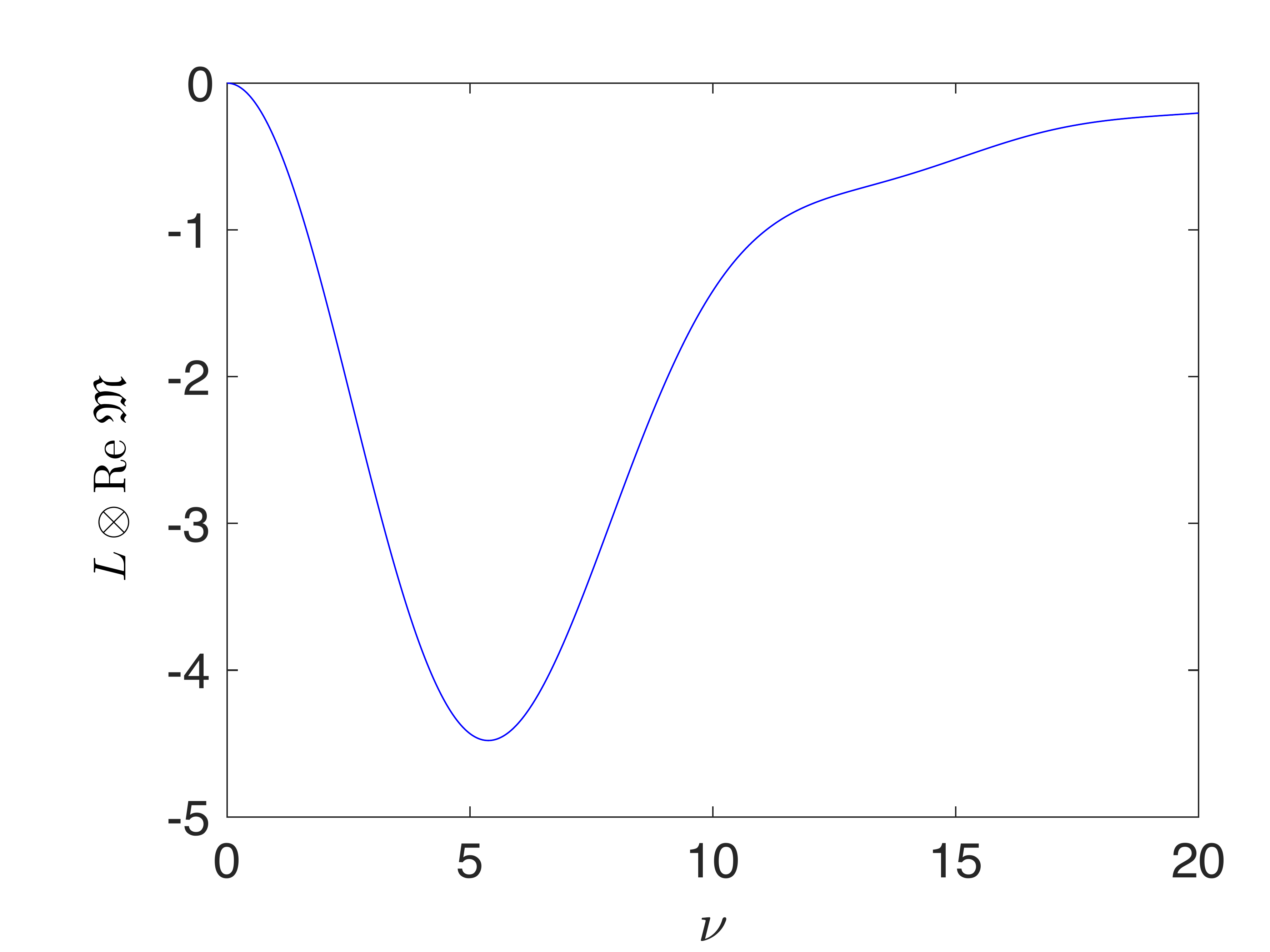}
\includegraphics[width=3in]{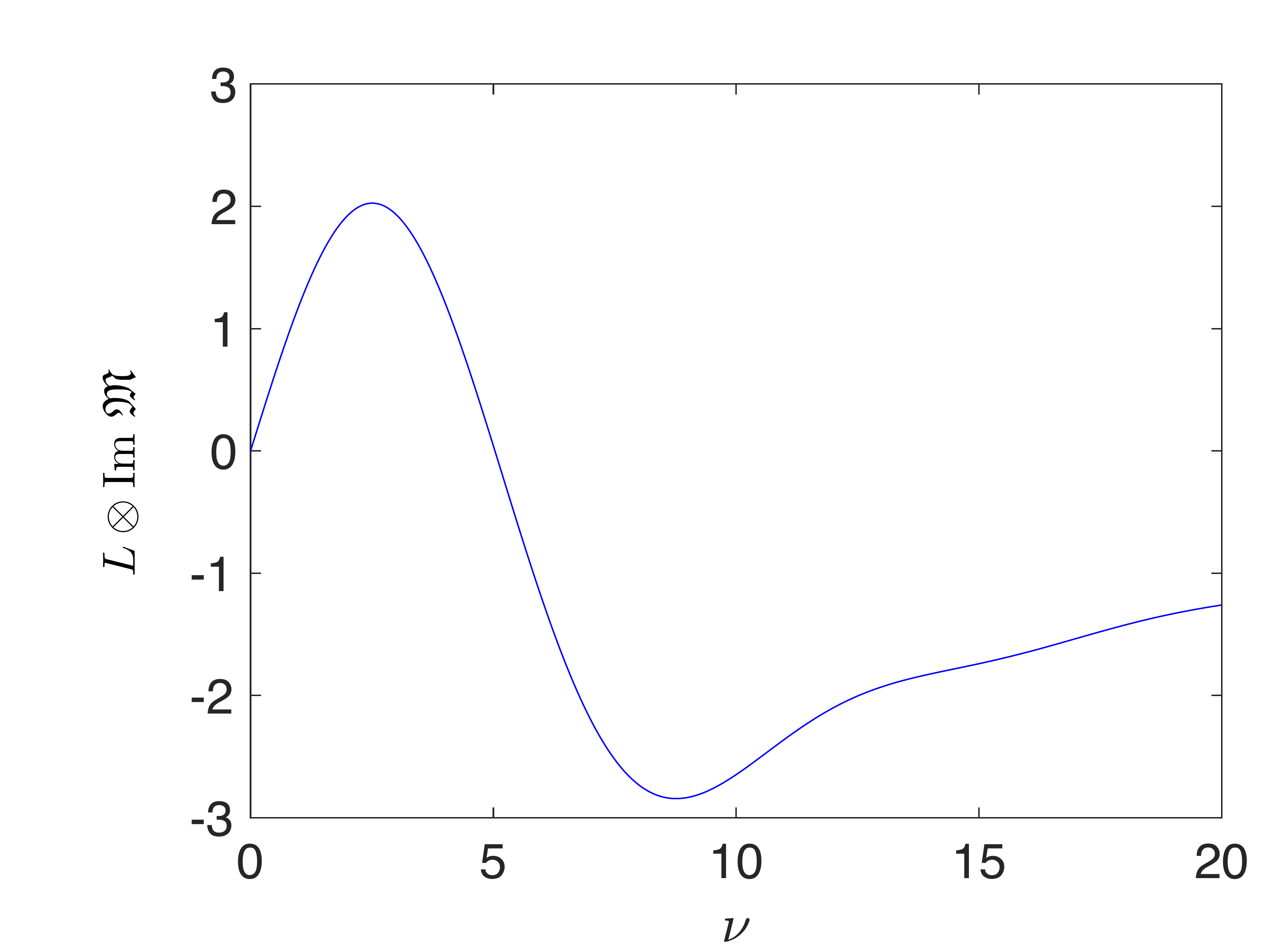}
\caption{The upper and lower figures show the convolutions for the evolution of Eq.~\protect\ref{eq:evo}, and  for the matching to $\overline{\rm MS}$ using the kernel of Eq.~\protect\ref{eq:lu}, respectively.
\label{fig:convo_theory}}
\end{figure}

All phenomenological PDF fits are performed in the $\overline{\rm MS}$
scheme either with NLO or NNLO truncation in the perturbative series
of the Wilson coefficients. Since the reduced pseudo-ITD is RGI, its
$z^2$ dependence is independent of any particular scheme, but its
dependence on this scale must match the $\mu$ dependence of the
$\overline{\rm MS}$ ITD. The $O(\alpha_s)$ perturbative $z^2$ evolution
equation is given by~\cite{Radyushkin:2017cyf}
\beq\label{eq:evo} z^2
\frac{ d\mathfrak{M}}{dz^2}(\nu,z^2) = -\frac{\alpha_s}{2\pi}C_F
\int_0^1 du\, B(u) \mathfrak{M}(u\nu,z^2)\,,
\eeq
where $B(u) = \left[
  \frac{1+u^2}{1-u}\right]_+$ is the Altarelli-Parisi kernel. This
equation is the pseudo-ITD's analog of the ITD's DGLAP evolution
equation.
Solving Eq.~\eqref{eq:evo} in the leading log approximation, as  in~\cite{Orginos:2017wcl}, the reduced pseudo-ITD at various scales $z^2$ can be evolved to the common scale $z_0^2$ by applying
\beq \label{eq:pseudo_evo}
 \mathfrak{M}(\nu,z_0^2) = \mathfrak{M}(\nu,z^2) + \ln \left (\frac{z^2}{z_0^2} \right )\frac{\alpha_s C_F}{2\pi} B \otimes \mathfrak{M}(\nu,z^2),
 \eeq
where
\beq
B \otimes \mathfrak{M}(\nu,z^2) = \int_0^1 du \bigg[\frac{1+u^2}{1-u}\bigg]_+ \mathfrak{M}(u\nu,z^2).
\eeq
As was done
in~\cite{Orginos:2017wcl}, one can estimate the effects of evolution
and matching by performing the convolution on a model reduced
pseudo-ITD. Consider the pseudo-ITD for the
model pseudo-PDF 
${\cal P}(x,z^2) = x^a (1-x)^b / B(a+1,b+1)$, where $B(x,y)$ is the Beta function, for $a=0.5$ and $b=3$. The convolutions for the DGLAP kernel are shown in Fig.~\ref{fig:convo_theory}.

Due to this logarithmic divergence in $z^2$, the determination of the PDF from the pseudo-ITD is not as straightforward as a simple limit of $z^2\to0$. 
At the leading-twist level, 
there exists a factorization relationship 
\beq
\mathfrak{M}(\nu,z^2) = \int_0^1 du~ K(u,z^2\mu^2, \alpha_s) Q(u \nu,\mu^2)\
\eeq
between the reduced pseudo-ITD $\mathfrak{M}(\nu,z^2) $ 
and the $\overline{\rm MS}$ ITD {$Q(\nu,\mu^2)$} defined as
 \be\label{eq:MSpdf} {Q}(\nu,\mu^2) = \int_{-1}^1 dx
e^{i\nu x} {f}(x,\mu^2) \,.  \ee 

The kernel $K$ for matching the pseudo-ITD to the $\overline{\rm MS}$ ITD has been calculated at NLO~\cite{Radyushkin:2018cvn,Zhang:2018ggy,Izubuchi:2018srq}
\beq\label{eq:pseudo_kernel}
K(u,z^2\mu^2, \alpha_s) = \delta(1-u) + \frac{\alpha_s C_F}{2\pi} \left[ \ln \left  (z^2\mu^2 \frac{e^{2\gamma_E +1}}4 \right ) B(u) + L(u) \right]   \,,
\eeq
where 
\beq\label{eq:lu}
L(u) = \left[ 4 \frac{\ln(1-u)}{1-u} - 2(1-u) \right]_+
\eeq
is a scale independent piece related to the specific choice of the $\overline{\rm MS}$ scheme. This gives the full NLO matching relationship
\beq\label{eq:pITD_to_ITD}
\mathfrak{M}(\nu,z^2) = Q(\nu,\mu^2) + \frac{\alpha_s C_F}{2\pi} \int_0^1 du \left[ \ln (z^2\mu^2 \frac{e^{2\gamma_E +1}}4) B(u) + L(u) \right] Q(u\nu,\mu^2)  \,.
\eeq
The convolution of $L(u)$ for the pseudo-PDF model is shown in Fig.~\ref{fig:convo_theory}. The scale dependent part of the kernel can be identified as the previously mentioned evolution process to a scale $z_0^2 = 4 e^{-2 \gamma_E-1} \mu^{-2}$. The reduced pseudo-ITD at many scales could be directly matched to the $\overline{\rm MS}$ ITD in a single step. The separation of the evolution and matching procedure can allow for the two steps to take into account the higher twist contamination in different ways. 

\subsection{Moments of pseudo-PDF}
\label{sec:mom}
Taylor expanding the right-hand-side of Eq.~\eqref{eq:ppdf} about $\nu=0$ provides access to the moments of the pseudo-PDF
\be
\mathfrak{M}(\nu,z^2) = \sum_{n=0}^\infty i^n \frac{\nu^2}{n!} b_n(z^2)\,,
\label{eq:pPDF_mom}
\ee where \be b_n(z^2) = \int_0^1 dx x^{n} {\cal P}(x,z^2) \,.  \ee These
moments 
can be related
 to the Mellin moments of the PDF through the OPE
of the reduced pseudo-ITD in the limit $z^2\Lambda_{\rm QCD} ^2 \ll
1$~\cite{Karpie:2018zaz}. The leading contribution in the OPE of the
reduced pseudo-ITD is given in terms of Mellin moments of the PDF,
$a_n(\mu^2)$, \be\label{eq:leading_ope} \mathfrak{M}(\nu,z^2) =
\sum_{n=0}^\infty i^n \frac{\nu^n}{n!}a_n(\mu^2) K_n(\mu^2 z^2) +
O(z^2) \,, \ee where $O(z^2)$ schematically represents terms at
sub-leading power in the twist expansion and $K_n(\mu^2 z^2)$ are the
Wilson coefficients of the moments. These Wilson coefficients are the
Mellin moments of the matching kernel in
Eq.~\eqref{eq:pseudo_kernel}. By comparing the $\nu$ dependencies of
Eq.~\eqref{eq:pPDF_mom} and Eq.~\eqref{eq:leading_ope}, one can find
the matching relationship between the pseudo-PDF's moments and the
PDF's moments, \beq b_n(z^2) = K_n(\mu^2 z^2) a_n(\mu^2)\,.  \eeq This
matching relationship is multiplicative unlike the convolution for
matching the pseudo-ITD to the ITD. These advantages of the
representation in Mellin space have been exploited to calculate high-order
evolution of the PDF~\cite{Vogt:2004mw,Moch:2004pa}.

As given in~\cite{Karpie:2018zaz}, the NLO Wilson Coefficients in Eq.~\eqref{eq:leading_ope} are given by
\begin{equation}\label{eq:pmom_match}
K_n(z^2\mu^2,\alpha_s) =  1  -  \frac{\alpha_s}{2\pi} C_F \left[\gamma_n \ln\left(z^2\mu^2\frac{e^{2\gamma_E +1}}{4}\right) + l_n\right]\,,
 \end{equation}
where
\begin{equation} 
\gamma_n = \int_0^1 du\, B(u) u^n= \frac{1}{(n+1) (n+2) } - \frac{1}{2}  
- 2 \sum_{k=2}^{n+1}\frac{1}{k}\,, 
\end{equation}
which  are the well known moments of the Altarelli-Parisi kernel, and

\begin{equation}
l_n = \int_0^1 du\, L(u) u^n=2\left[ \left(\sum_{k=1}^n \frac{1}{k}\right)^2 + \sum_{k=1}^n \frac{1}{k^2}
+\frac12 - \frac{1}{(n+1)(n+2)} \right]
\,.
\end{equation}
With the Wilson coefficients computed, we can now obtain the $\overline{\rm MS}$ moments up to ${\cal O}(\alpha_s^2,z^2)$ directly from the reduced function ${\mathfrak M}(\nu,z^2)$ as
\begin{equation}
a_{n}(\mu^2) = (-i)^n\frac{1}{K_n(z^2\mu^2,\alpha_s)}\left.\frac{\partial^n\,{\mathfrak M}(\nu,z^2)}{\partial \nu^n} \right|_{\nu=0}
 + O (z^2\Lambda^2_{\rm QCD},\alpha_s^2)\,.
 \label{eq:msbarMOM}
\end{equation}

There are other analogous proposals for the calculation of moments of parton distributions from Lattice QCD, which do not use local twist-2 matrix elements. Using two spatially separated current operators,~\cite{Detmold:2005gg,Chambers:2017dov}, one can extract the moments of the DA or of the PDF. These ideas, as well as the above technique for determining moments, go under the name of  ``OPE without OPE'', where one calculates a non-local operator in order to estimate the moments defined by local operators~\cite{Martinelli:1998hz}. This procedure is particularly useful when the local operators are subject to some systematic difficulty such as the power divergent mixing of the twist-2 operators for PDF moments. In principle, one could calculate all PDF moments from the pseudo-ITD, though this fact is not necessarily true in practice due to the discretized values of $z$ and $p$ which are available to lattice calculations.

\section{Details of the Lattice Calculation}\label{dlc}
The numerical calculation is performed on three different ensembles of lattice QCD configurations. The ensembles that are used in this article were generated by the JLab/W\&M
collaboration~\cite{lattices} employing 2+1 flavors of stout-smeared clover Wilson fermions and a tree-level
tadpole-improved Symanzik gauge action. 

In the fermionic action, one iteration of stout smearing was used
with the weight $\rho = 0.125$ for the staples. A direct consequence of this smearing
is that the tadpole corrected tree-level clover coefficient
$c_{\rm SW}$ used is very close to the non-perturbative value determined,
a posteriori, employing the Schr\"odinger functional
method~\cite{lattices}. The parameters used for the three ensembles are listed in Tab.~\ref{tab:lat}. The lattice spacings, $a$, are estimated using the Wilson flow scale $w_0$ described in Ref \cite{w0}.

\begin{table}
\begin{tabular}{ l | c c | c c | c c | c c }
ID & ~$a$(fm)~ & ~$M_\pi$(MeV) & ~$\beta~$ & ~$c_{\rm SW}$~ & ~$am_l$~ & ~$am_s$~ & $L^3 \times T$ & $N_{cfg}$\\
\hline\hline
$a127m415$ & 0.127(2) & 415(23) & 6.1 & 1.24930971 & -0.2800 & -0.2450 & $24^3 \times 64$ & $2147$ \\
\hline
$a127m415L$ & 0.127(2) & 415(23) & 6.1 & 1.24930971 & -0.2800 & -0.2450 & $32^3 \times 96$ & $2560$\\
\hline
$a094m390$ & 0.094(1) & 390(71) & 6.3 & 1.20536588 & -0.2350 & -0.2050 & $32^3 \times 64$ & $417$\\
\hline \hline
\end{tabular}
\label{tab:lat}
\caption{\footnotesize Parameters for the lattices generated by the JLab/W\&M collaboration \cite{lattices} using 2+1 flavors of stout-smeared clover Wilson fermions and a tree-level tadpole-improved Symanzik gauge action. The lattice spacings, $a$, are estimated using the Wilson flow scale $w_0$. Stout smearing implemented in the fermion action makes the tadpole corrected tree-level clover coefficient $c_{\rm SW}$ used to be very close to the value determined non-pertubatively with the Schr\"odinger functional method. The $a127m415L$ contains 10 independent streams of 256 configurations each. The other ensembles only contain a single stream.}

\end{table}

\subsection{Momentum smearing}
In order to improve the overlap of the interpolating fields on the nucleon ground state, smearing procedures are performed on the quark fields. Standard Gaussian smearing, which was used in the previous study of pseudo-ITDs, can help improve the overlap with the state at rest, but deteriorates the overlap with states of higher momenta. The momentum smearing procedure \cite{Bali:2016lva} changes the smearing operation to improve the overlap of the interpolating field to nucleon ground states with arbitrary momenta. The quark fields are transformed by
\beq
\tilde{q}(x) =q(x) + \rho \sum_k \bigg(2 q(x)-e^{i\left(\frac{2\pi}L\right)\vec{\zeta} \cdot \hat k} U_k(x) q(x+k)-e^{i\left(\frac{2\pi}L\right)\vec{\zeta} \cdot \hat k} U_k(x-k) q(x-k)\bigg) \, ,
\eeq
where $\rho$ and $\mathbf{\zeta}$ are tunable parameters. This transformation has the same form as used in standard Gaussian smearing with an extra phase, $e^{i \mathbf{\zeta} \cdot  \hat{k}}$ multiplying the gauge links. Here, $\rho$ plays the same role as in Gaussian smearing and $\zeta$ determines which momentum states the quark field will predominantly overlap with. 

As suggested in previous works, this procedure was implemented using the existing iterative Gaussian smearing routines, but using a set of rotated gauge links $\tilde{U}_k(x) = e^{i \mathbf{\zeta} \cdot \hat k} U_k(x)$ in order to account for the phase. Unlike previous works with the momentum smearing technique, the momentum smearing parameter, $\mathbf{\zeta}$, was not chosen to be dependent on the nucleon momentum. 
Also, unlike the sequential source technique, the matrix element extraction based upon the Feynman-Hellmann theorem allows for calculating several nucleon momentum states without additional propagator inversions. Fixing the smearing parameter as a fraction of the momentum would greatly increase the cost of this calculation by requiring different propagator calculations for each momentum  used. The parameters were chosen to overlap with the higher momenta states which had a poor signal using standard Gaussian smearing. The choice of $\zeta$ was made by maximizing the signal-to-noise ratio of the 2-point correlation function for the desired range of momenta.

\begin{figure}[h]  \centering

\includegraphics[width=0.49\textwidth]{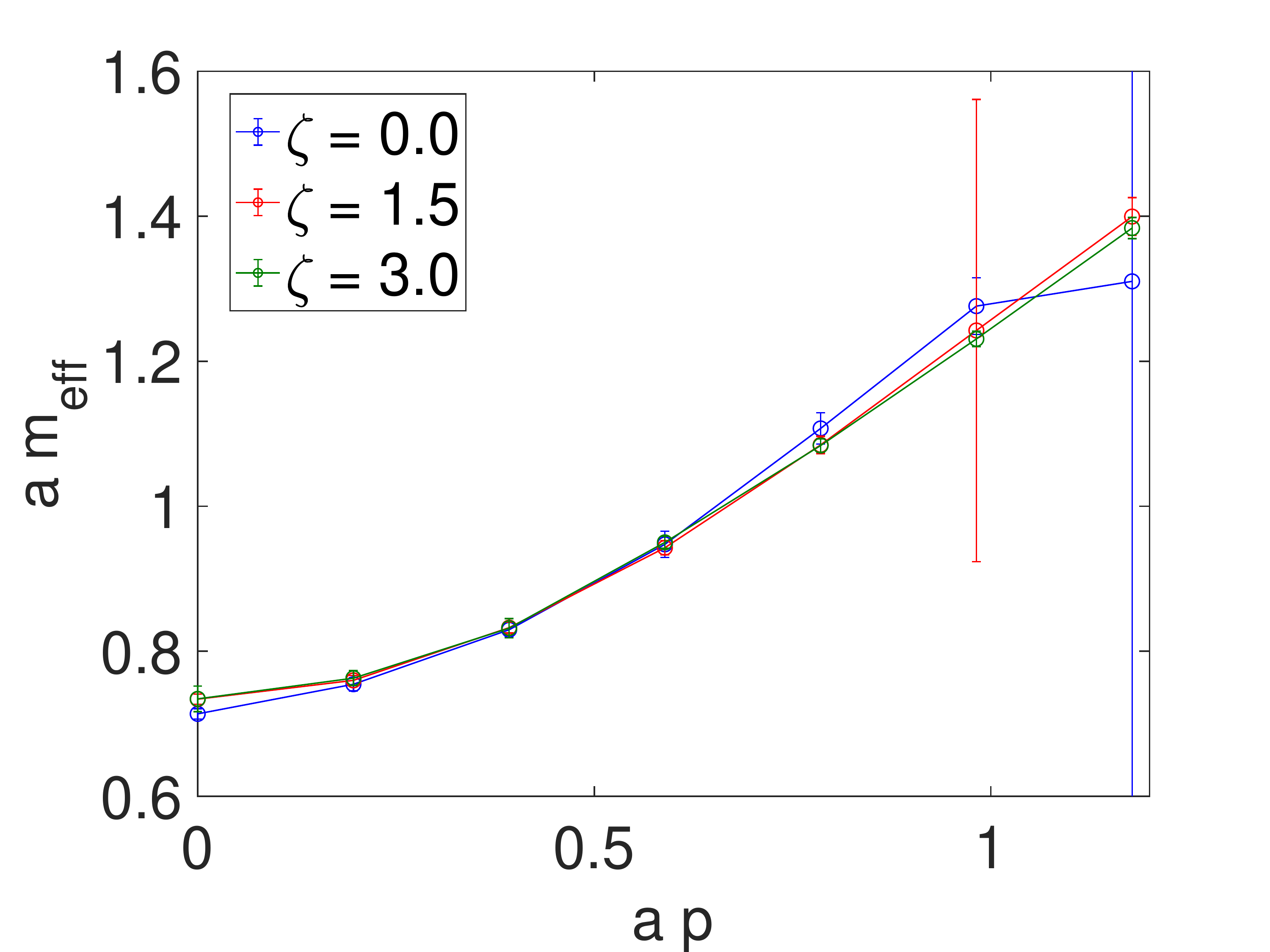}
\includegraphics[width=0.49\textwidth]{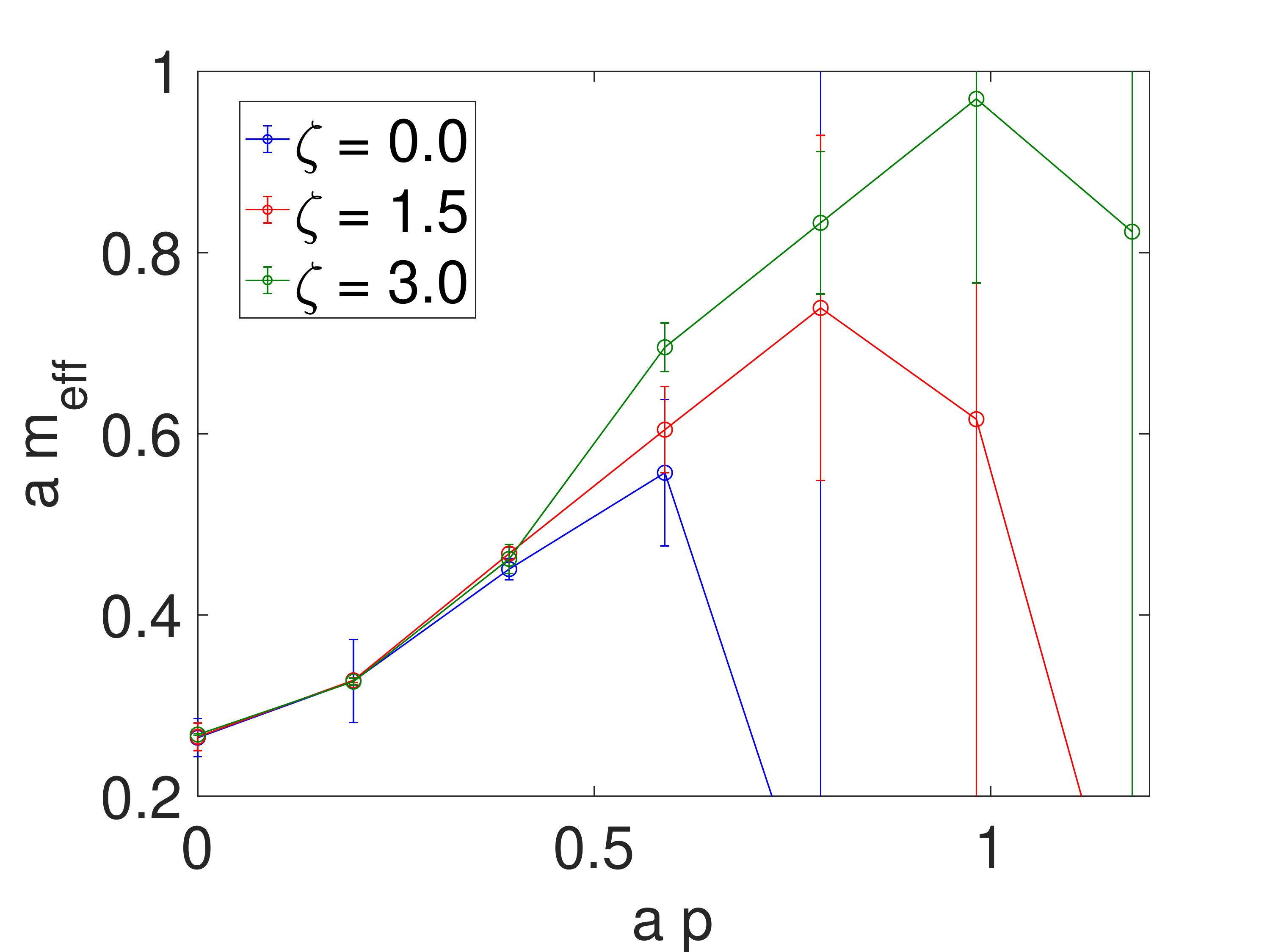}
    \caption{ The effective mass for a nucleon (left) and pion (right) two point functions at various momenta from the ensemble $a127m415L$.  The signal-to-noise ratio for the effective mass, particularly for the pion, decays without momentum smearing. Even without tuning $\zeta$ for each momentum, the momentum smearing procedure significantly improves the signal and allows for precision determination of high momentum effective masses which were not attainable with Gaussian smearing alone. }   \label{fig:eff_mass}
\end{figure}

In order to demonstrate the efficacy of the momentum smearing procedure, the effective masses of the pion and nucleon were calculated on the $a127m415L$ ensemble, plotted in Fig.~\ref{fig:eff_mass}. Each of these momentum smearings was set to improve the signal-to-noise ratio for a different range of nucleon momentum states. As can be seen the smearing without the momentum phase performs poorly for the highest momentum states, the intermediate $\zeta$ improves the signal for the middle range of momenta, and the largest $\zeta$ performs significantly better at the highest momentum states than the other two. It should be noted that momentum smearing only alleviates one of the sources of decaying signal-to-noise ratio. Momentum smearing improves the overlap of the ground state with the operator at non-zero momentum which allows for more time sources to be available before machine precision issues arise. The other source is the variance inherent in the correlation functions, which decays exponentially in $T$ with a rate determined by the hadron's energy
\beq
R_{\rm S/N} = \frac{C_2(T)}{{\rm std}\left[C_2(T)\right]} \sim e^{-(E - \frac{n_q}2 m_\pi)T},
\eeq
where $n_q$ is the number of quarks and anti-quarks in the interpolating operator. This variance will not be affected by the momentum smearing procedures, because it is an inherent property of the theory. 

\subsection{Feynman-Hellmann Matrix element extraction}
The extraction of the matrix element is performed with a method based on the Feynman-Hellmann theorem. Hadron matrix elements can be calculated by studying the time dependence of the ratio of 3-point and 2-point correlation functions. We refer the reader to~\cite{Bouchard:2016heu, Orginos:2017wcl} for more details on the implementation of this method. For completeness, the key points are highlighted below.

The Feynman-Hellmann theorem relates matrix elements of energy eigenstates to derivatives of the corresponding energy eigenvalue with respect to a parameter of the theory,
\be \frac{d E_n}{d \lambda} = \langle n | \frac{d H_\lambda}{d\lambda}  | n \rangle \,. \ee 
In order to calculate matrix elements of arbitrary operators, following the procedure described in \cite{Bouchard:2016heu}, we consider a change of the action by adding the following term,
\be S_\lambda = \lambda \int d^4x j(x) \,, \ee
where $j(x)$ is some operator of interest.
The vacuum state is labeled by $| \lambda \rangle$. The 2-point correlation functions are defined as
\be C_\lambda(T) = \langle \lambda | O(T) O^\dagger(0) | \lambda \rangle = \frac{1}{Z_\lambda} \int D\Phi e^{-(S+S_\lambda)} O(T)  O^\dagger(0) \,, \ee
where $\Phi$ represents collectively the gauge and fermion fields of the theory, $O(t)$ is an interpolating field for a desired hadron, and $Z_\lambda$ is the partition function defined by 
\be
Z_\lambda = \int D\Phi e^{-(S+S_\lambda)}.
\ee

These three objects, $|\lambda \rangle$, $C_\lambda$, and $Z_\lambda$, become the true QCD vacuum, correlation function, and partition function respectively in the $\lambda \to 0$ limit.  In this formulation, the hadron matrix element of the operator $j(x)$ can be computed through the Feynman-Hellmann theorem as
\be \frac{\partial m^{h_n}_\lambda}{\partial \lambda} \big|_{\lambda=0} = \int dt' \langle h_n|J(t')|h_n\rangle \,, \ee
where $J(t) = \int d^3x j(\vec x, t)$ and $|h_n\rangle$ is the $n^{th}$ hadron state with the quantum numbers of $O(t)$ and mass $m^{h_n}$. In the large Euclidean-time limit, the ground state mass can be approximated by the effective mass 
\be m_\lambda^{\rm{eff}}(T,\tau) = \frac 1 \tau \ln \bigg(\frac{C_\lambda(T)}{C_\lambda(T+\tau)} \bigg) \xrightarrow[T\to\infty]{} \frac 1 \tau \ln(e^{m^{0}_\lambda \tau}) \,. \ee
The derivative of the effective mass can be shown to be
\be \label{eq:meffder}
\frac{\partial m_\lambda^{\rm{eff}}(T,\tau)}{\partial \lambda} \big|_{\lambda=0} = \frac 1 \tau (R(T+\tau) - R(T)) \, ,\ee
where \be R(t) = \frac{\int dt' \langle \Omega | T\{ O(T) J(t') O^\dagger (0)\} | \Omega \rangle}{C(T)}\, .\ee
The contributions to the time ordered matrix element in the numerator of $R(t)$ from the regions where the current is not inserted between the hadron states were shown in \cite{Bouchard:2016heu} to be suppressed for lattices with large time extent $N_t$ in the difference of the two terms which appear in the RHS of Eq.~\eqref{eq:meffder}. At large Euclidean time  $t$, Eq.~\eqref{eq:meffder} isolates the matrix element of interest up to exponentially small corrections,
\be
\frac{\partial m_\lambda^{\rm{eff}}(T,\tau)}{\partial \lambda} \big|_{\lambda = 0} = \frac{ \langle h^0 | J | h^0 \rangle}{2 E^h_0} \big(1 + A e^{-\Delta T} + B T e^{-\Delta T} + O(e^{-\Delta' T})\big)  \,.
\ee
where $A$ and $B$ are fit parameters related to the matrix elements of the lowest excited state with mass gap $\Delta$. The last term, $O(e^{-\Delta' T})$ represents the neglected higher state corrections. It should be noted that the higher state effects of this method are significantly smaller than the typical ratio method where excited state effects are of order $O(e^{-\Delta T /2})$.

This method of matrix element extraction has a number of advantages over the more common techniques. First, the summation over the
operator insertion time eliminates one of the independent variables of
the correlation function. This reduction allows for a clear
identification of excited state contamination.

Without the summation over operator insertion time, points with several insertion
  times are required to visually identify a plateau.  Furthermore,
  several source/sink separations are needed to unambiguously
  determine the ground-state matrix element.  A second advantage,
demonstrated in~\cite{Chang:2018uxx}, is that the matrix element
extraction can begin at much earlier times in contrast to what is
possible from other calculations, whose excited state effects are much
larger. This advantage is particularly important for physical mass
calculations where the excited state effects require long
time extents in order to be controlled.

\subsection{Desired lattice correlation functions}
All correlation functions were calculated with randomly determined source points. The correlation functions with smeared source operators and both point and smeared sink operators are constructed and each of these smearings will be performed for 3 different values of the momentum smearing parameter $\zeta$. All of these correlation functions will be simultaneously analyzed to extract the matrix element. The nucleon states were boosted up to a maximum momenta of $p_{\rm{max}} = \pi / 2a$, except for the ensemble $a127m415L$ in which the maximum momentum was $p_{\rm{max}}= 3 \pi / 8a$. Within these ranges of momenta, the continuum energy dispersion relation is still reasonably satisfied by the lattice calculation within errors as can be seen in Fig~\ref{fig:eff_mass}. The momentum smearing technique is used to allow for more precise access to the high momenta correlation functions.

To calculate the matrix element for the nucleon Ioffe time distribution, the relevant 2-point correlation function is defined by
\be
C_2(p,T) = \langle N_p(T) \overline{N}_p(0)\rangle,
\ee
where $N_p$ is a helicity averaged interpolating field of a nucleon with momentum $p$ and $T$ is the Euclidean time separation between the interpolating operators for the nucleon creation and annihilation operators. The quark fields in $N_p$ have all been smeared using Gaussian momentum smearing in order to improve the overlap with the boosted ground state. The relevant 3-point correlation function is defined by
\be C_3(p,z,T) =  \langle N_p(T) O_{\gamma_4}(z) \overline{N}_p(0)\rangle,\ee
where $O_{\Gamma}(z) =  \bar \psi(0) \Gamma W(0;z) \tau_3 \psi(z)$. In
this extraction method, the time of this operator insertion is summed
over.  The  flavor isospin Pauli matrix $\tau_3$ is   used  in order to create the
iso-vector quark combination, $u-d$. This step is performed to avoid
the potentially costly additional calculation of disconnected
diagrams. The effective bare matrix element, $M(z\cdot p , z^2,T)$, is
defined by 
\be
M^{\rm{eff}}(p,z, T) = (2E_N) \big( R(p,z,T+1) - R(p,z,T) \big ),
\ee
where $E_N$ is the nucleon energy and \be R(p,z,T) = C_3(p,z,T) / C_2(p,T). \ee
The factor of $2E_N$ in the definition of the bare matrix element will be used to cancel the factor of $2p_4$ in the pseudo-ITD definition from Eq.~\eqref{eq:pseudo_lor_decomp}, such that the difference of ratios will be an effective bare pseudo-ITD
\beq
\mathcal{M}^{\rm eff}(\nu,z^2,T) = \big( R(p,z,T+1) - R(p,z,T) \big)\,.
\eeq
The separation of the quark fields and the nucleon momentum are both chosen along the $\hat z$ axis, $z=z_3$ and $p=p_3$. The bare matrix element is determined through the large time asymptotics of the effective matrix element, 
\be \mathcal{M}^0(\nu, z^2) = \lim_{T\to \infty} \mathcal{M}^{\rm{eff}}(\nu, z^2,T).\ee
As found in~\cite{Bouchard:2016heu}, the bare matrix element extracted with this method has different excited state effects than in a typical calculation of a three point correlation function. The bare matrix element will have contamination from higher state effects proportional to $e^{-\Delta T}$ and  $T e^{-\Delta T}$ where $\Delta$ is the energy gap between nucleon's the ground state and the first excited state. These terms will be included in the matrix element fit to control these effects in the low $T$ region.
 
\subsection{Fits and Excited States Effects}
The simplest method to extract the reduced pseudo-ITD from the effective matrix element is to analyze the large time limit with a mean value in a region where the effective reduced matrix element has a plateau. These data have plateau regions, especially at low momenta, where this procedure could be performed. A more sophisticated method can be used to take into account the effects from the lowest excited states which would provide a more justified account of the systematic errors. The effective bare matrix element has excited-state contamination of the form
\be
\mathcal{M}^{\rm{eff}}(\nu, z^2, T) = \mathcal{M}^0(\nu,z^2)( 1  + A_p(z^2) e^{-\Delta_p T} + B_p(z^2) T e^{-\Delta_p T}) \ , 
\label{eq:fitform}
\ee
where $A_p(z^2)$ and $B_p(z^2)$ are the contributions from the matrix elements containing the first excited state and $\Delta_p$ is the effective energy gap between the ground state and the first excited state. The coefficients of the excited state term are, in general,  correlation function dependent, i.e. different for each set of smearing parameters. 

Further simplifications of this functional form are possible to reduce
the number of fit parameters. One could adopt the lattice energy
dispersion relation to fix $\Delta_p$ to $\Delta_0$ removing another
parameter. The rest frame energy could be extracted from two point
functions prior to the study of the matrix element or left as a fit
parameter common to all matrix elements. The latter option, requiring
a simultaneous fit of many matrix elements for the sake of fixing a
single parameter, may not be too practical. Another choice is to
perform the spectroscopy fits on the  2-point functions to find the
energy gaps and hold those values fixed in a fit for the coefficients
of the exponentials.

\begin{figure}[ht]
\includegraphics[width=3in]{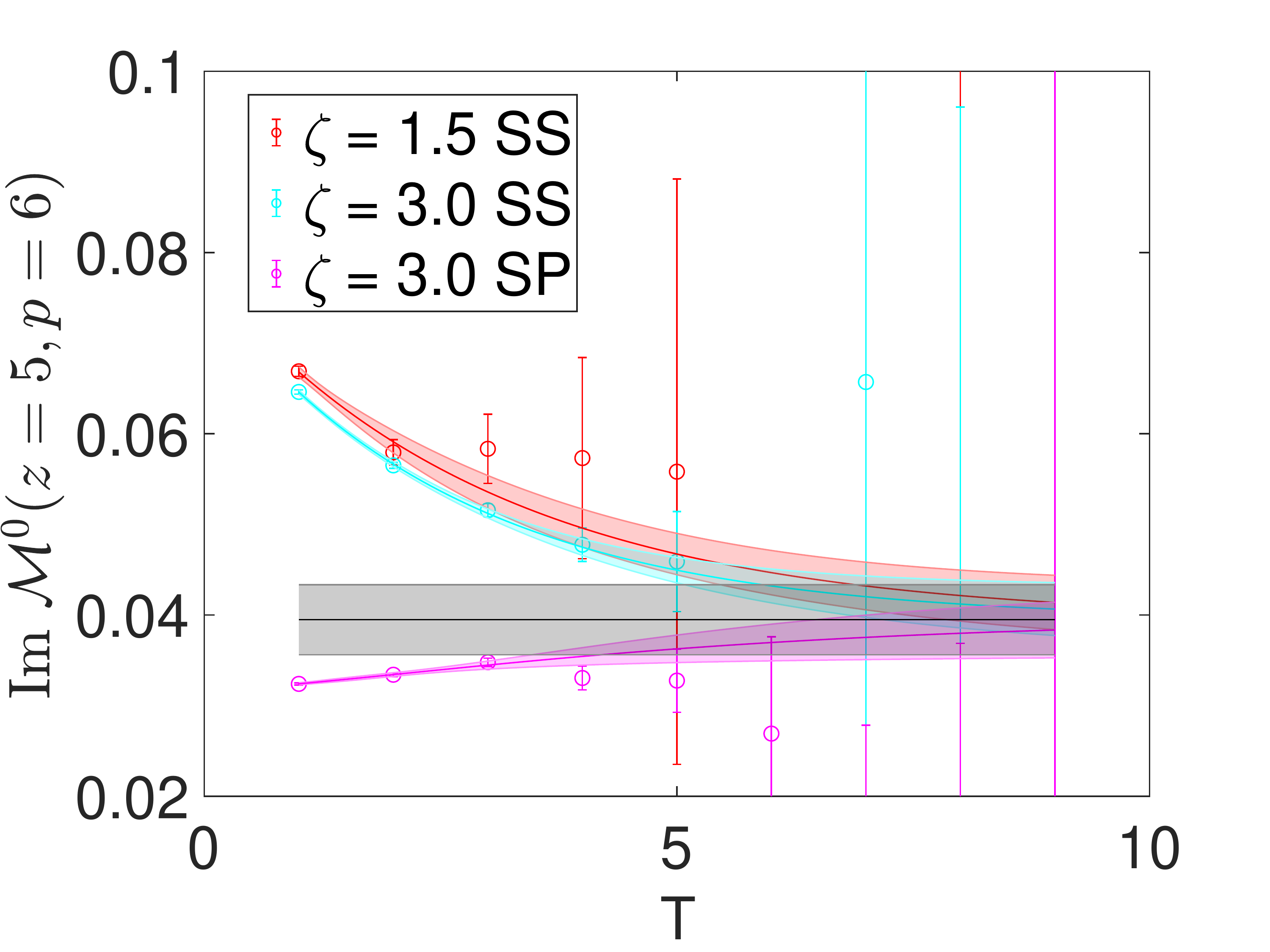}
\includegraphics[width=3in]{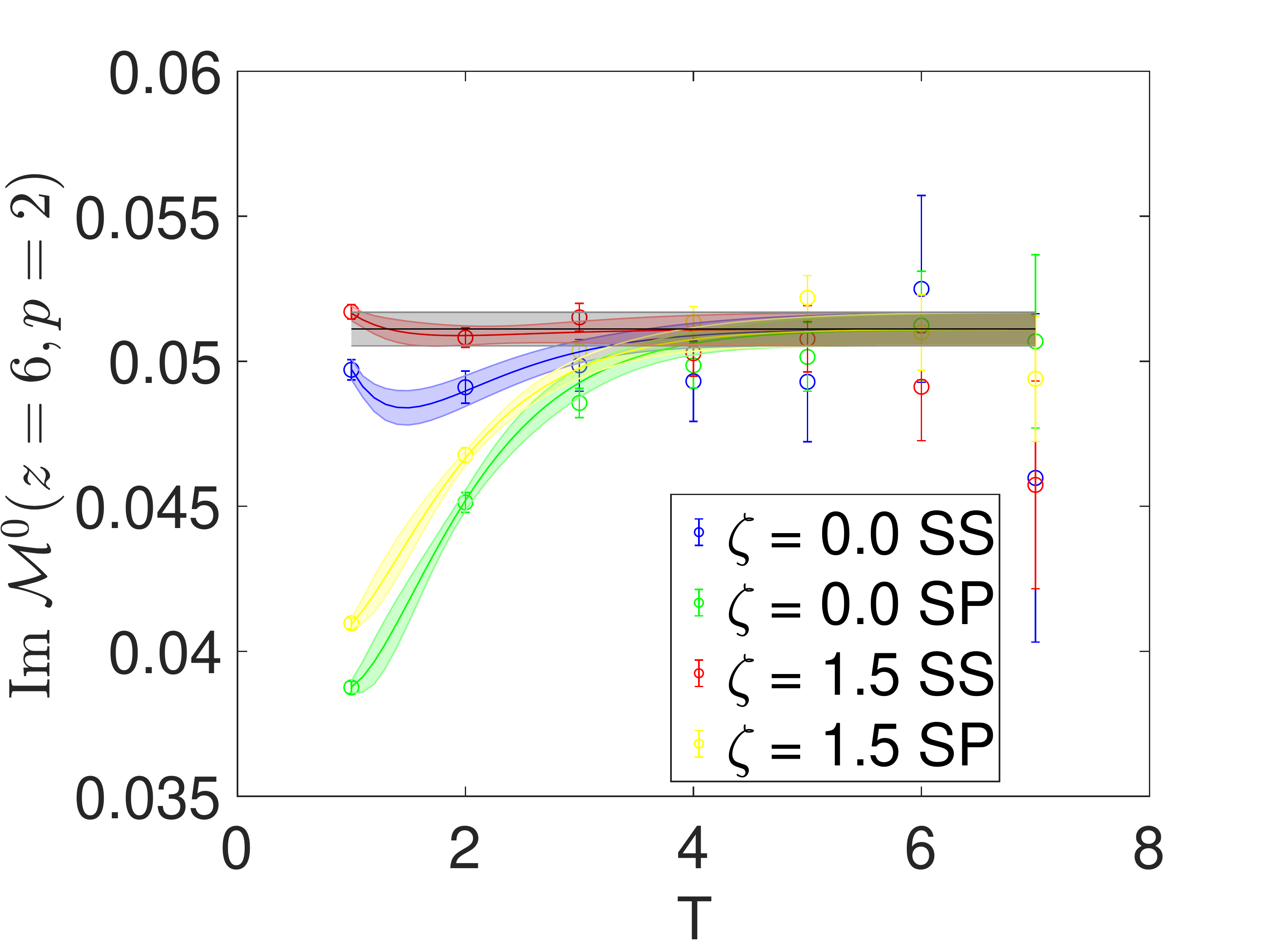}
\caption{Typical fits of the reduced Ioffe time pseudo-distribution from the ensemble $a127m415$. The left plot corresponds to the imaginary component with $z_3=5 a$ and $p_3 = 6 (2 \pi/L)$. The right plot corresponds to the imaginary component with $z_3=6 a$ and $p_3 = 2 (2 \pi/L)$. The color points and bands correspond to the different correlation functions used in the fit and the resulting fit respectively. The grey band corresponds to the extracted matrix element. }
\label{fig:fit_coarse}
\end{figure}

\begin{figure}[ht]
\includegraphics[width=3in]{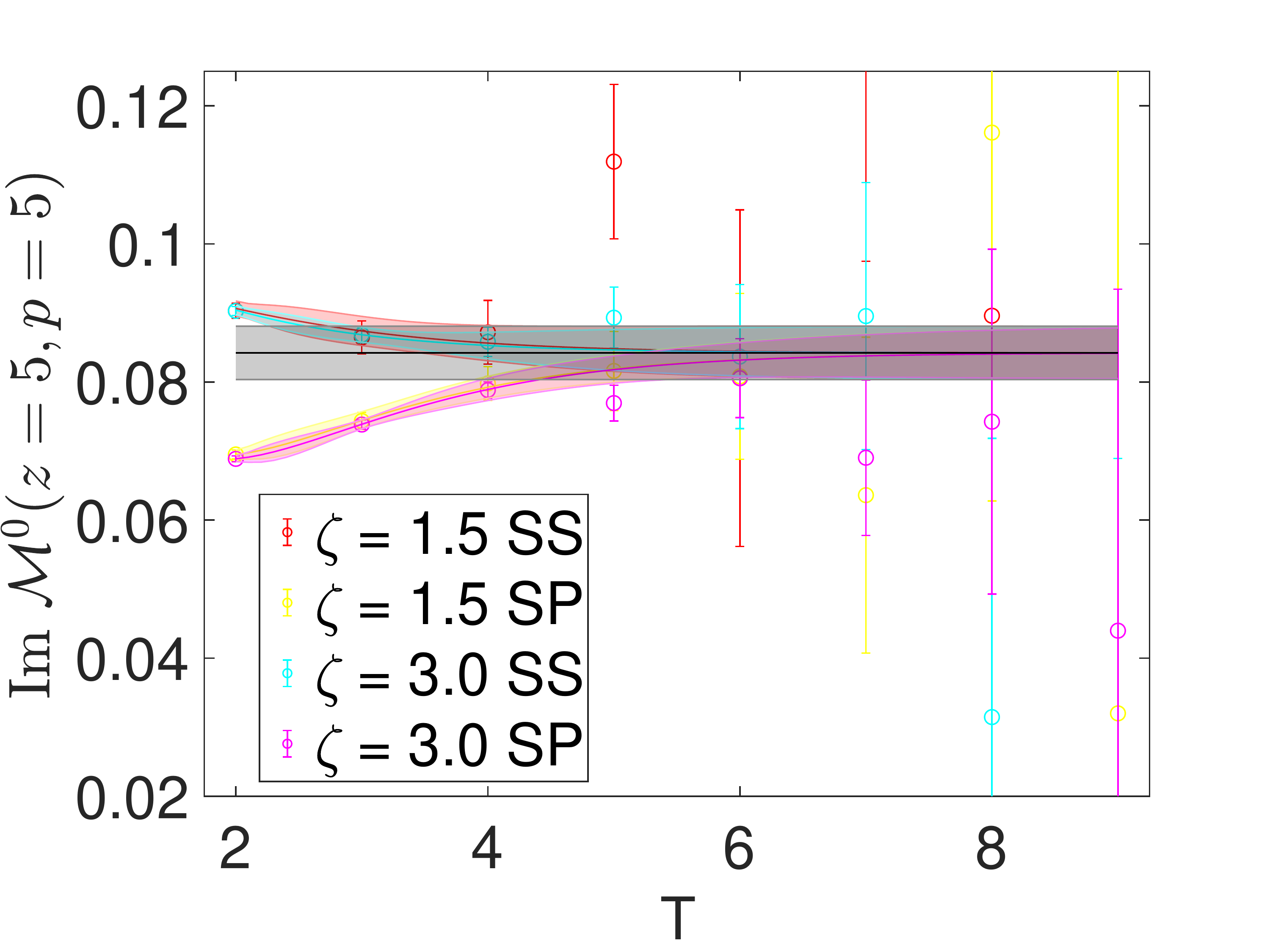}
\includegraphics[width=3in]{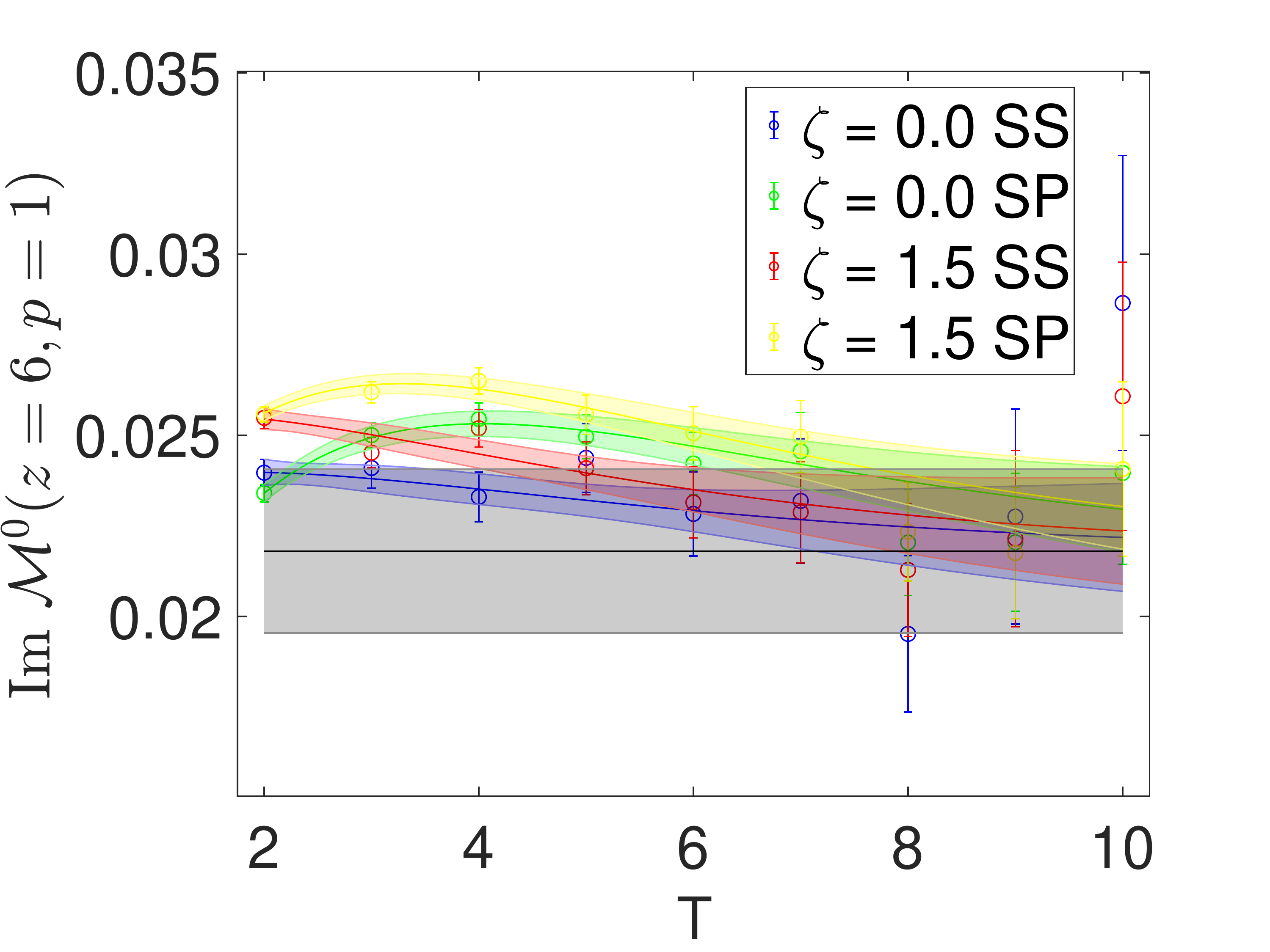}
\caption{Typical fits of the reduced Ioffe time pseudo-distribution from the ensemble $a127m415L$. The left plot corresponds to the imaginary component with $z_3=5 a$ and $p_3 = 5 (2 \pi/L)$. The right plot corresponds to the imaginary component with $z_3=6 a$ and $p_3 = 1 (2 \pi/L)$. The color points and bands correspond to the different correlation functions used in the fit and the resulting fit respectively. The grey band corresponds to the extracted matrix element. }
\label{fig:fit_big}
\end{figure}

\begin{figure}[ht]
\includegraphics[width=3in]{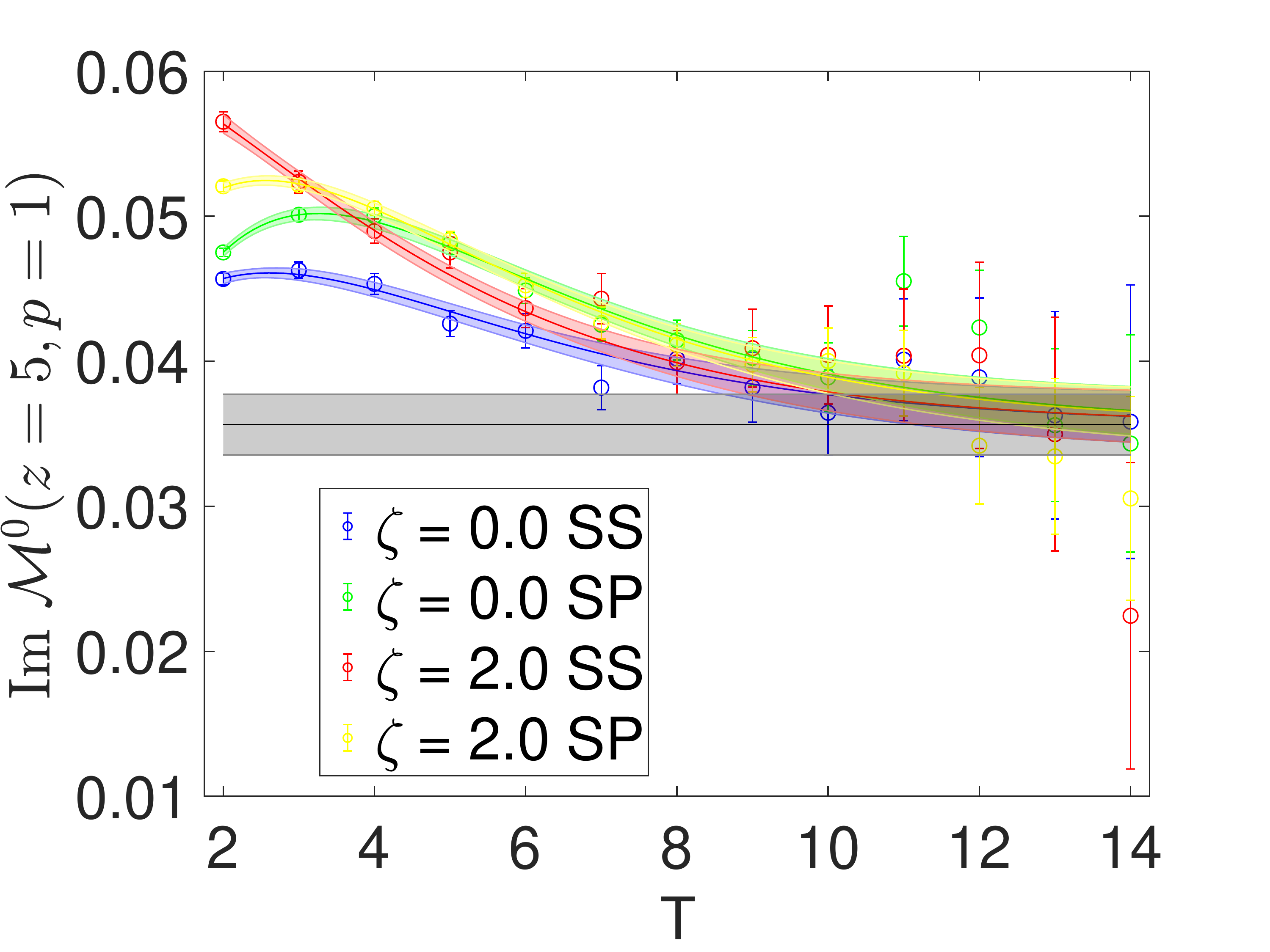}
\includegraphics[width=3in]{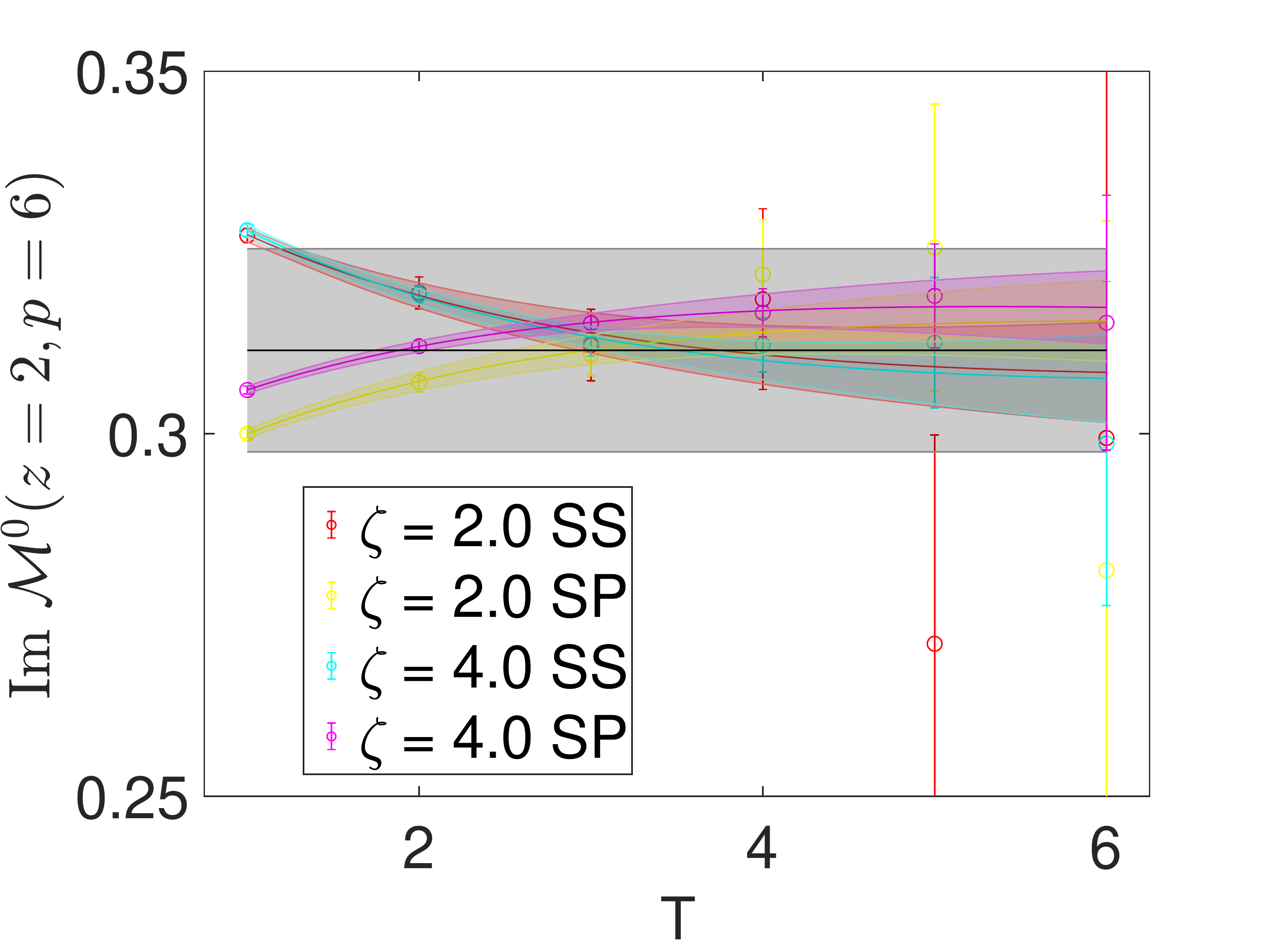}
\caption{Typical fits of the reduced Ioffe time pseudo-distribution from the ensemble $a094m390$. The left plot corresponds to the real component with $z_3=5 a$ and $p_3 = 1 (2 \pi/L)$. The right plot corresponds to the imaginary component with $z_3=2 a$ and $p_3 = 6 (2 \pi/L)$. The color points and bands correspond to the different correlation functions used in the fit and the resulting fit respectively. The grey band corresponds to the extracted matrix element. }
\label{fig:fit_fine}
\end{figure}

In this work, a fit of the data to Eq.~\eqref{eq:fitform} is used
simultaneously for each correlation function, holding the ground state
matrix element and effective energy gap fixed. Specifically, a fit is
performed on $N$ different effective bare matrix elements with
different smearing setups, $\mathcal{M}_j^{\rm{eff}}$, to the form
\beq
\mathcal{M}_j^{\rm {eff}}(\nu, z^2, T) = \mathcal{M}^0(\nu,z^2)\left( 1  + e^{-\Delta_p T}   \left[ A^{(j)}_p(z^2)  + B^{(j)}_p(z^2) T \right] \right),
\label{eq:full_fit_form}
\eeq
with $2N+2$ fit parameters where $j = 1\dots N$ labels the different
smearings. The fit parameters will be chosen with a weighted $\chi^2$
minimization which employs a block diagonal covariance matrix, where
the covariances between different correlation functions are
neglected. Each momentum smearing parameters are only helpful
for a certain range of momentum states. The choice of correlation functions
used in the fit, as well as the $T$ range of the fit, are varied to
minimize the $\chi^2$ per degree of freedom. All statistical errors
and covariances are estimated using the jackknife resampling
technique. An example of these fits are plotted in
Figs.~\ref{fig:fit_coarse}-\ref{fig:fit_fine}.
Tabs.~\ref{tab:ioffe_coarse}-\ref{tab:ioffe_fine2} contain the results
of the bare matrix elements and their standard deviations.

\begin{table}
\centering
\renewcommand{\arraystretch}{0.825}
\begin{tabular}{| c  c | c  | c | c | }
\hline
$z$($a$)  & $p$($2\pi/L$) & $\nu$ & Re $\mathcal{M}^0\pm\Delta$ Re $\mathcal{M}^0$ & Im $\mathcal{M}^0\pm\Delta$ Im $\mathcal{M}^0$ \\
\hline\hline
~ 0 ~ & ~ 0 ~ &~ 0 ~ & ~	1.251(6)	 ~ & ~ 	0(0)	 ~   \\ \hline
~ 1 ~ & ~ 0 ~ &~ 0 ~ & ~	1.030(5)	 ~ & ~ 	0(0)	 ~   \\ \hline
~ 2 ~ & ~ 0 ~ &~ 0 ~ & ~	0.716(4)	 ~ & ~ 	0(0)	 ~   \\ \hline
~ 3 ~ & ~ 0 ~ &~ 0 ~ & ~	0.4600(26)	 ~ & ~ 	0(0)	 ~   \\ \hline
~ 4 ~ & ~ 0 ~ &~ 0 ~ & ~	0.2808(24)	 ~ & ~ 	0(0)	 ~   \\ \hline
~ 5 ~ & ~ 0 ~ &~ 0 ~ & ~	0.1644(15)	 ~ & ~ 	0(0)	 ~   \\ \hline
~ 6 ~ & ~ 0 ~ &~ 0 ~ & ~	0.0935(10)	 ~ & ~ 	0(0)	 ~   \\ \hline
~ 0 ~ & ~ 1 ~ &~ 0 ~ & ~	1.245(7)	 ~ & ~ 	0(0)	 ~   \\ \hline
~ 1 ~ & ~ 1 ~ &~ 0.261799 ~ & ~	1.023(5)	 ~ & ~ 	0.0482(21)	 ~   \\ \hline
~ 2 ~ & ~ 1 ~ &~ 0.523599 ~ & ~	0.706(4)	 ~ & ~ 	0.067(5)	 ~   \\ \hline
~ 3 ~ & ~ 1 ~ &~ 0.785398 ~ & ~	0.4468(26)	 ~ & ~ 	0.067(4)	 ~   \\ \hline
~ 4 ~ & ~ 1 ~ &~ 1.047198 ~ & ~	0.2671(18)	 ~ & ~ 	0.056(3)	 ~   \\ \hline
~ 5 ~ & ~ 1 ~ &~ 1.308997 ~ & ~	0.1522(14)	 ~ & ~ 	0.0417(27)	 ~   \\ \hline
~ 6 ~ & ~ 1 ~ &~ 1.570796 ~ & ~	0.0835(11)	 ~ & ~ 	0.0284(22)	 ~   \\ \hline
~ 0 ~ & ~ 2 ~ &~ 0 ~ & ~	1.228(7)	 ~ & ~ 	0(0)	 ~   \\ \hline
~ 1 ~ & ~ 2 ~ &~ 0.523599 ~ & ~	1.001(6)	 ~ & ~ 	0.094(7)	 ~   \\ \hline
~ 2 ~ & ~ 2 ~ &~ 1.047198 ~ & ~	0.673(5)	 ~ & ~ 	0.133(10)	 ~   \\ \hline
~ 3 ~ & ~ 2 ~ &~ 1.570796 ~ & ~	0.408(4)	 ~ & ~ 	0.129(10)	 ~   \\ \hline
~ 4 ~ & ~ 2 ~ &~ 2.094395 ~ & ~	0.2293(26)	 ~ & ~ 	0.108(6)	 ~   \\ \hline
~ 5 ~ & ~ 2 ~ &~ 2.617994 ~ & ~	0.1209(20)	 ~ & ~ 	0.076(5)	 ~   \\ \hline
~ 6 ~ & ~ 2 ~ &~ 3.141593 ~ & ~	0.0600(16)	 ~ & ~ 	0.0511(6)	 ~   \\ \hline
~ 0 ~ & ~ 3 ~ &~ 0 ~ & ~	1.196(9)	 ~ & ~ 	0(0)	 ~   \\ \hline
~ 1 ~ & ~ 3 ~ &~ 0.785398 ~ & ~	0.960(8)	 ~ & ~ 	0.144(15)	 ~   \\ \hline
~ 2 ~ & ~ 3 ~ &~ 1.570796 ~ & ~	0.616(7)	 ~ & ~ 	0.190(29)	 ~   \\ \hline
~ 3 ~ & ~ 3 ~ &~ 2.356194 ~ & ~	0.347(6)	 ~ & ~ 	0.185(16)	 ~   \\ \hline
~ 4 ~ & ~ 3 ~ &~ 3.141593 ~ & ~	0.175(6)	 ~ & ~ 	0.142(10)	 ~   \\ \hline
~ 5 ~ & ~ 3 ~ &~ 3.926991 ~ & ~	0.080(6)	 ~ & ~ 	0.0893(15)	 ~   \\ \hline
~ 6 ~ & ~ 3 ~ &~ 4.712389 ~ & ~	0.031(7)	 ~ & ~ 	0.0496(13)	 ~   \\ \hline
~ 0 ~ & ~ 4 ~ &~ 0 ~ & ~	1.149(12)	 ~ & ~ 	0(0)	 ~   \\ \hline
~ 1 ~ & ~ 4 ~ &~ 1.047198 ~ & ~	0.902(11)	 ~ & ~ 	0.193(23)	 ~   \\ \hline
~ 2 ~ & ~ 4 ~ &~ 2.094395 ~ & ~	0.539(11)	 ~ & ~ 	0.258(19)	 ~   \\ \hline
~ 3 ~ & ~ 4 ~ &~ 3.141593 ~ & ~	0.269(11)	 ~ & ~ 	0.221(8)	 ~   \\ \hline
~ 4 ~ & ~ 4 ~ &~ 4.188790 ~ & ~	0.115(12)	 ~ & ~ 	0.143(3)	 ~   \\ \hline
~ 5 ~ & ~ 4 ~ &~ 5.235988 ~ & ~	0.034(5)	 ~ & ~ 	0.0779(22)	 ~   \\ \hline
~ 6 ~ & ~ 4 ~ &~ 6.283185 ~ & ~	0.0089(14)	 ~ & ~ 	0.0376(20)	 ~   \\ \hline
~ 0 ~ & ~ 5 ~ &~ 0 ~ & ~	1.123(5)	 ~ & ~ 	0(0)	 ~   \\ \hline
~ 1 ~ & ~ 5 ~ &~ 1.308997 ~ & ~	0.848(5)	 ~ & ~ 	0.244(16)	 ~   \\ \hline
~ 2 ~ & ~ 5 ~ &~ 2.617994 ~ & ~	0.440(7)	 ~ & ~ 	0.290(14)	 ~   \\ \hline
~ 3 ~ & ~ 5 ~ &~ 3.926991 ~ & ~	0.187(19)	 ~ & ~ 	0.229(5)	 ~   \\ \hline
~ 4 ~ & ~ 5 ~ &~ 5.235988 ~ & ~	0.066(10)	 ~ & ~ 	0.130(6)	 ~   \\ \hline
~ 5 ~ & ~ 5 ~ &~ 6.544985 ~ & ~	0.021(4)	 ~ & ~ 	0.055(4)	 ~   \\ \hline
~ 6 ~ & ~ 5 ~ &~ 7.853982 ~ & ~	0.0079(23)	 ~ & ~ 	0.0251(17)	 ~   \\ \hline
~ 0 ~ & ~ 6 ~ &~ 0 ~ & ~	1.01(5)	 ~ & ~ 	0(0)	 ~   \\ \hline
~ 1 ~ & ~ 6 ~ &~ 1.570796 ~ & ~	0.742(4)	 ~ & ~ 	0.281(13)	 ~   \\ \hline
~ 2 ~ & ~ 6 ~ &~ 3.141593 ~ & ~	0.357(21)	 ~ & ~ 	0.323(8)	 ~   \\ \hline
~ 3 ~ & ~ 6 ~ &~ 4.712389 ~ & ~	0.126(15)	 ~ & ~ 	0.227(16)	 ~   \\ \hline
~ 4 ~ & ~ 6 ~ &~ 6.283185 ~ & ~	0.054(11)	 ~ & ~ 	0.090(11)	 ~   \\ \hline
~ 5 ~ & ~ 6 ~ &~ 7.853982 ~ & ~	0.024(6)	 ~ & ~ 	0.039(4)	 ~   \\ \hline
~ 6 ~ & ~ 6 ~ &~ 9.424778 ~ & ~	0.0103(24)	 ~ & ~ 	0.0190(9)	 ~   \\ \hline
\end{tabular}
\renewcommand{\arraystretch}{1}

\caption{\footnotesize The bare matrix elements and their standard deviations from the ensemble $a127m415$. The imaginary component is assumed to be 0 for $\nu=0$.\label{tab:ioffe_coarse}}
\end{table}

\begin{table}
\centering
\renewcommand{\arraystretch}{0.825}
\begin{tabular}{| c  c | c  | c | c | }
\hline
$z$($a$)  & $p$($2\pi/L$) & $\nu$ & Re $\mathcal{M}^0 \pm\Delta$ Re $\mathcal{M}^0$ & Im $\mathcal{M}^0\pm\Delta$ Im $\mathcal{M}^0$ \\
\hline\hline
~ 0 ~ & ~ 0 ~ &~ 0 ~ & ~	1.272(15)	 ~ & ~ 	0(0)	 ~   \\ \hline
~ 1 ~ & ~ 0 ~ &~ 0 ~ & ~	1.050(12)	 ~ & ~ 	0(0)	 ~   \\ \hline
~ 2 ~ & ~ 0 ~ &~ 0 ~ & ~	0.730(8)	 ~ & ~ 	0(0)	 ~   \\ \hline
~ 3 ~ & ~ 0 ~ &~ 0 ~ & ~	0.468(9)	 ~ & ~ 	0(0)	 ~   \\ \hline
~ 4 ~ & ~ 0 ~ &~ 0 ~ & ~	0.284(3)	 ~ & ~ 	0(0)	 ~   \\ \hline
~ 5 ~ & ~ 0 ~ &~ 0 ~ & ~	0.1663(21)	 ~ & ~ 	0(0)	 ~   \\ \hline
~ 6 ~ & ~ 0 ~ &~ 0 ~ & ~	0.0941(15)	 ~ & ~ 	0(0)	 ~   \\ \hline
~ 7 ~ & ~ 0 ~ &~ 0 ~ & ~	0.0519(19)	 ~ & ~ 	0(0)	 ~   \\ \hline
~ 8 ~ & ~ 0 ~ &~ 0 ~ & ~	0.0281(16)	 ~ & ~ 	0(0)	 ~   \\ \hline
~ 0 ~ & ~ 1 ~ &~ 0 ~ & ~	1.270(13)	 ~ & ~ 	0(0)	 ~   \\ \hline
~ 1 ~ & ~ 1 ~ &~ 0.196350 ~ & ~	1.044(11)	 ~ & ~ 	0.033(5)	 ~   \\ \hline
~ 2 ~ & ~ 1 ~ &~ 0.392699 ~ & ~	0.722(8)	 ~ & ~ 	0.049(5)	 ~   \\ \hline
~ 3 ~ & ~ 1 ~ &~ 0.589049 ~ & ~	0.459(5)	 ~ & ~ 	0.049(5)	 ~   \\ \hline
~ 4 ~ & ~ 1 ~ &~ 0.785398 ~ & ~	0.276(3)	 ~ & ~ 	0.042(6)	 ~   \\ \hline
~ 5 ~ & ~ 1 ~ &~ 0.981748 ~ & ~	0.1589(21)	 ~ & ~ 	0.032(5)	 ~   \\ \hline
~ 6 ~ & ~ 1 ~ &~ 1.178097 ~ & ~	0.0882(14)	 ~ & ~ 	0.0218(23)	 ~   \\ \hline
~ 7 ~ & ~ 1 ~ &~ 1.374447 ~ & ~	0.048(10)	 ~ & ~ 	0.0146(15)	 ~   \\ \hline
~ 8 ~ & ~ 1 ~ &~ 1.570796 ~ & ~	0.0250(7)	 ~ & ~ 	0.0093(10)	 ~   \\ \hline
~ 0 ~ & ~ 2 ~ &~ 0 ~ & ~	1.263(14)	 ~ & ~ 	0(0)	 ~   \\ \hline
~ 1 ~ & ~ 2 ~ &~ 0.392699 ~ & ~	1.034(12)	 ~ & ~ 	0.0710(18)	 ~   \\ \hline
~ 2 ~ & ~ 2 ~ &~ 0.785398 ~ & ~	0.706(8)	 ~ & ~ 	0.102(22)	 ~   \\ \hline
~ 3 ~ & ~ 2 ~ &~ 1.178097 ~ & ~	0.439(6)	 ~ & ~ 	0.101(17)	 ~   \\ \hline
~ 4 ~ & ~ 2 ~ &~ 1.570796 ~ & ~	0.255(4)	 ~ & ~ 	0.082(14)	 ~   \\ \hline
~ 5 ~ & ~ 2 ~ &~ 1.963495 ~ & ~	0.1405(26)	 ~ & ~ 	0.056(10)	 ~   \\ \hline
~ 6 ~ & ~ 2 ~ &~ 2.356194 ~ & ~	0.0736(17)	 ~ & ~ 	0.039(6)	 ~   \\ \hline
~ 7 ~ & ~ 2 ~ &~ 2.748894 ~ & ~	0.0369(12)	 ~ & ~ 	0.025(3)	 ~   \\ \hline
~ 8 ~ & ~ 2 ~ &~ 3.141593 ~ & ~	0.0176(8)	 ~ & ~ 	0.015(3)	 ~   \\ \hline
~ 0 ~ & ~ 3 ~ &~ 0 ~ & ~	1.250(16)	 ~ & ~ 	0(0)	 ~   \\ \hline
~ 1 ~ & ~ 3 ~ &~ 0.589049 ~ & ~	1.016(13)	 ~ & ~ 	0.114(10)	 ~   \\ \hline
~ 2 ~ & ~ 3 ~ &~ 1.178097 ~ & ~	0.676(10)	 ~ & ~ 	0.161(14)	 ~   \\ \hline
~ 3 ~ & ~ 3 ~ &~ 1.767146 ~ & ~	0.404(7)	 ~ & ~ 	0.154(16)	 ~   \\ \hline
~ 4 ~ & ~ 3 ~ &~ 2.356194 ~ & ~	0.221(5)	 ~ & ~ 	0.121(15)	 ~   \\ \hline
~ 5 ~ & ~ 3 ~ &~ 2.945243 ~ & ~	0.113(4)	 ~ & ~ 	0.085(11)	 ~   \\ \hline
~ 6 ~ & ~ 3 ~ &~ 3.534292 ~ & ~	0.054(4)	 ~ & ~ 	0.0535(8)	 ~   \\ \hline
~ 7 ~ & ~ 3 ~ &~ 4.123340 ~ & ~	0.0236(27)	 ~ & ~ 	0.0302(9)	 ~   \\ \hline
~ 8 ~ & ~ 3 ~ &~ 4.712389 ~ & ~	0.0087(15)	 ~ & ~ 	0.0160(19)	 ~   \\ \hline
 \end{tabular}
\renewcommand{\arraystretch}{1}

\caption{\footnotesize The bare matrix elements and their standard deviations from the ensemble $a127m415L$. The imaginary component is assumed to be 0 for $\nu=0$.\label{tab:ioffe_big}}
\end{table}

\begin{table}
\centering
\renewcommand{\arraystretch}{0.825}
\begin{tabular}{| c | c | c  |  c |  c | }
\hline
$z$($a$)  & $p$($2\pi/L$) & $\nu$ & Re $\mathcal{M}^0 \pm\Delta$ Re $\mathcal{M}^0$ & Im $\mathcal{M}^0\pm\Delta$ Im $\mathcal{M}^0$ \\
\hline\hline
~ 0 ~ & ~ 4 ~ &~ 0 ~ & ~	1.222(20)	 ~ & ~ 	0(0)	 ~   \\ \hline
~ 1 ~ & ~ 4 ~ &~ 0.785398 ~ & ~	0.981(17)	 ~ & ~ 	0.160(9)	 ~   \\ \hline
~ 2 ~ & ~ 4 ~ &~ 1.570796 ~ & ~	0.630(12)	 ~ & ~ 	0.220(13)	 ~   \\ \hline
~ 3 ~ & ~ 4 ~ &~ 2.356194 ~ & ~	0.352(9)	 ~ & ~ 	0.202(11)	 ~   \\ \hline
~ 4 ~ & ~ 4 ~ &~ 3.141593 ~ & ~	0.174(6)	 ~ & ~ 	0.147(19)	 ~   \\ \hline
~ 5 ~ & ~ 4 ~ &~ 3.926991 ~ & ~	0.076(5)	 ~ & ~ 	0.0913(25)	 ~   \\ \hline
~ 6 ~ & ~ 4 ~ &~ 4.712389 ~ & ~	0.029(4)	 ~ & ~ 	0.0513(16)	 ~   \\ \hline
~ 7 ~ & ~ 4 ~ &~ 5.497787 ~ & ~	0.0087(25)	 ~ & ~ 	0.0264(12)	 ~   \\ \hline
~ 8 ~ & ~ 4 ~ &~ 6.283185 ~ & ~	0.0014(8)	 ~ & ~ 	0.0124(8)	 ~   \\ \hline
~ 0 ~ & ~ 5 ~ &~ 0 ~ & ~	1.185(20)	 ~ & ~ 	0(0)	 ~   \\ \hline
~ 1 ~ & ~ 5 ~ &~ 0.981748 ~ & ~	0.936(17)	 ~ & ~ 	0.204(15)	 ~   \\ \hline
~ 2 ~ & ~ 5 ~ &~ 1.963495 ~ & ~	0.567(15)	 ~ & ~ 	0.267(18)	 ~   \\ \hline
~ 3 ~ & ~ 5 ~ &~ 2.945243 ~ & ~	0.278(9)	 ~ & ~ 	0.225(25)	 ~   \\ \hline
~ 4 ~ & ~ 5 ~ &~ 3.926991 ~ & ~	0.121(11)	 ~ & ~ 	0.148(6)	 ~   \\ \hline
~ 5 ~ & ~ 5 ~ &~ 4.908739 ~ & ~	0.040(5)	 ~ & ~ 	0.084(4)	 ~   \\ \hline
~ 6 ~ & ~ 5 ~ &~ 5.890486 ~ & ~	0.009(10)	 ~ & ~ 	0.0422(21)	 ~   \\ \hline
~ 7 ~ & ~ 5 ~ &~ 6.872234 ~ & ~	-0.002(6)	 ~ & ~ 	0.0185(8)	 ~   \\ \hline
~ 8 ~ & ~ 5 ~ &~ 7.853982 ~ & ~	-0.014(25)	 ~ & ~ 	0.0082(6)	 ~   \\ \hline
~ 0 ~ & ~ 6 ~ &~ 0 ~ & ~	1.16(3)	 ~ & ~ 	0(0)	 ~   \\ \hline
~ 1 ~ & ~ 6 ~ &~ 1.178097 ~ & ~	0.90(3)	 ~ & ~ 	0.249(19)	 ~   \\ \hline
~ 2 ~ & ~ 6 ~ &~ 2.356194 ~ & ~	0.512(28)	 ~ & ~ 	0.300(12)	 ~   \\ \hline
~ 3 ~ & ~ 6 ~ &~ 3.534292 ~ & ~	0.225(22)	 ~ & ~ 	0.231(15)	 ~   \\ \hline
~ 4 ~ & ~ 6 ~ &~ 4.712389 ~ & ~	0.078(13)	 ~ & ~ 	0.138(18)	 ~   \\ \hline
~ 5 ~ & ~ 6 ~ &~ 5.890486 ~ & ~	0.021(5)	 ~ & ~ 	0.068(12)	 ~   \\ \hline
~ 6 ~ & ~ 6 ~ &~ 7.068583 ~ & ~	0.001(9)	 ~ & ~ 	0.029(6)	 ~   \\ \hline
~ 7 ~ & ~ 6 ~ &~ 8.246681 ~ & ~	-0.003(5)	 ~ & ~ 	0.0131(17)	 ~   \\ \hline
~ 8 ~ & ~ 6 ~ &~ 9.424778 ~ & ~	-0.004(6)	 ~ & ~ 	0.0055(7)	 ~   \\ \hline
 \hline
\end{tabular}
\renewcommand{\arraystretch}{1}

\caption{\footnotesize The bare matrix elements and their standard deviations from the ensemble $a127m415L$. The imaginary component is assumed to be 0 for $\nu=0$.\label{tab:ioffe_big2}}
\end{table}

\begin{table}
\centering
\renewcommand{\arraystretch}{0.825}
\begin{tabular}{| c |  c | c  |  c |  c | }
\hline
$z$($a$)  & $p$($2\pi/L$) & $\nu$ & Re $\mathcal{M}^0 \pm \Delta$ Re $\mathcal{M}^0$ & Im $\mathcal{M}^0\pm\Delta$ Im $\mathcal{M}^0$ \\
\hline\hline
~ 0 ~ & ~ 0 ~ &~ 0 ~ & ~	1.187(18)	 ~ & ~ 	0(0)	 ~   \\ \hline
~ 1 ~ & ~ 0 ~ &~ 0 ~ & ~	0.994(15)	 ~ & ~ 	0(0)	 ~   \\ \hline
~ 2 ~ & ~ 0 ~ &~ 0 ~ & ~	0.713(11)	 ~ & ~ 	0(0)	 ~   \\ \hline
~ 3 ~ & ~ 0 ~ &~ 0 ~ & ~	0.477(8)	 ~ & ~ 	0(0)	 ~   \\ \hline
~ 4 ~ & ~ 0 ~ &~ 0 ~ & ~	0.306(5)	 ~ & ~ 	0(0)	 ~   \\ \hline
~ 5 ~ & ~ 0 ~ &~ 0 ~ & ~	0.191(3)	 ~ & ~ 	0(0)	 ~   \\ \hline
~ 6 ~ & ~ 0 ~ &~ 0 ~ & ~	0.1164(23)	 ~ & ~ 	0(0)	 ~   \\ \hline
~ 7 ~ & ~ 0 ~ &~ 0 ~ & ~	0.0700(17)	 ~ & ~ 	0(0)	 ~   \\ \hline
~ 8 ~ & ~ 0 ~ &~ 0 ~ & ~	0.0414(13)	 ~ & ~ 	0(0)	 ~   \\ \hline
~ 0 ~ & ~ 1 ~ &~ 0 ~ & ~	1.187(21)	 ~ & ~ 	0(0)	 ~   \\ \hline
~ 1 ~ & ~ 1 ~ &~ 0.196350 ~ & ~	0.993(17)	 ~ & ~ 	0.0317(17)	 ~   \\ \hline
~ 2 ~ & ~ 1 ~ &~ 0.392699 ~ & ~	0.711(12)	 ~ & ~ 	0.0475(26)	 ~   \\ \hline
~ 3 ~ & ~ 1 ~ &~ 0.589049 ~ & ~	0.474(8)	 ~ & ~ 	0.0498(28)	 ~   \\ \hline
~ 4 ~ & ~ 1 ~ &~ 0.785398 ~ & ~	0.301(5)	 ~ & ~ 	0.0443(25)	 ~   \\ \hline
~ 5 ~ & ~ 1 ~ &~ 0.981748 ~ & ~	0.185(3)	 ~ & ~ 	0.0356(21)	 ~   \\ \hline
~ 6 ~ & ~ 1 ~ &~ 1.178097 ~ & ~	0.1112(23)	 ~ & ~ 	0.0267(17)	 ~   \\ \hline
~ 7 ~ & ~ 1 ~ &~ 1.374447 ~ & ~	0.0654(17)	 ~ & ~ 	0.0189(14)	 ~   \\ \hline
~ 8 ~ & ~ 1 ~ &~ 1.570796 ~ & ~	0.0380(13)	 ~ & ~ 	0.0126(13)	 ~   \\ \hline
~ 0 ~ & ~ 2 ~ &~ 0 ~ & ~	1.14(7)	 ~ & ~ 	0(0)	 ~   \\ \hline
~ 1 ~ & ~ 2 ~ &~ 0.392699 ~ & ~	0.95(6)	 ~ & ~ 	0.056(9)	 ~   \\ \hline
~ 2 ~ & ~ 2 ~ &~ 0.785398 ~ & ~	0.691(28)	 ~ & ~ 	0.084(13)	 ~   \\ \hline
~ 3 ~ & ~ 2 ~ &~ 1.178097 ~ & ~	0.450(16)	 ~ & ~ 	0.088(12)	 ~   \\ \hline
~ 4 ~ & ~ 2 ~ &~ 1.570796 ~ & ~	0.280(8)	 ~ & ~ 	0.077(10)	 ~   \\ \hline
~ 5 ~ & ~ 2 ~ &~ 1.963495 ~ & ~	0.165(4)	 ~ & ~ 	0.060(8)	 ~   \\ \hline
~ 6 ~ & ~ 2 ~ &~ 2.356194 ~ & ~	0.093(4)	 ~ & ~ 	0.045(7)	 ~   \\ \hline
~ 7 ~ & ~ 2 ~ &~ 2.748894 ~ & ~	0.050(6)	 ~ & ~ 	0.032(5)	 ~   \\ \hline
~ 8 ~ & ~ 2 ~ &~ 3.141593 ~ & ~	0.0268(13)	 ~ & ~ 	0.020(4)	 ~   \\ \hline
~ 0 ~ & ~ 3 ~ &~ 0 ~ & ~	1.14(6)	 ~ & ~ 	0(0)	 ~   \\ \hline
~ 1 ~ & ~ 3 ~ &~ 0.589049 ~ & ~	0.95(5)	 ~ & ~ 	0.092(22)	 ~   \\ \hline
~ 2 ~ & ~ 3 ~ &~ 1.178097 ~ & ~	0.658(22)	 ~ & ~ 	0.14(3)	 ~   \\ \hline
~ 3 ~ & ~ 3 ~ &~ 1.767146 ~ & ~	0.411(10)	 ~ & ~ 	0.148(16)	 ~   \\ \hline
~ 4 ~ & ~ 3 ~ &~ 2.356194 ~ & ~	0.240(26)	 ~ & ~ 	0.124(14)	 ~   \\ \hline
~ 5 ~ & ~ 3 ~ &~ 2.945243 ~ & ~	0.131(12)	 ~ & ~ 	0.092(11)	 ~   \\ \hline
~ 6 ~ & ~ 3 ~ &~ 3.534292 ~ & ~	0.065(3)	 ~ & ~ 	0.060(8)	 ~   \\ \hline
~ 7 ~ & ~ 3 ~ &~ 4.123340 ~ & ~	0.027(4)	 ~ & ~ 	0.0441(8)	 ~   \\ \hline
~ 8 ~ & ~ 3 ~ &~ 4.712389 ~ & ~	0.011(3)	 ~ & ~ 	0.0256(7)	 ~   \\ \hline

 \end{tabular}
\renewcommand{\arraystretch}{1}

\caption{\footnotesize The bare matrix elements and their standard deviations from the ensemble $a094m390$. The imaginary component is assumed to be 0 for $\nu=0$.\label{tab:ioffe_fine}}
\end{table}

\begin{table}
\centering
\renewcommand{\arraystretch}{0.825}
\begin{tabular}{| c | c | c  |  c |  c | }
\hline
$z$($a$)  & $p$($2\pi/L$) & $\nu$ & Re $\mathcal{M}^0\pm\Delta$ Re $\mathcal{M}^0$ & Im $\mathcal{M}^0\pm \Delta$ Im $\mathcal{M}^0$ \\
\hline\hline
~ 0 ~ & ~ 4 ~ &~ 0 ~ & ~	1.164(9)	~ & ~ 0(0)	~   \\ \hline
~ 1 ~ & ~ 4 ~ &~ 0.785398 ~ & ~	0.944(8)	~ & ~ 0.173(6)	~   \\ \hline
~ 2 ~ & ~ 4 ~ &~ 1.570796 ~ & ~	0.616(7)	~ & ~ 0.249(9)	~   \\ \hline
~ 3 ~ & ~ 4 ~ &~ 2.356194 ~ & ~	0.350(6)	~ & ~ 0.2310(11)	~   \\ \hline
~ 4 ~ & ~ 4 ~ &~ 3.141593 ~ & ~	0.175(6)	~ & ~ 0.1723(13)	~   \\ \hline
~ 5 ~ & ~ 4 ~ &~ 3.926991 ~ & ~	0.074(5)	~ & ~ 0.1134(11)	~   \\ \hline
~ 6 ~ & ~ 4 ~ &~ 4.712389 ~ & ~	0.022(5)	~ & ~ 0.0670(10)	~   \\ \hline
~ 7 ~ & ~ 4 ~ &~ 5.497787 ~ & ~	0.002(5)	~ & ~ 0.0369(12)	~   \\ \hline
~ 8 ~ & ~ 4 ~ &~ 6.283185 ~ & ~	-0.001(10)	~ & ~ 0.0185(10)	~   \\ \hline
~ 0 ~ & ~ 5 ~ &~ 0 ~ & ~	1.129(9)	~ & ~ 0(0)	~   \\ \hline
~ 1 ~ & ~ 5 ~ &~ 0.981748 ~ & ~	0.895(12)	~ & ~ 0.207(13)	~   \\ \hline
~ 2 ~ & ~ 5 ~ &~ 1.963495 ~ & ~	0.542(9)	~ & ~ 0.299(10)	~   \\ \hline
~ 3 ~ & ~ 5 ~ &~ 2.945243 ~ & ~	0.262(14)	~ & ~ 0.2488(23)	~   \\ \hline
~ 4 ~ & ~ 5 ~ &~ 3.926991 ~ & ~	0.101(14)	~ & ~ 0.1678(27)	~   \\ \hline
~ 5 ~ & ~ 5 ~ &~ 4.908739 ~ & ~	0.028(17)	~ & ~ 0.0937(29)	~   \\ \hline
~ 6 ~ & ~ 5 ~ &~ 5.890486 ~ & ~	0.0086(5)	~ & ~ 0.0445(26)	~   \\ \hline
~ 7 ~ & ~ 5 ~ &~ 6.872234 ~ & ~	-0.011(27)	~ & ~ 0.0183(23)	~   \\ \hline
~ 8 ~ & ~ 5 ~ &~ 7.853982 ~ & ~	-0.006(9)	~ & ~ 0.0066(23)	~   \\ \hline
~ 0 ~ & ~ 6 ~ &~ 0 ~ & ~	1.112(10)	~ & ~ 0(0)	~   \\ \hline
~ 1 ~ & ~ 6 ~ &~ 1.178097 ~ & ~	0.855(14)	~ & ~ 0.25(3)	~   \\ \hline
~ 2 ~ & ~ 6 ~ &~ 2.356194 ~ & ~	0.516(3)	~ & ~ 0.312(14)	~   \\ \hline
~ 3 ~ & ~ 6 ~ &~ 3.534292 ~ & ~	0.17(3)	~ & ~ 0.242(6)	~   \\ \hline
~ 4 ~ & ~ 6 ~ &~ 4.712389 ~ & ~	0.06(3)	~ & ~ 0.158(12)	~   \\ \hline
~ 5 ~ & ~ 6 ~ &~ 5.890486 ~ & ~	0.0197(14)	~ & ~ 0.078(14)	~   \\ \hline
~ 6 ~ & ~ 6 ~ &~ 7.068583 ~ & ~	0.001(6)	~ & ~ 0.026(12)	~   \\ \hline
~ 7 ~ & ~ 6 ~ &~ 8.246681 ~ & ~	0.004(6)	~ & ~ 0.0127(20)	~   \\ \hline
~ 8 ~ & ~ 6 ~ &~ 9.424778 ~ & ~	0.002(3)	~ & ~ 0.005(5)	~   \\ \hline
\end{tabular}
\renewcommand{\arraystretch}{1}

\caption{\footnotesize The bare matrix elements and their standard deviations from the ensemble $a094m390$. The imaginary component is assumed to be 0 for $\nu=0$.\label{tab:ioffe_fine2}}
\end{table}

\subsection{Cancellation of Renormalization Constants}

With a lattice regulator (unlike in dimensional regularization),  the
operator $O_{\rm WL}(z)$ has a power divergence in $\frac za$. The
handling of this power divergence in lattice QCD renormalization
schemes, such as the popular RI-MOM scheme, and the associated
matching relationships have generated a large amount of discussion,
see~\cite{Alexandrou:2019lfo} for a comparison of methods. In a
lattice QCD calculation, renormalization constants require a separate
calculation. Alternatively, when possible one can form ratios of matrix elements where the UV divergences cancel.
 In this spirit, a ratio, which has the same leading order in the OPE as the pseudo-ITD,
will be constructed where all renormalization constants cancel.

The local vector current, $M^0(z\cdot p , z^2)|_{z=0}$, where
$M^0$ is the bare matrix element can be used to ensure quark number conservation. In the continuum limit, where the
vector current is conserved, this matrix element should be equal to
1. Due to lattice artifacts, the local vector current is not
conserved and it possesses an $a p$ dependence, but has a finite
renormalization constant. This renormalization constant does have the
property $Z_V \to 1$ in the continuum limit $a\to 0$. Again the
leading order behavior of the OPE for the ratio
\beq
\mathcal{M}_V^0 (\nu,z^2 ) = \frac{\mathcal{M}^0(\nu,z^2)}{\mathcal{M}^0(\nu , z^2)|_{z=0}},
\eeq
will match the original pseudo-ITD. This ratio still contains the logarithmic and power divergences associated with the Wilson line operator.

\begin{figure}[ht]\centering
\includegraphics[width=0.32\textwidth]{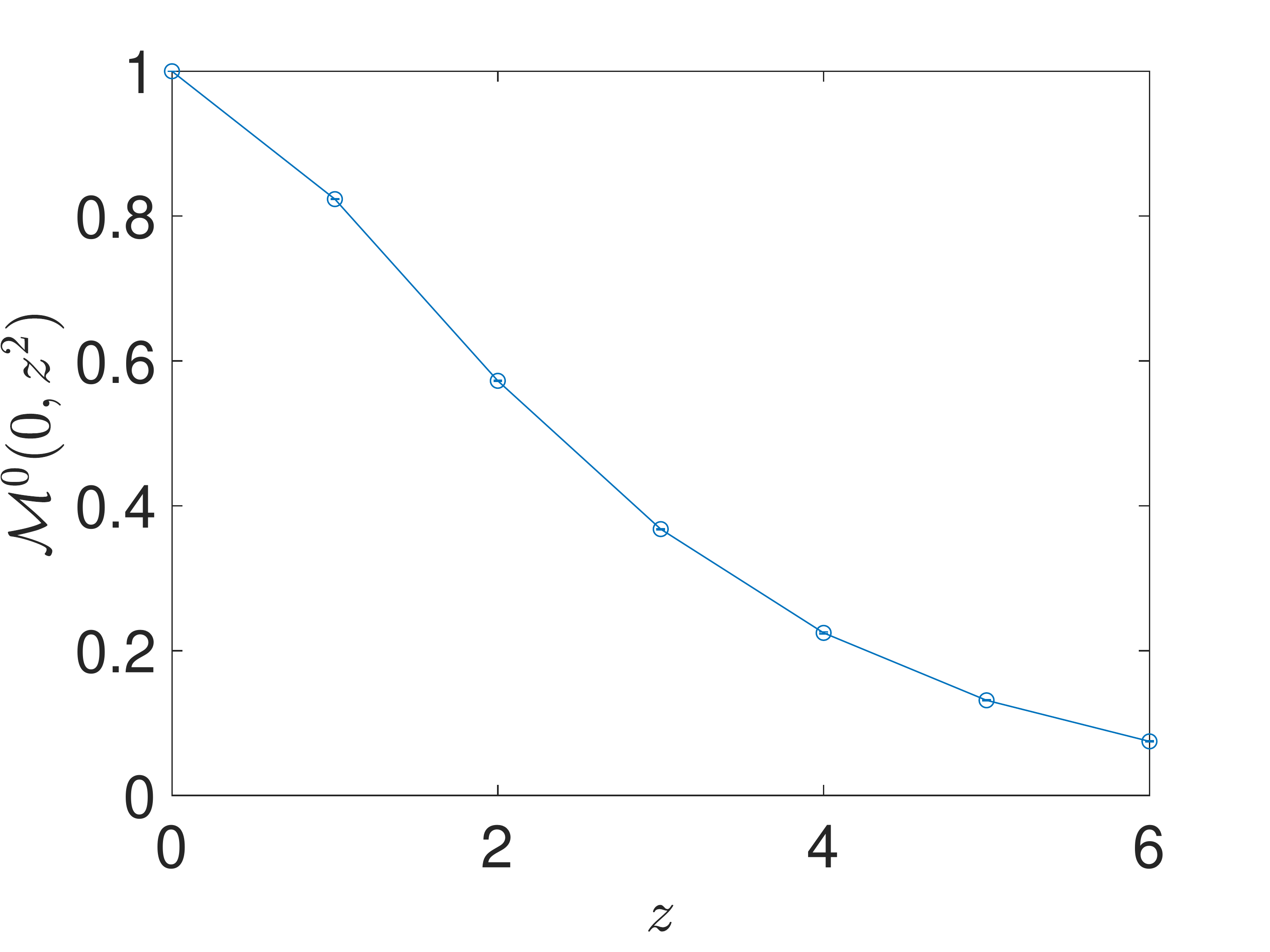}
\includegraphics[width=0.32\textwidth]{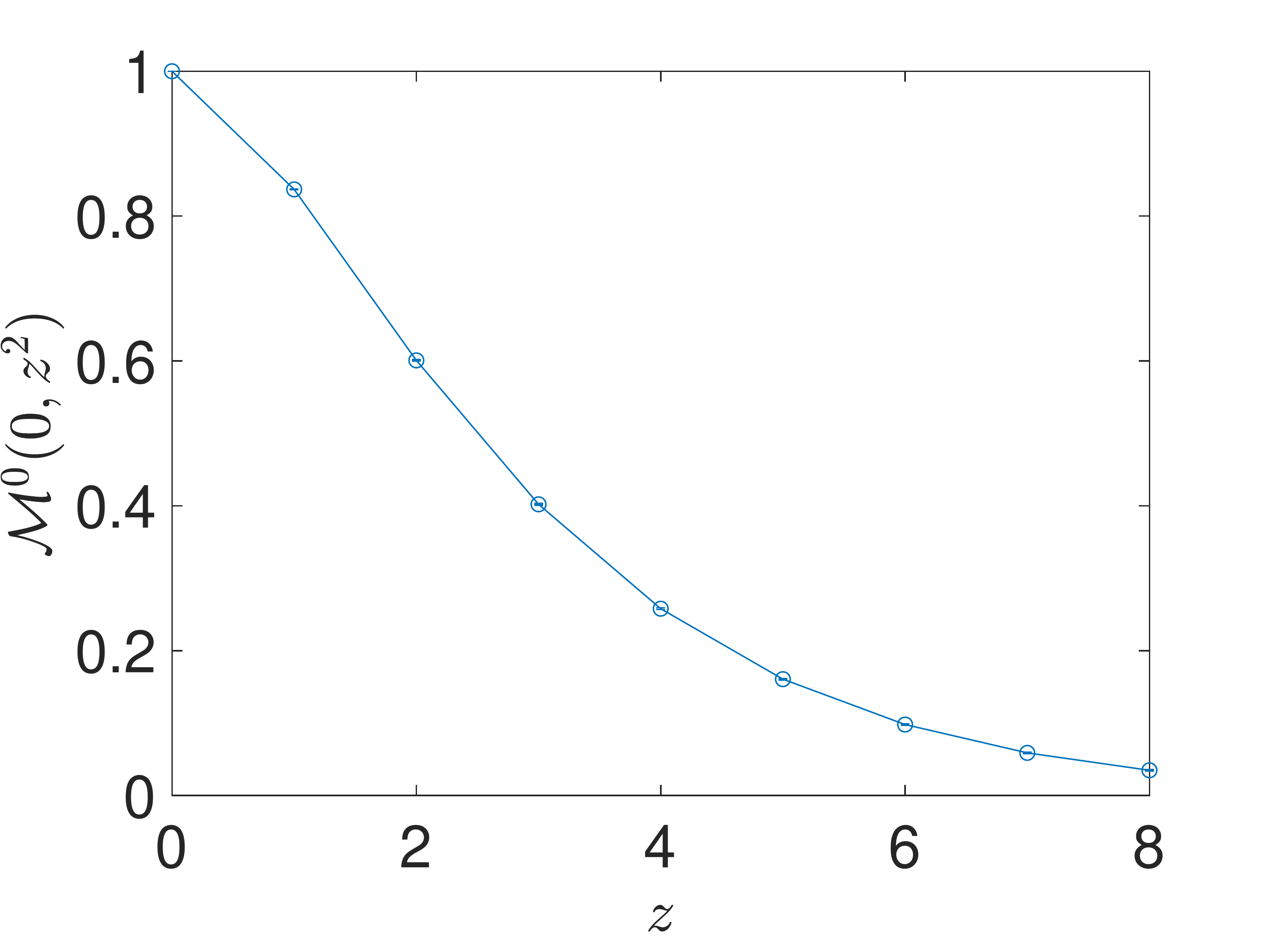}
\includegraphics[width=0.32\textwidth]{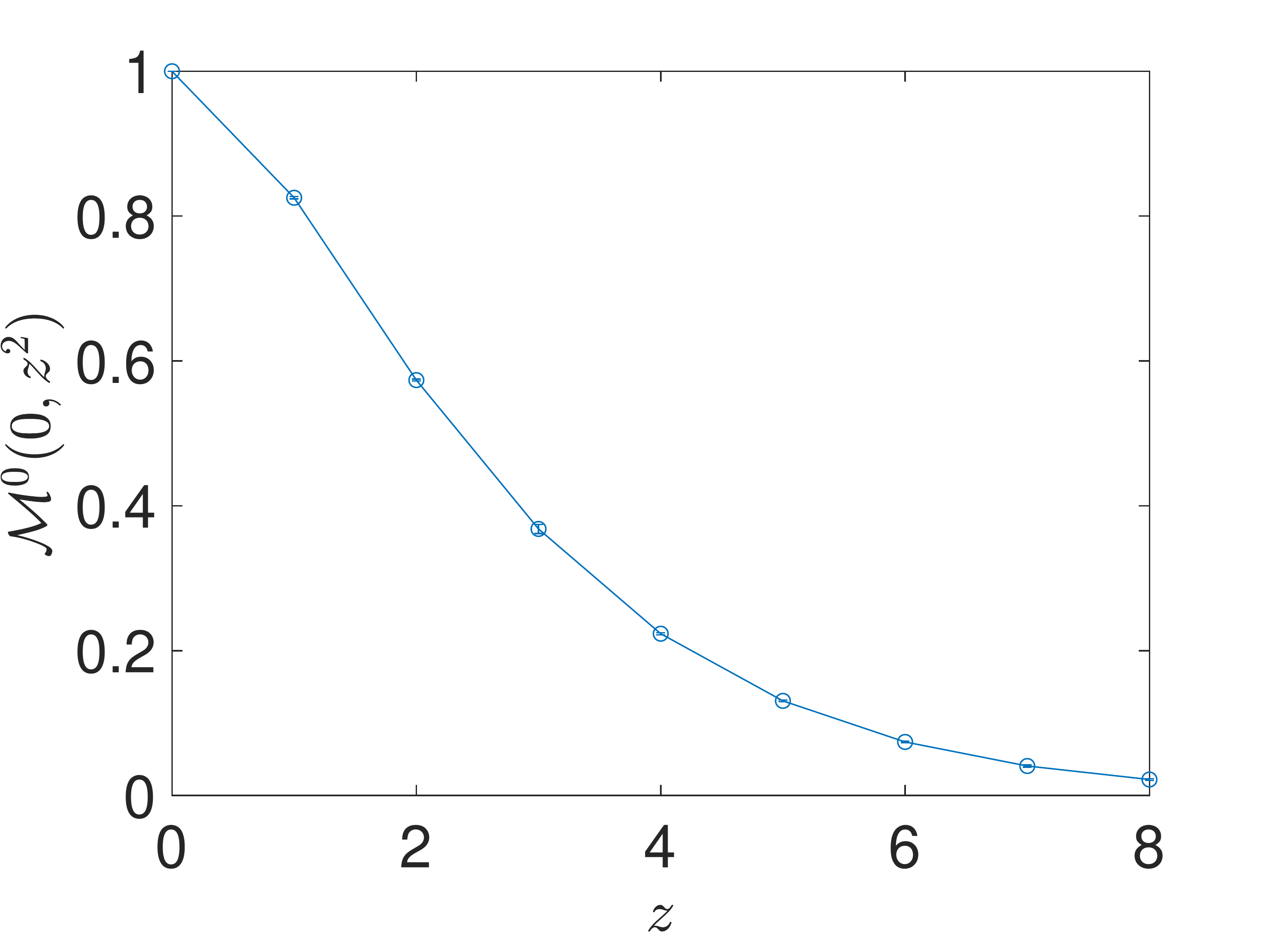}

\caption{The rest frame distribution $\mathcal{M}_V^0(z\cdot p , z^2)|_{p=0}$. The left plot is from the ensemble $a127m415$, the middle plot is from the ensemble $a094m390$, and the right plot is from the ensemble $a127m415L$. The cusp as $z\to0$ is the signal for the power divergences which occur in perturbation theory. The large $z$ limit reveals the exponential nature of the Wilson line renormalization constant. }

\label{fig:rest_frame}
\end{figure}

The UV divergences of the Wilson line operator will be
canceled by forming ratios which have in the denominator the rest
frame matrix element $\mathcal{M}_V^0(z\cdot p , z^2)|_{p=0}$. The
imaginary component is consistent with zero for all $z$. The real
component results, which are plotted in Fig.~\ref{fig:rest_frame}, have
the exponential behavior expected from the non-perturbative effects
generated by the Wilson line operator. The low $z/a$ region exhibits a
cusp as $z\to 0$, which is a signal for the power divergence. For the
pseudo-ITD, the matching kernel is unity at $\nu=0$, meaning the rest
frame matrix element will be the integral of the PDF, which for the
iso-vector flavor quark combination is 1, up to potential higher twist
and discretization errors. The leading order behavior of the OPE for
this ratio will be the same as for the pseudo-ITD, so that this matrix
element satisfies the properties required for the reduced pseudo-ITD.

Finally the reduced matrix element is defined by the double ratio 
\beq \label{eq:double_ratio}
 \mathfrak{M}(\nu,z^2) = \bigg(\frac{\mathcal{M}^0(\nu,z^2)}{\mathcal{M}^0(\nu,0)|_{z=0}}\bigg)/ \bigg(\frac{\mathcal{M}^0(0,z^2)|_{p=0}}{ \mathcal{M}^0(0,0)|_{p=0,z=0}}\bigg) \,.
\eeq
This double ratio not only takes care of cancelling the multiplicative renormalization constants. It also has the desired goal of cancelling some $O(z^2)$ higher twist contaminations. This feature was demonstrated in the quenched approximation~\cite{Orginos:2017kos}, where at fixed Ioffe time and large $z^2$, the reduced pseudo-ITD was independent of $z^2$ instead of showing 
a 
polynomial behavior. The reduced pseudo-ITD is a renormalization scheme independent quantity which can be matched directly to the $\overline{\rm MS}$ light cone PDF through the OPE-based  Eq.~\eqref{eq:pITD_to_ITD}. 

The form of this double ratio has an additional advantage of being exactly equal to unity at $\nu=0$ with no possible higher twist effects and lattice spacing errors. This feature explicitly sets the iso-vector quark PDF sum rule, $\int_0^1 dx \left(u(x) - d(x)\right) = 1$. Any error in the sum rule for a PDF determined from this reduced pseudo-ITD must be an artifact of the procedure for calculating the PDF.

\subsection{Reduced pseudo-ITD results}
In this section we discuss the results that we have obtained for the
reduced Ioffe time pseudo-distribution. The real and imaginary
components of the ITD, and therefore the continuum reduced pseudo-ITD, can
be analyzed separately yielding additional insight to the
structure of the hadron. The real (CP even) component describes the valence quark
distribution
\begin{eqnarray}
{\rm Re}~ Q(\nu,\mu^2) & =&\int_0^1 dx \cos(\nu x) \big(q(x,\mu^2)-\bar{q}(x,\mu^2)\big)\nonumber \\& =& \int_0^1 dx \cos(\nu x) q_v(x,\mu^2)\,.
\label{eq:pdf_fit_real}
\end{eqnarray}
The imaginary (CP odd) component describes the sum of the quark and anti-quark distributions 
\begin{eqnarray}
{\rm Im} ~Q(\nu,\mu^2)& =&\int_0^1 dx \sin(\nu x) \big(q(x,\mu^2) + \bar{q}(x,\mu^2)\big)\nonumber\\& =& \int_0^1 dx \sin(\nu x) \big(q_v(x,\mu^2) + 2 \bar{q}(x,\mu^2)\big)\,.
\label{eq:pdf_fit_imag}
\end{eqnarray}
A combined analysis of these two components will allow for isolating the valence quark and sea/anti-quark contributions to the reduced pseudo-ITD. 

\begin{figure}[ht]
\includegraphics[width=3in]{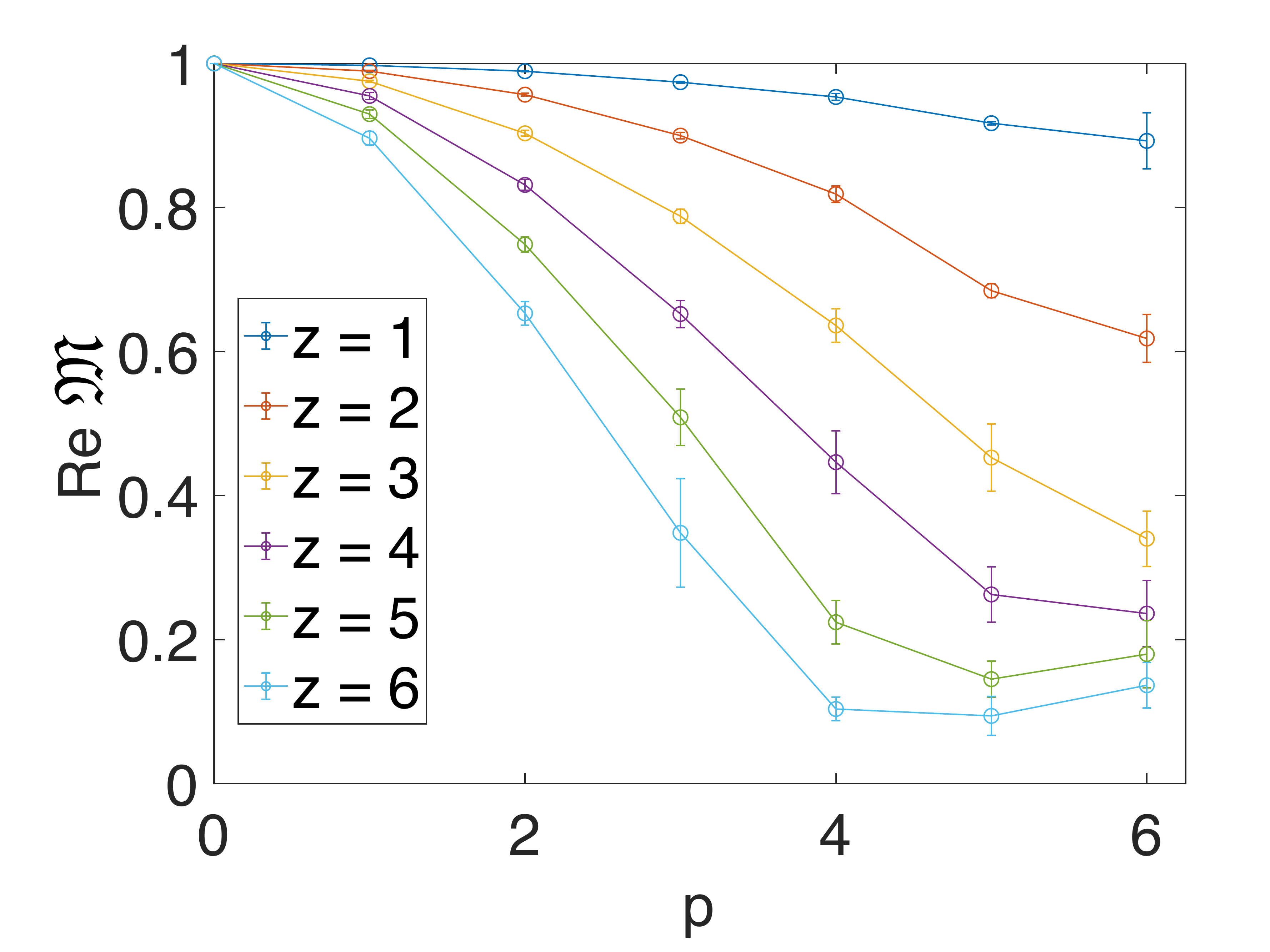}
\includegraphics[width=3in]{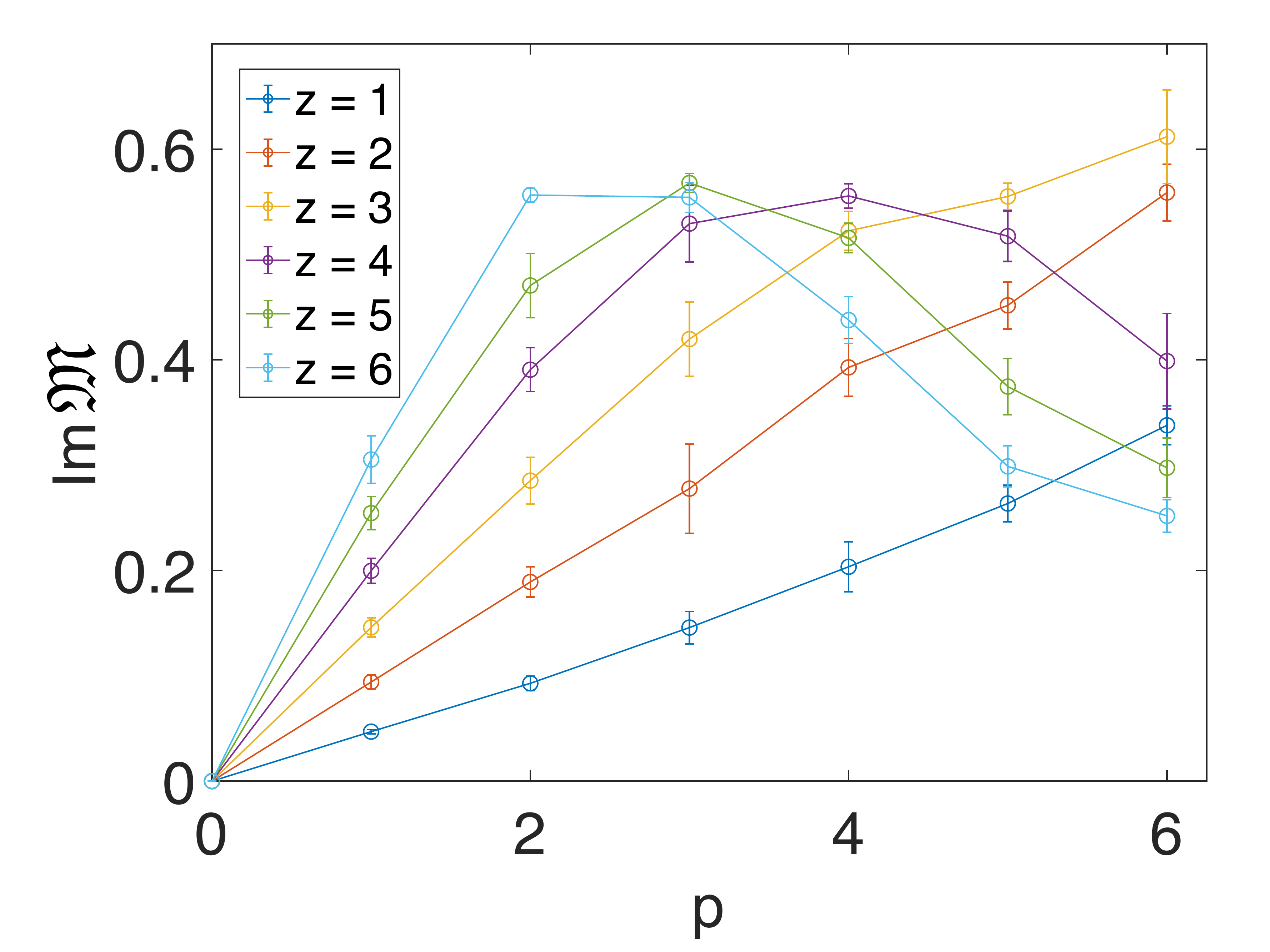}
\includegraphics[width=3in]{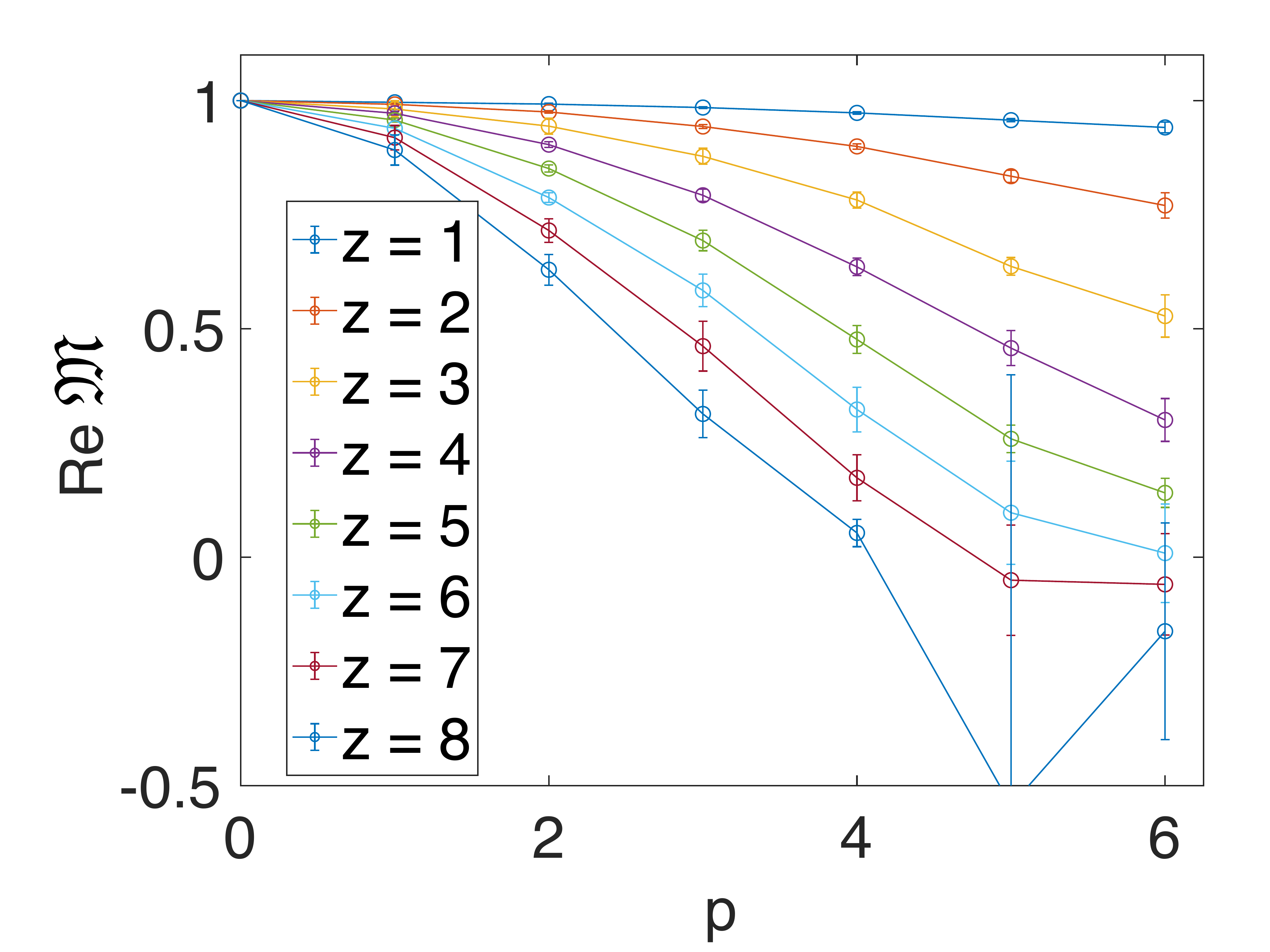}
\includegraphics[width=3in]{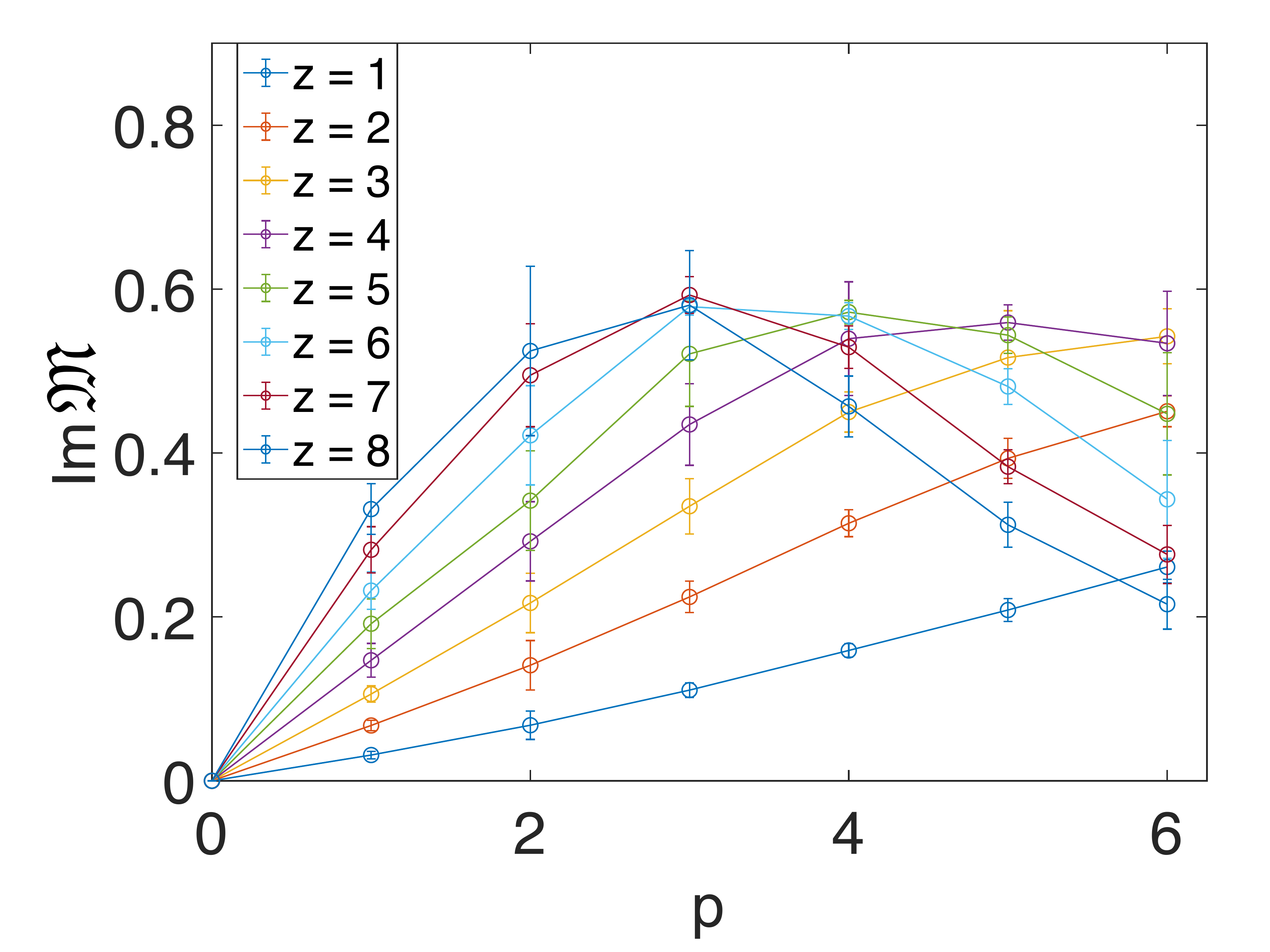}
\includegraphics[width=3in]{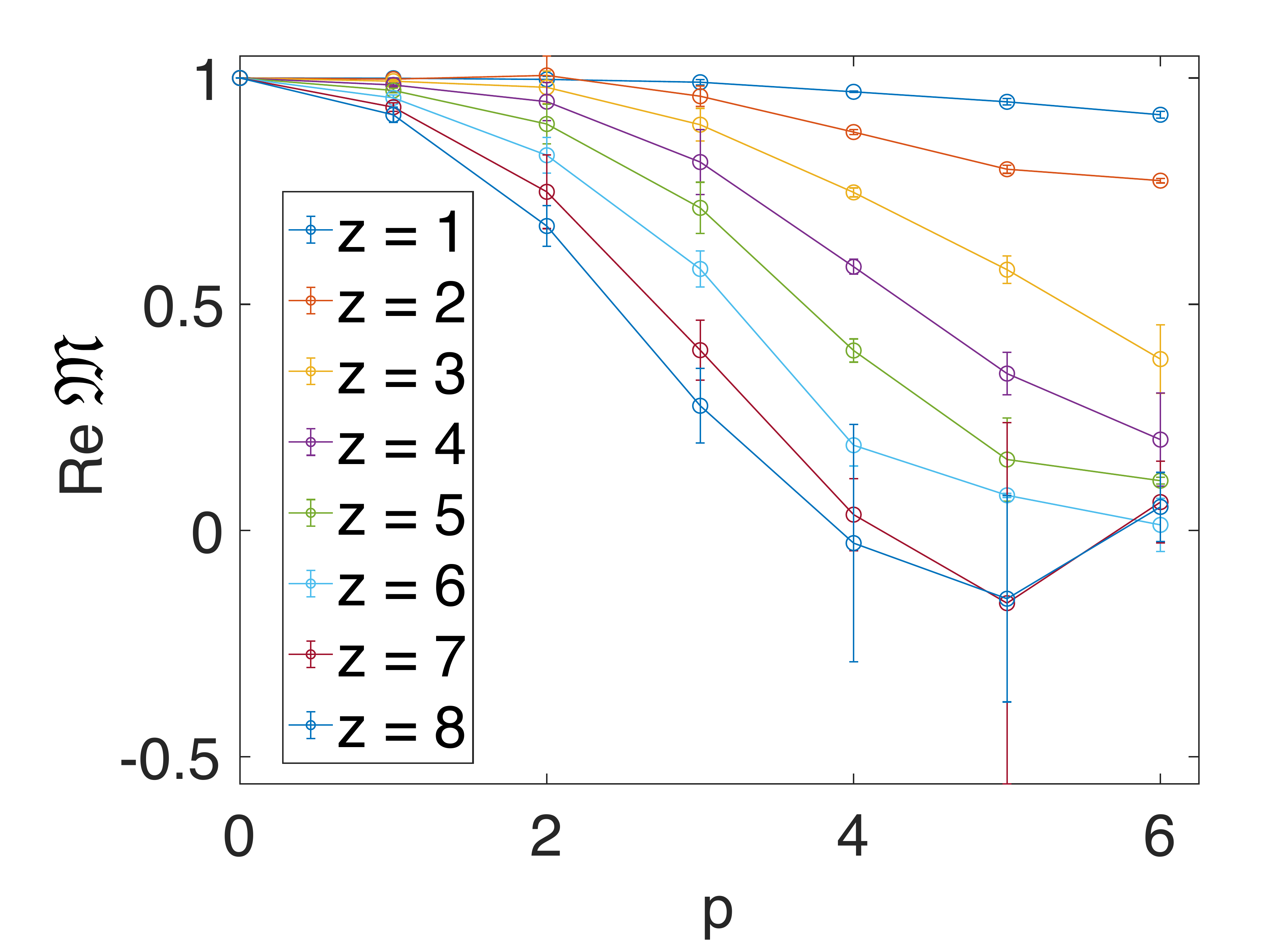}
\includegraphics[width=3in]{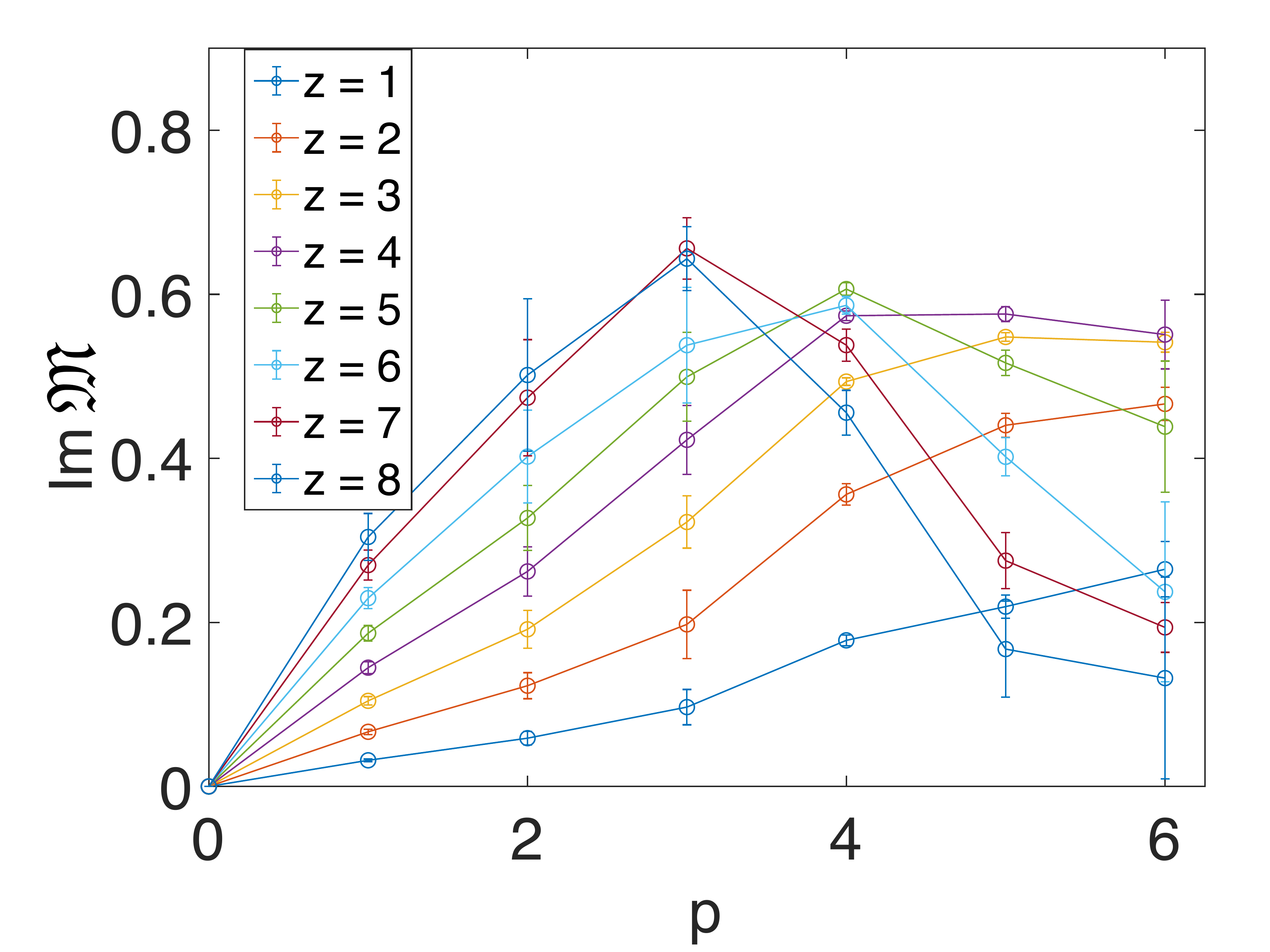}
\caption{The reduced pseudo-ITD as a function of $p$. Here the momenta and displacements are labeled in lattice units, $p_3= 2 \pi p / L$ and $z_3 = z a$. The left plots are the real component and the right are the imaginary component. The top plots are from the ensemble $a127m415$, the middle plots are from the ensemble $a127m415L$, and the bottom plots are from the ensemble $a094m390$. }
\label{fig:Ioffez}
\end{figure}

\begin{figure}[ht]
\includegraphics[width=3in]{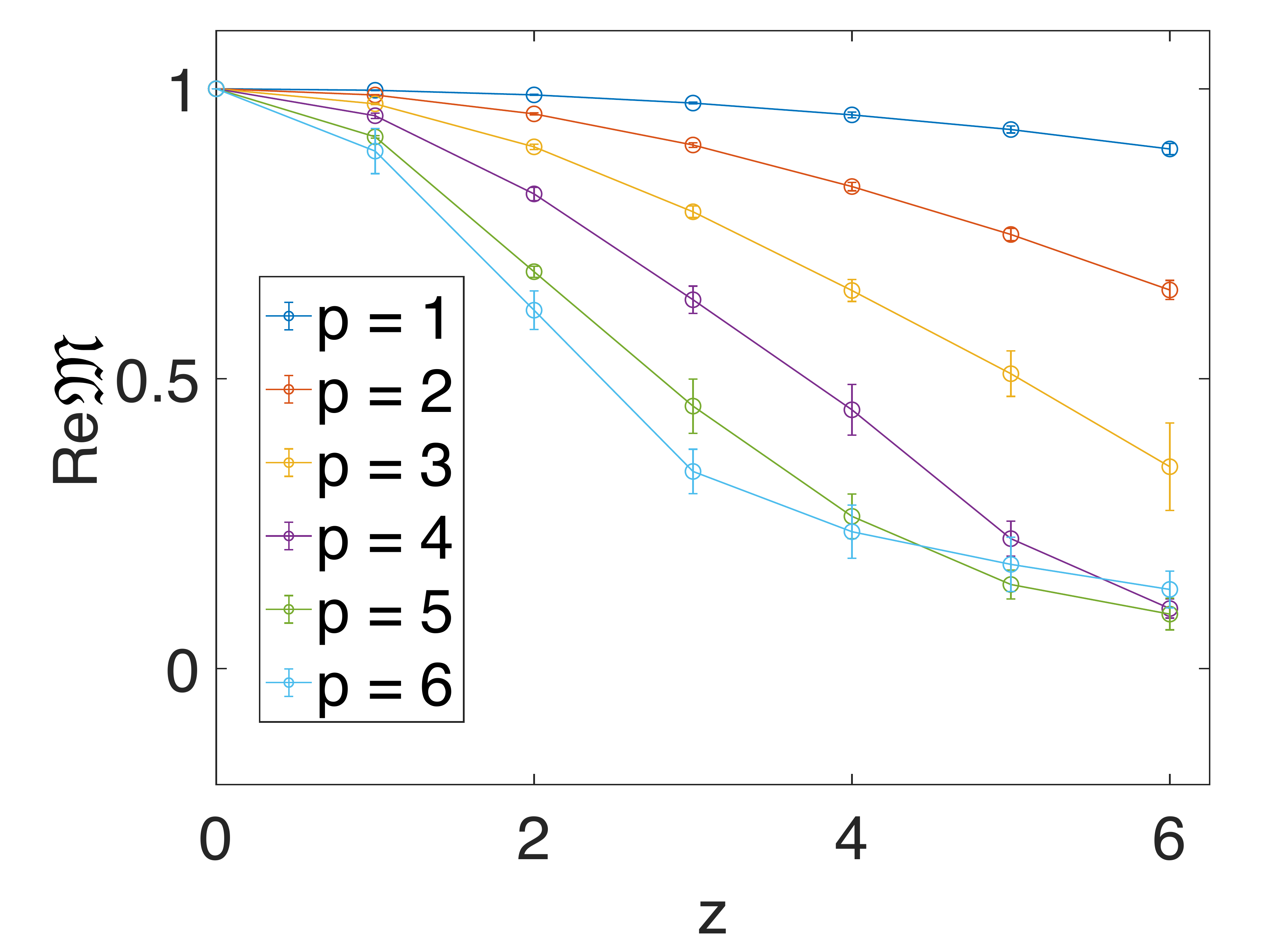}
\includegraphics[width=3in]{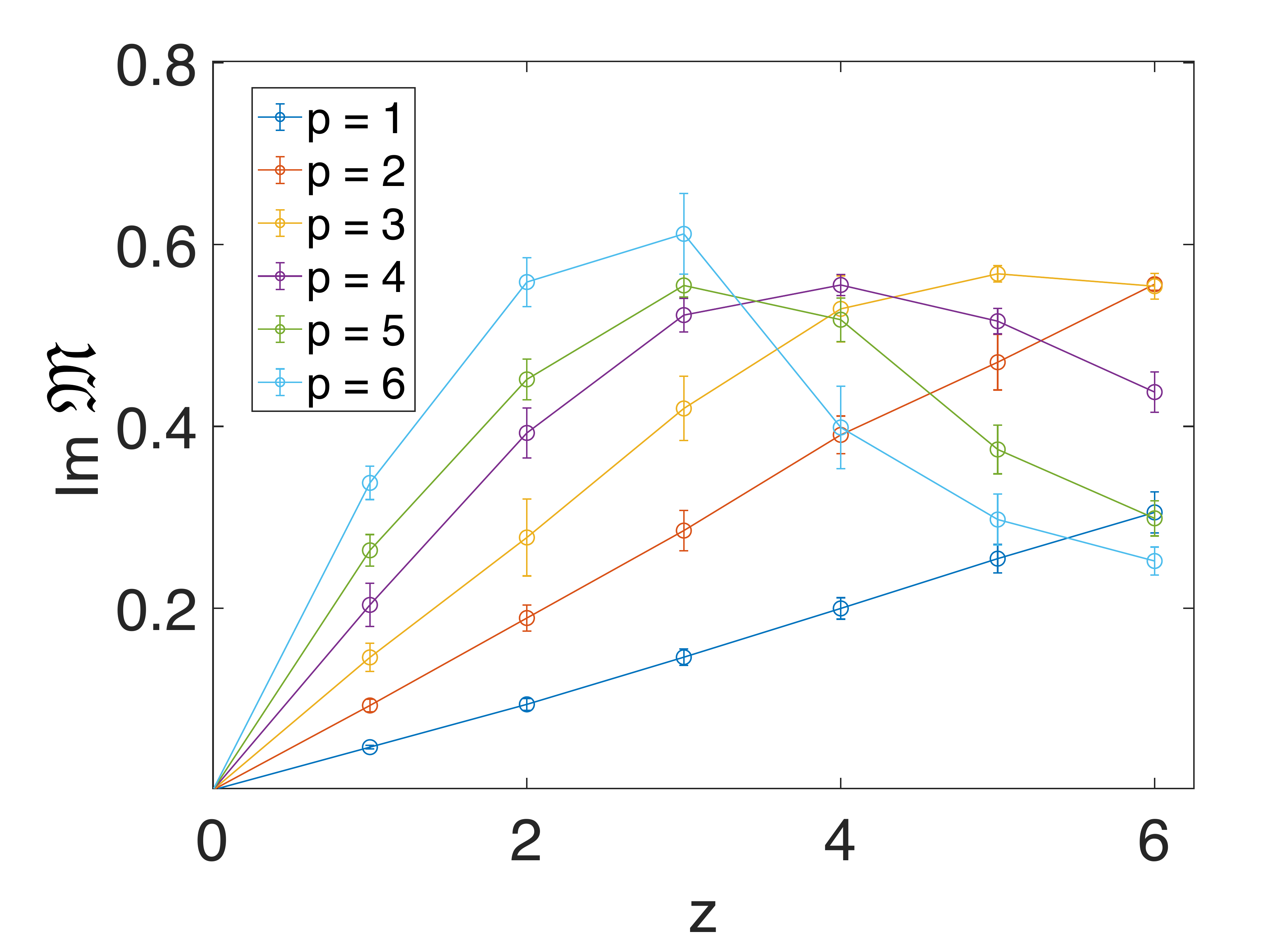}
\includegraphics[width=3in]{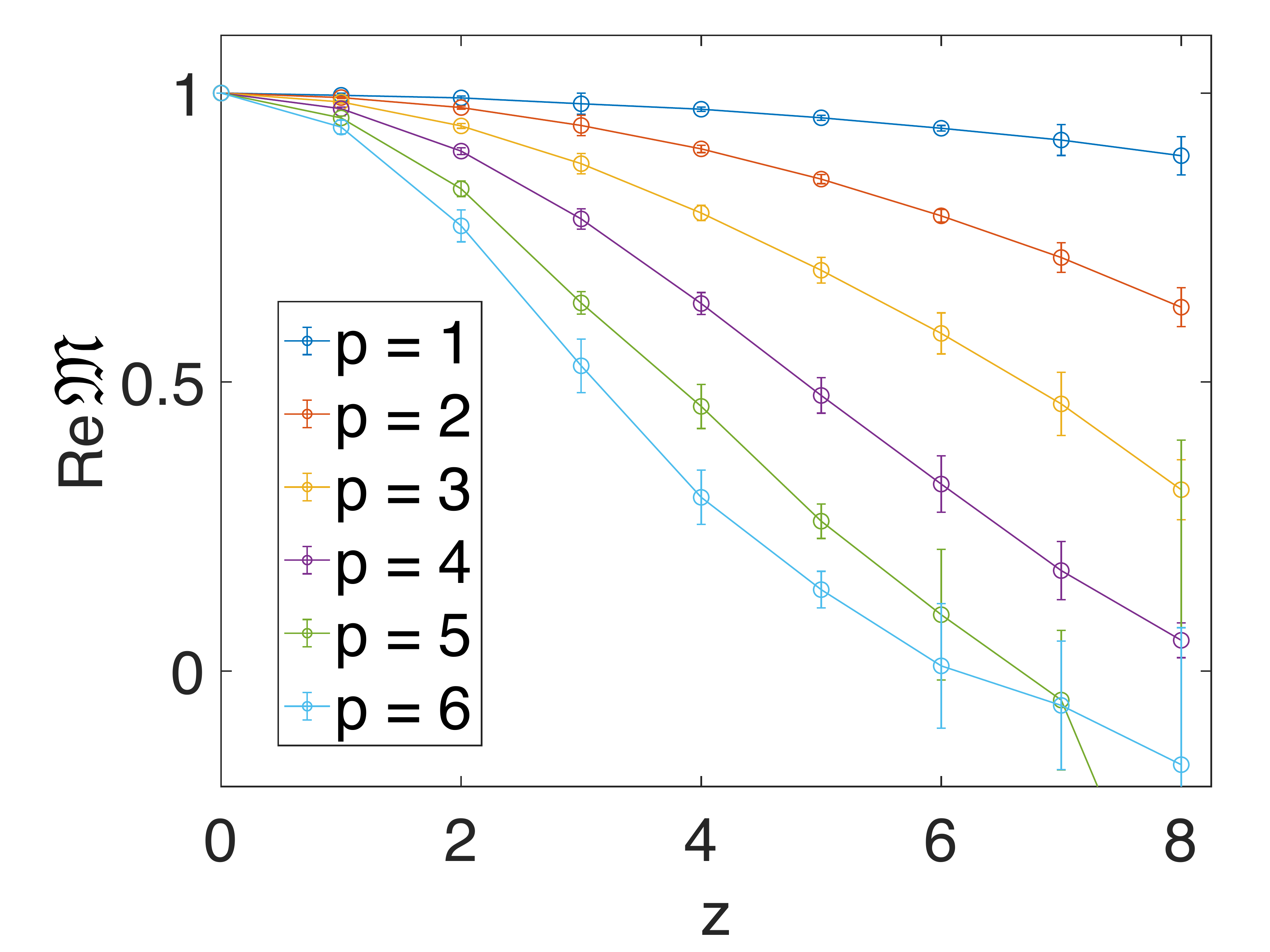}
\includegraphics[width=3in]{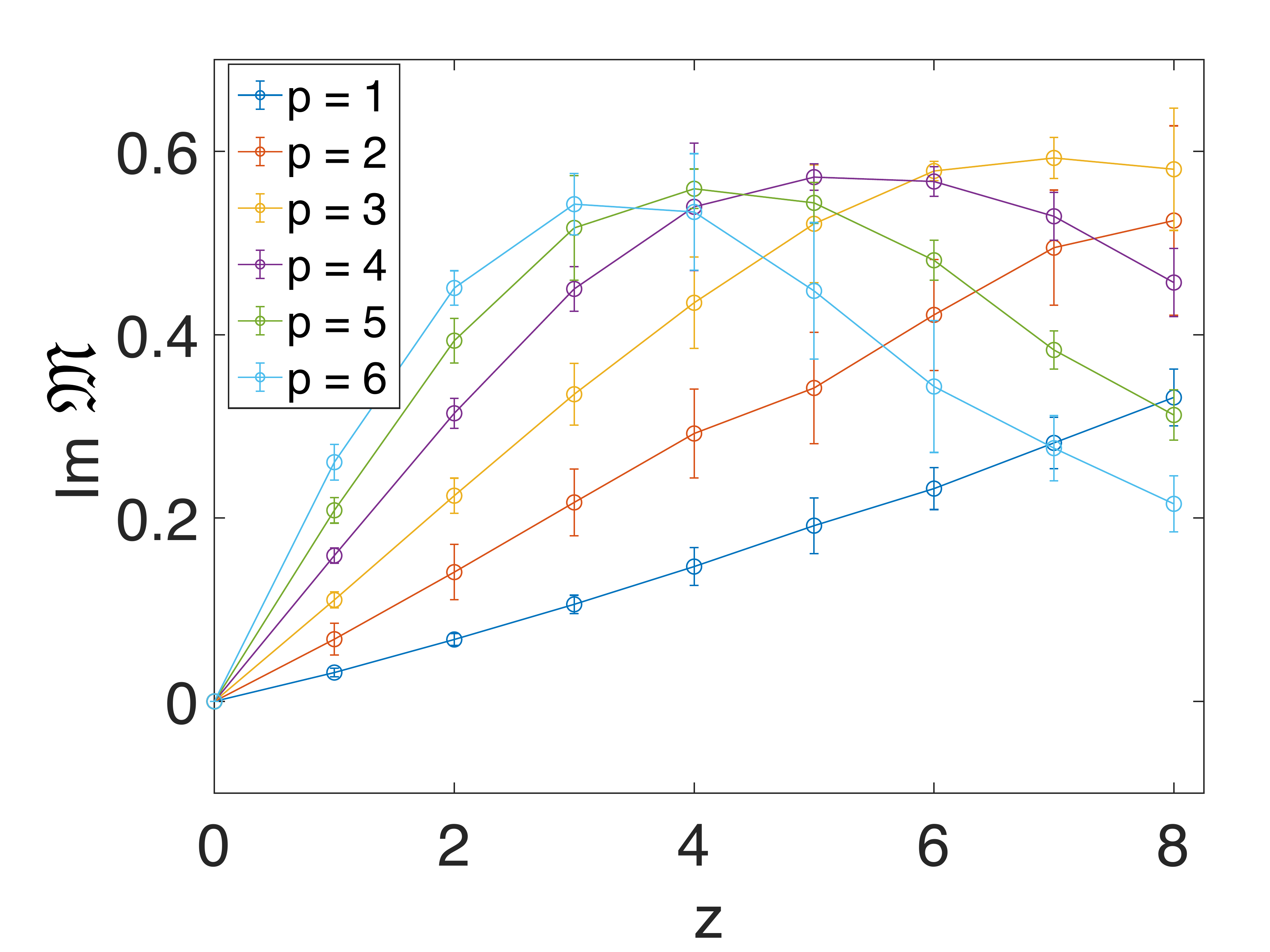}
\includegraphics[width=3in]{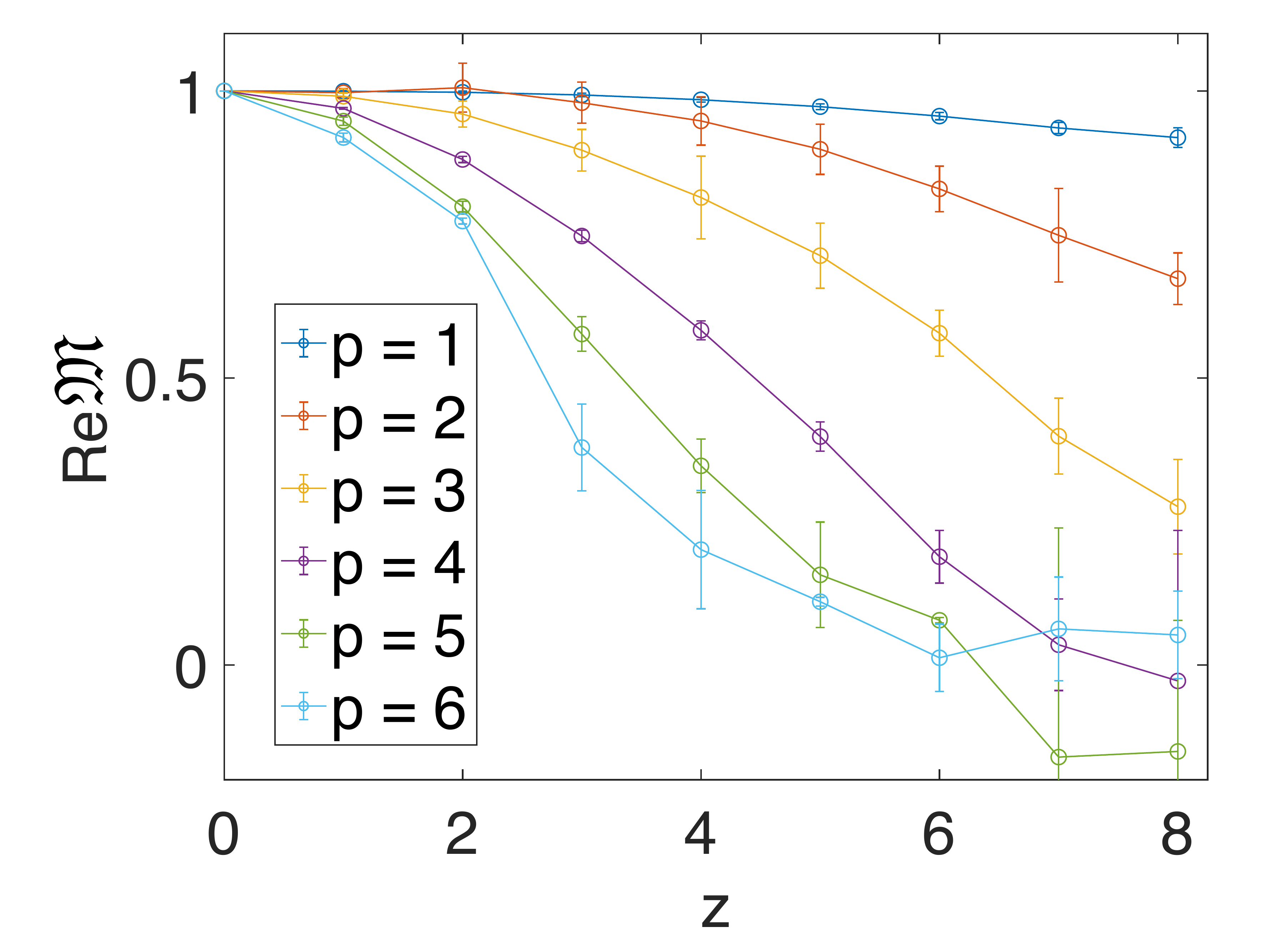}
\includegraphics[width=3in]{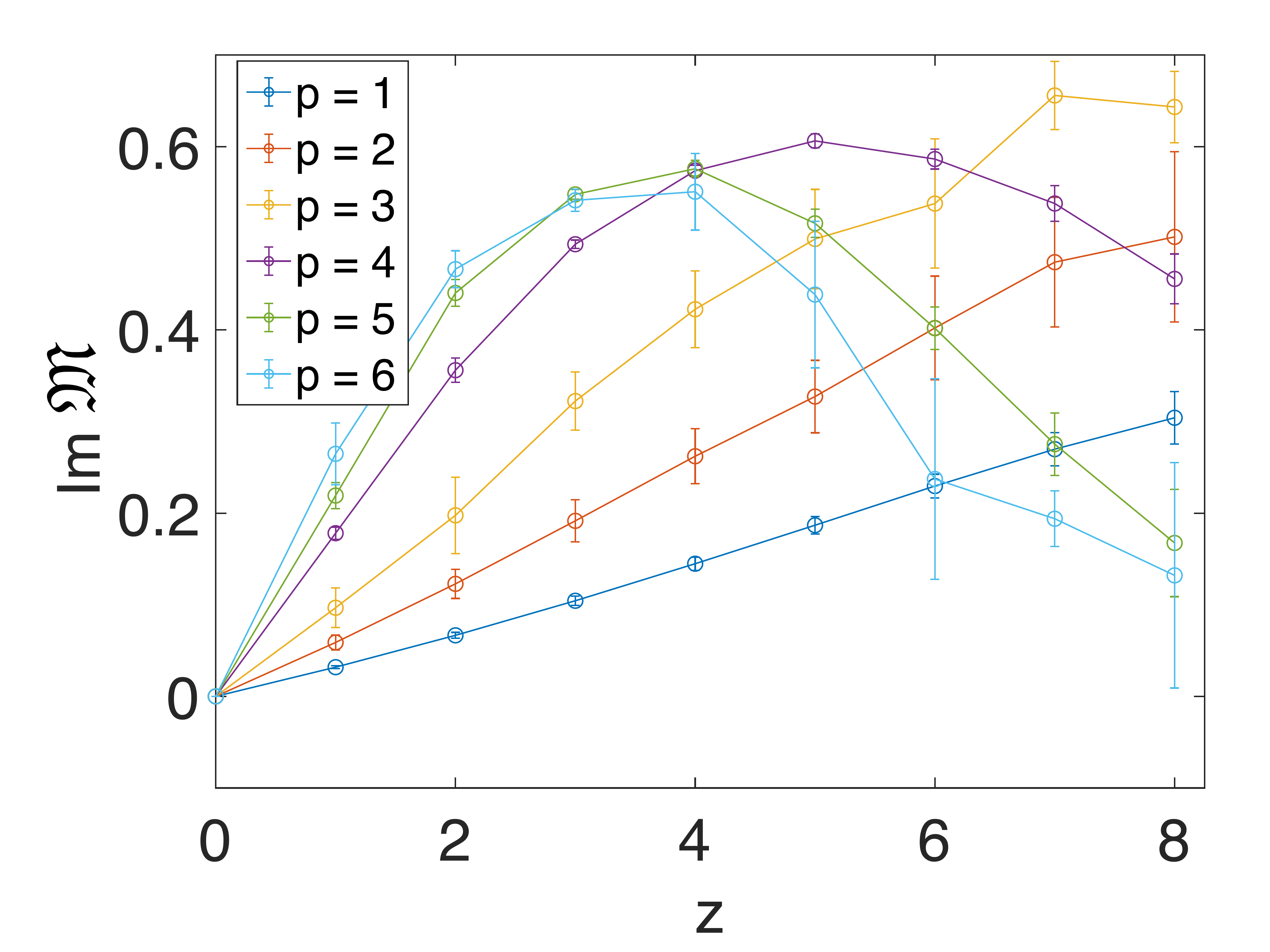}
\caption{The reduced pseudo-ITD as a function of $z$. Here the momenta and displacements are labeled in lattice units, $p_3= 2 \pi p / L$ and $z_3 = z a$. The left plots are the real component and the right are the imaginary component. The top plots are from the ensemble $a127m415$, the middle plots are from the ensemble $a127m415L$, and the bottom plots are from the ensemble $a094m390$. }
\label{fig:Ioffep}
\end{figure}

\begin{figure}[ht]
\includegraphics[width=3in]{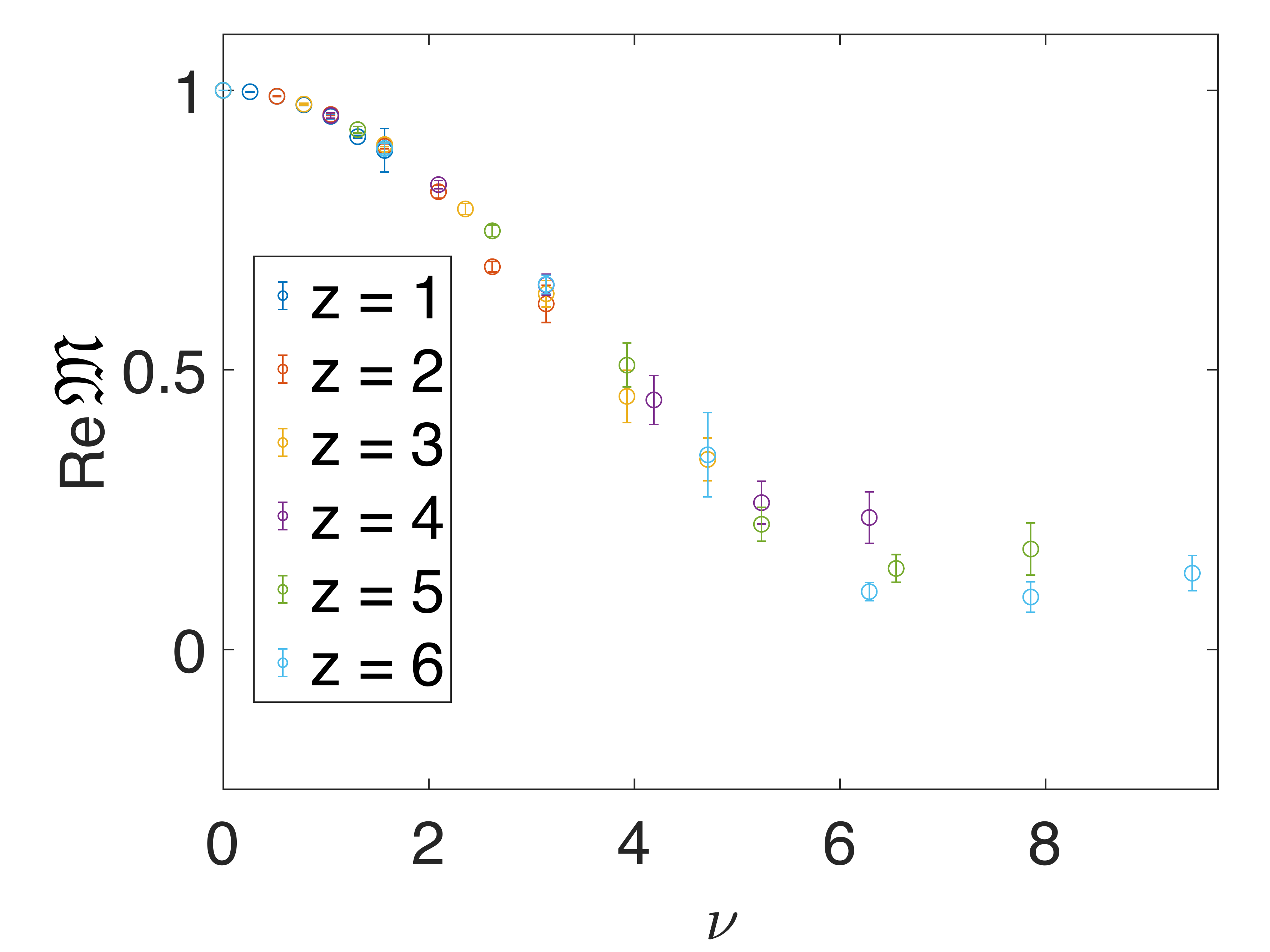}
\includegraphics[width=3in]{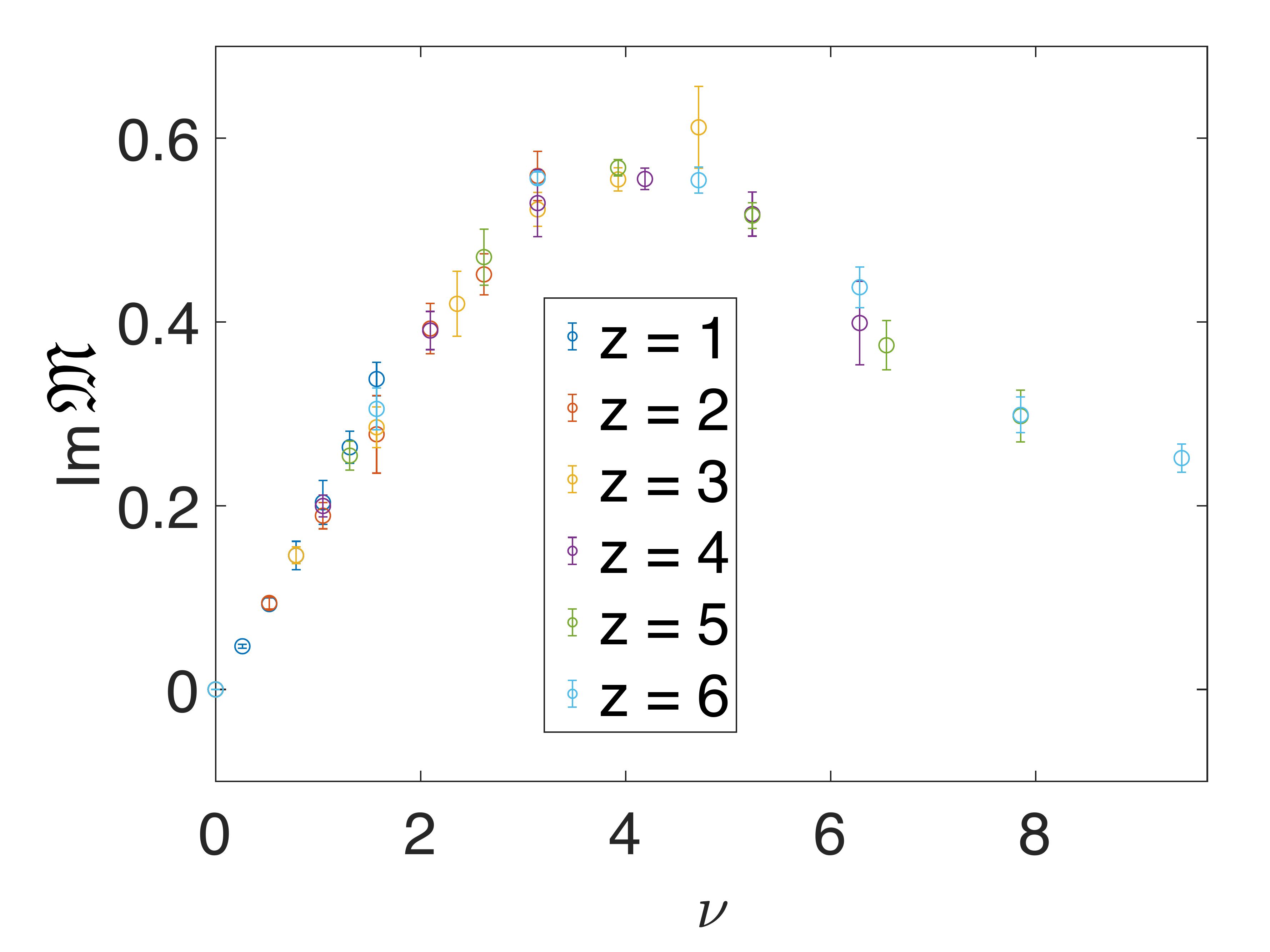}
\caption{The real and imaginary components of the reduced pseudo-ITD on the ensemble $a127m415$ as a function of $\nu$.}
\label{fig:Ioffenucoarse}
\end{figure}

\begin{figure}[ht]
\includegraphics[width=3in]{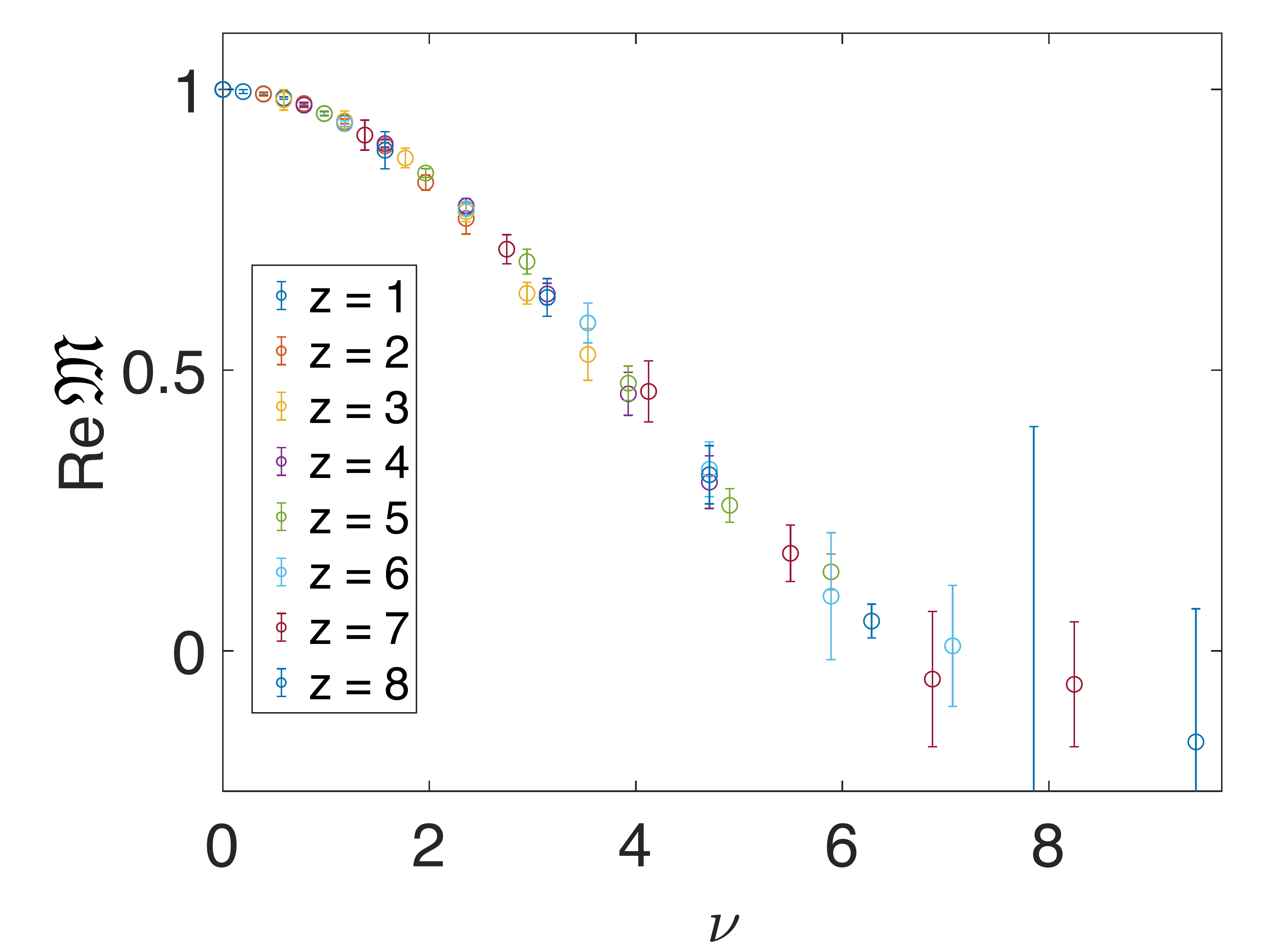}
\includegraphics[width=3in]{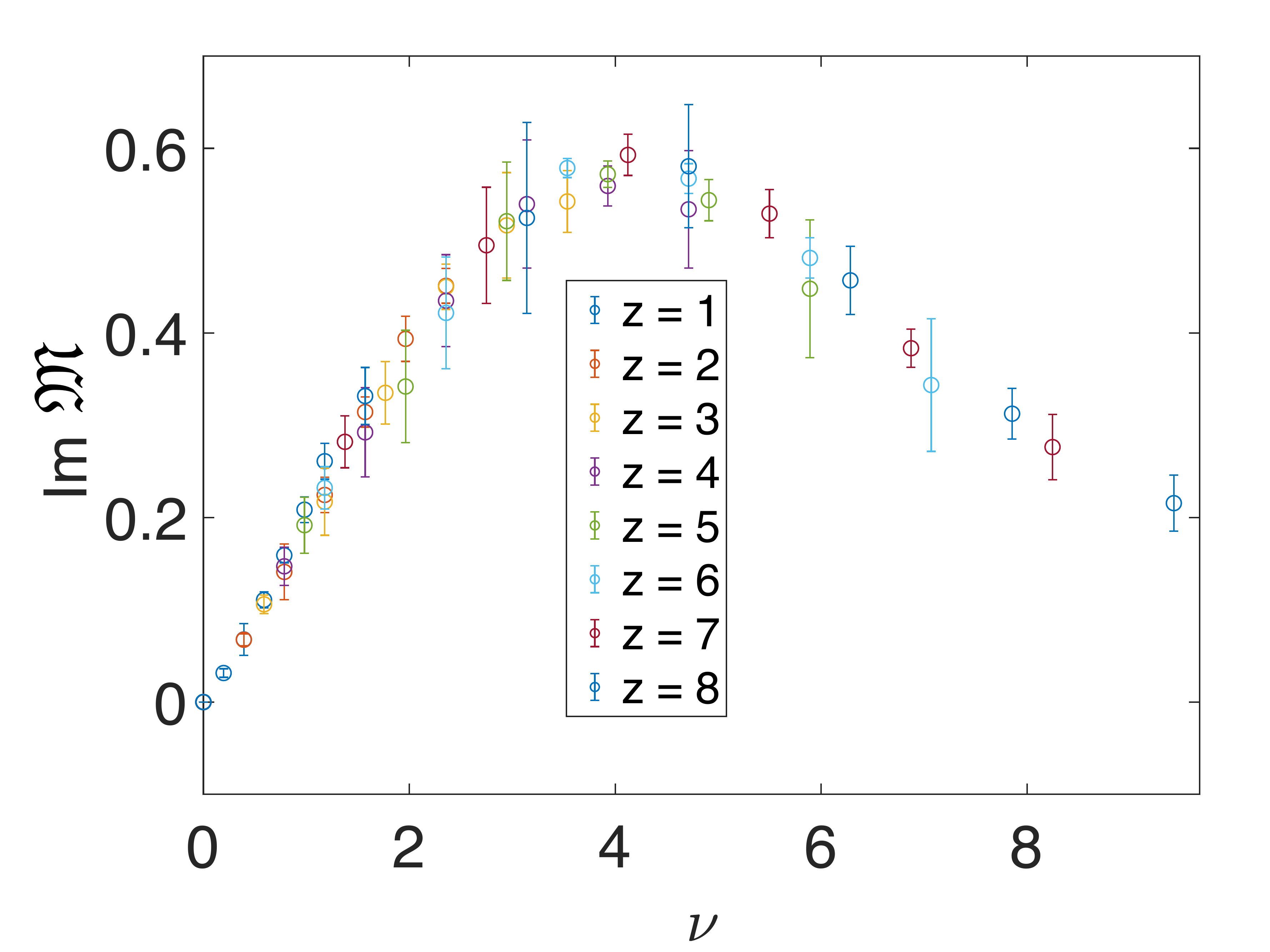}
\caption{The real and imaginary components of the reduced pseudo-ITD on the ensemble $a127m415L$ as a function of $\nu$.}
\label{fig:Ioffenubig}
\end{figure}

\begin{figure}[ht]
\includegraphics[width=3in]{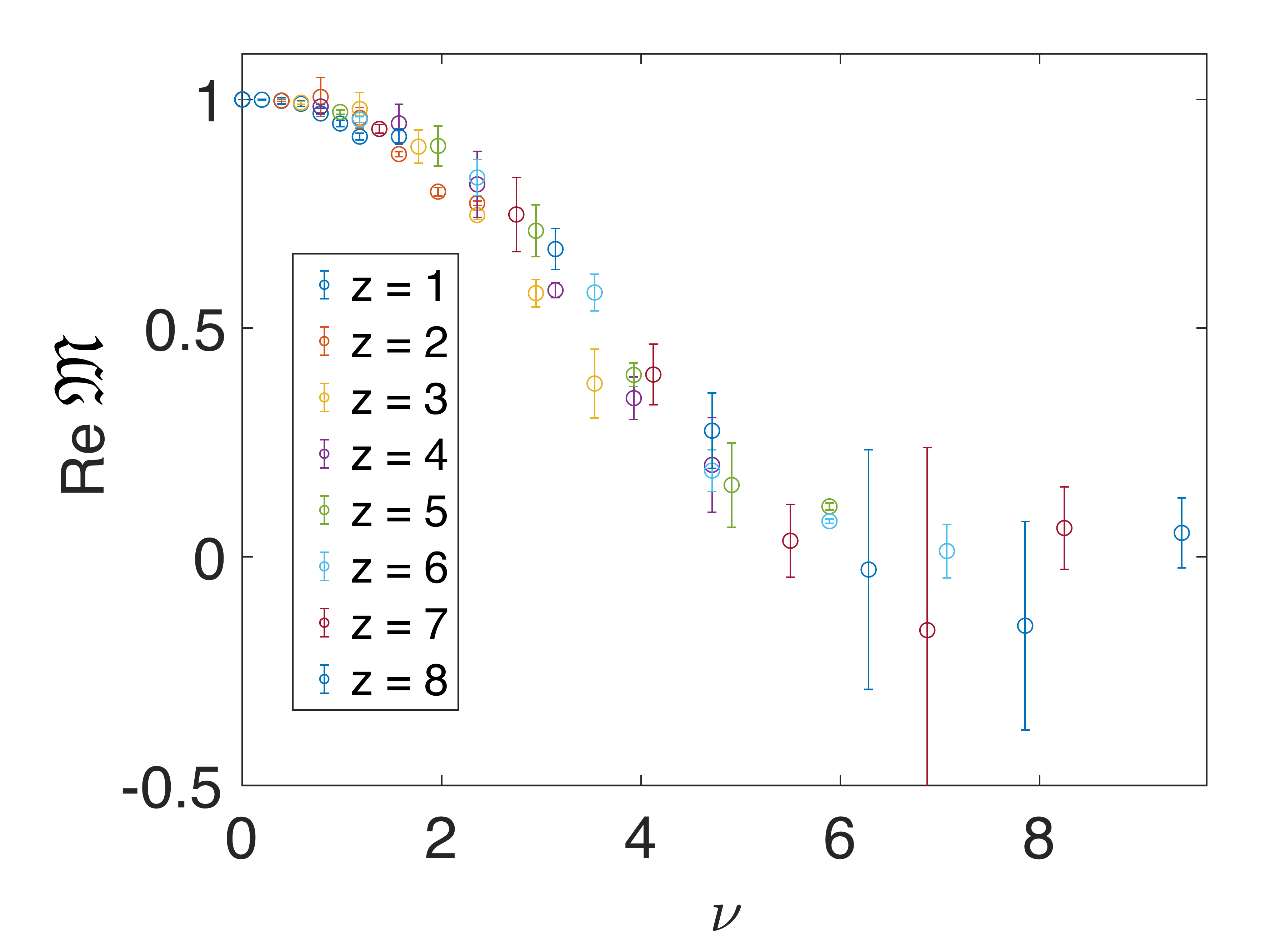}
\includegraphics[width=3in]{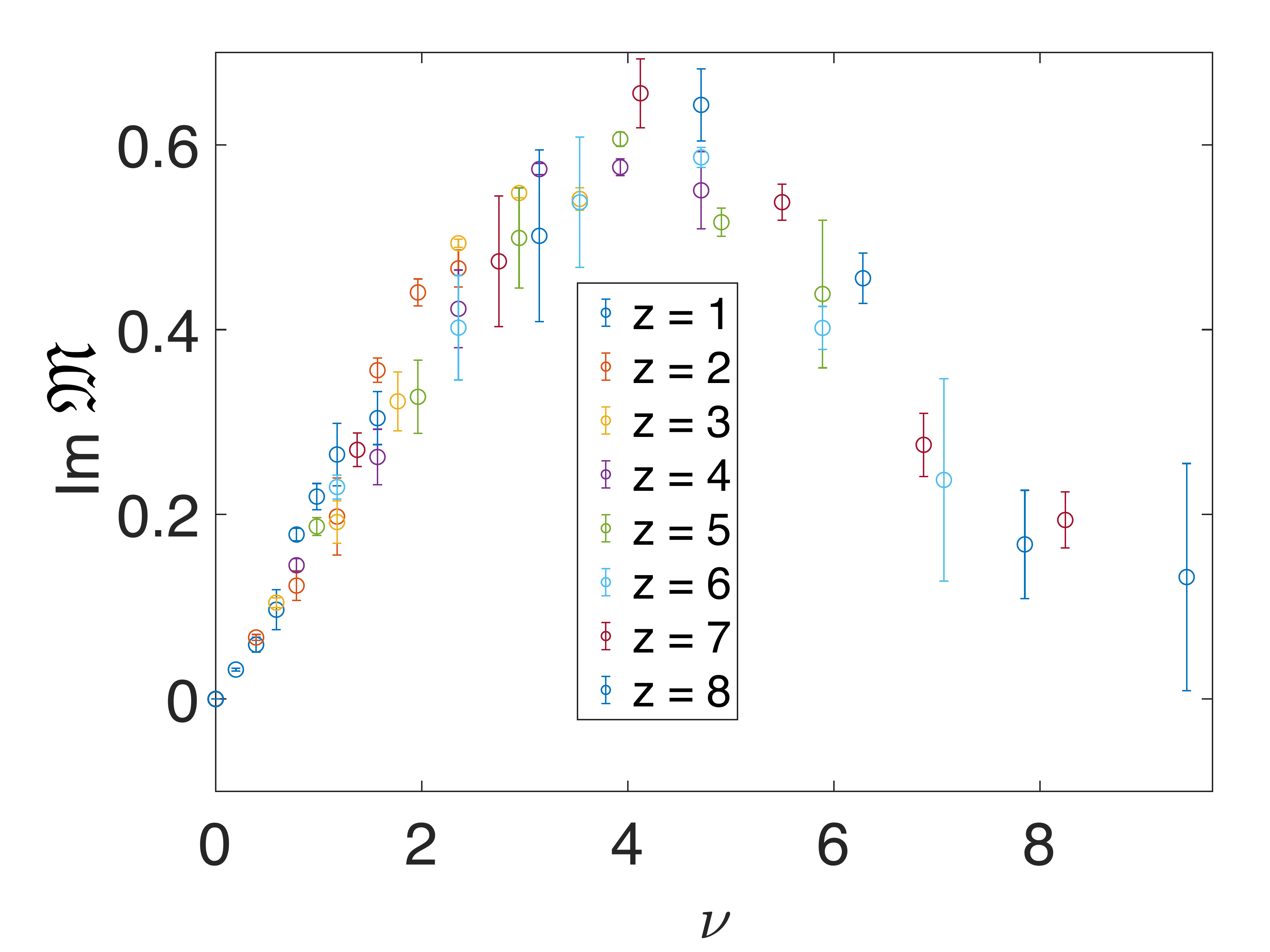}
\caption{The real and imaginary components of the reduced pseudo-ITD on the ensemble $a094m390$ as a function of $\nu$.}
\label{fig:Ioffenufine}
\end{figure}

In Figs.~\ref{fig:Ioffez} and~\ref{fig:Ioffep} we show the reduced
pseudo-ITD as a function of $p_3$ and $z_3$ respectively. The
real-component curves all have a Gaussian shape which suggests that
the renormalization of the Wilson line was indeed canceled. The curves
look similar, but their width  decreases  with increasing momentum. If, 
instead,  the real-component Ioffe time pseudo-distribution is plotted
as a function of the Ioffe time (see  Figs.~\ref{fig:Ioffenucoarse} -
\ref{fig:Ioffenufine}), then the data appear on a more universal curve
which is nearly independent of $z_3^2$. In the absence of higher twist effects, this feature was to be
expected since the perturbative $z^2$ dependence of $\mathfrak{M}(\nu,z^2)$ only
begins at $O(\alpha_s)$.

\subsection{Perturbative evolution and matching to $\overline{\rm MS}$}

Even though the data follow a nearly $z_3$ independent curve, the
DGLAP evolution of the PDF dictates a perturbatively calculable
dependence on the scale $z^2$. Understanding this $z^2$ dependence is
particularly necessary for comparing the data to phenomenological fits
which are renormalized and given at a single scale. The $\overline{\rm MS}$ matching procedures could be performed
in a single step by applying the kernel in
Eq.~\eqref{eq:pseudo_kernel} to each set of data with different $z_3$
independently. It is also possible to separate the $z^2$ evolution
from the $\overline{\rm MS}$ matching steps.  As long as the steps are of
the same order in $\alpha_s$, the one step and two step matching
relationships should result in the exact same final $\overline{\rm MS}$
ITD, though they may have different systematic errors.

Above a certain length scale, the perturbative evolution of the data
ceases and a separation of the $\nu$ and $z^2$ variables
appears, up to the neglected higher twist effects which are partly
canceled by the ratio. This separation of variables results in a $z^2$
independence for the reduced pseudo-ITD for large $z^2$. For the
evolution of the data points in this regime, the initial scale could
be treated as the scale when evolution first appeared to stop. In the
quenched approximation, this scale was found to be $z^{-1} \lesssim
400$ MeV~\cite{Orginos:2017kos}. This scale is particularly low for
using a perturbative evolution, and a non-perturbative evolution
method would be preferable. Failing to account for this cessation of
perturbative evolution would cause the longest distance points to be
evolved away from the universal curve.

In order to perform the convolutions in Eq.~\eqref{eq:pseudo_kernel}, the reduced pseudo-ITD for constant $z^2$ are fit to a sixth degree polynomial and subsequently integrated over. The real and imaginary components are fit to the even and odd powers in the polynomial respectively, with three free parameters each
\bea\label{eq:pseudo-poly}
{\rm Re} ~\mathfrak{M}(\nu) \sim 1 + c_2 \nu^2 + c_4 \nu^4 + c_6 \nu^6 \,, \nonumber \\
{\rm Im}~ \mathfrak{M}(\nu) \sim c_1\nu + c_3\nu^3 + c_5\nu^5 \,.
\eea

\begin{figure}[ht]\centering
\includegraphics[width=0.495\textwidth]{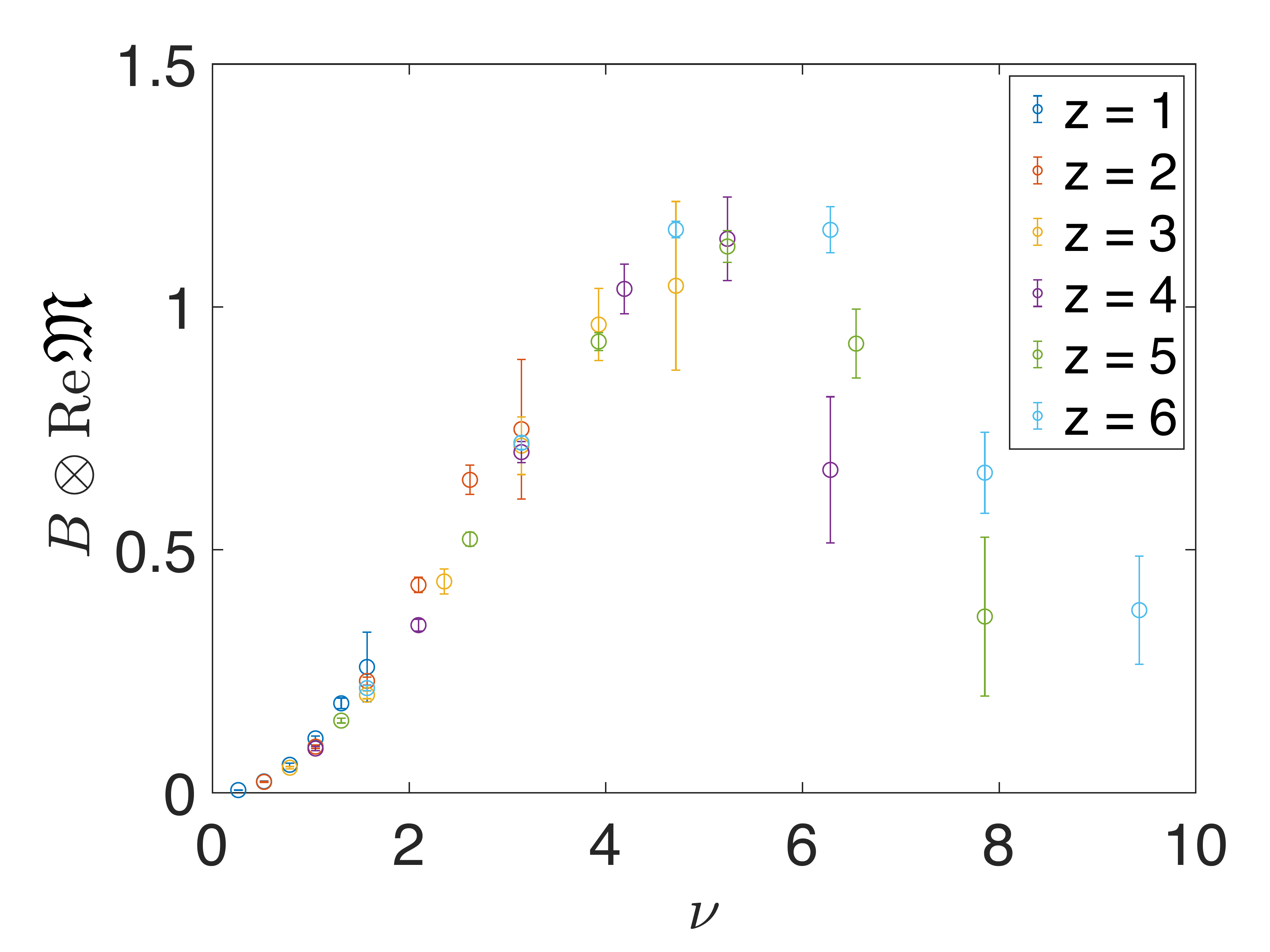}
\includegraphics[width=0.495\textwidth]{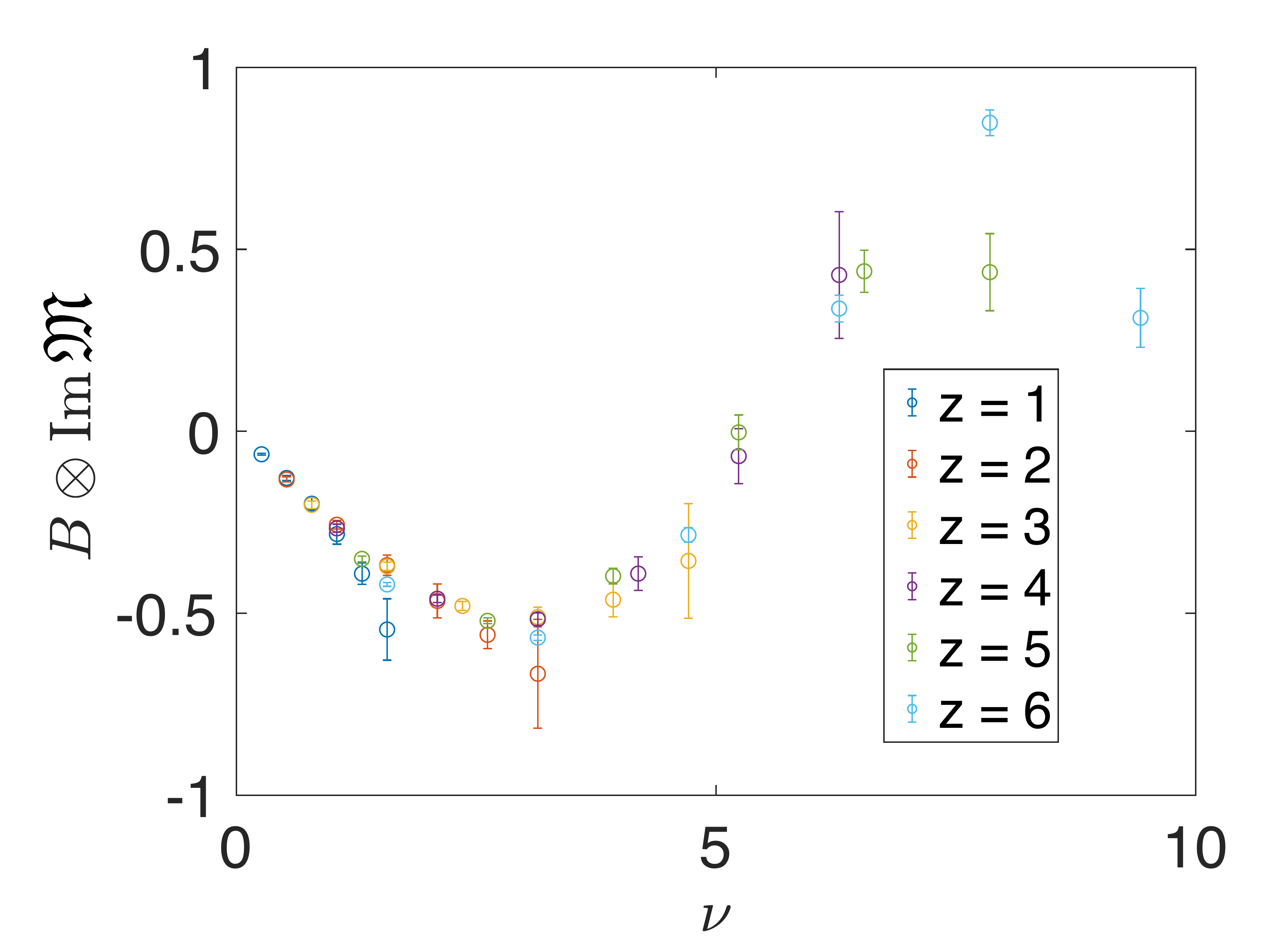}
\includegraphics[width=0.495\textwidth]{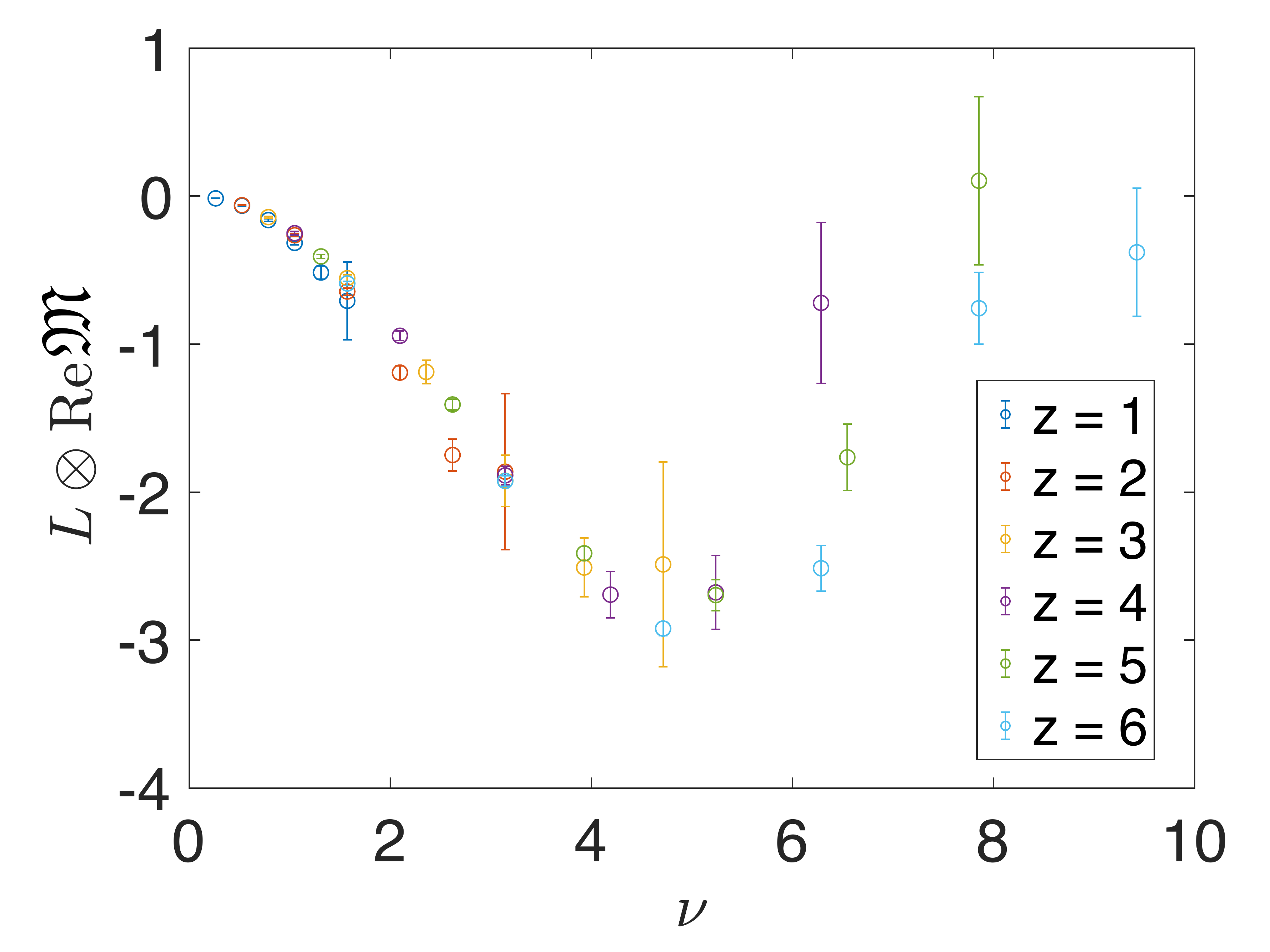}
\includegraphics[width=0.495\textwidth]{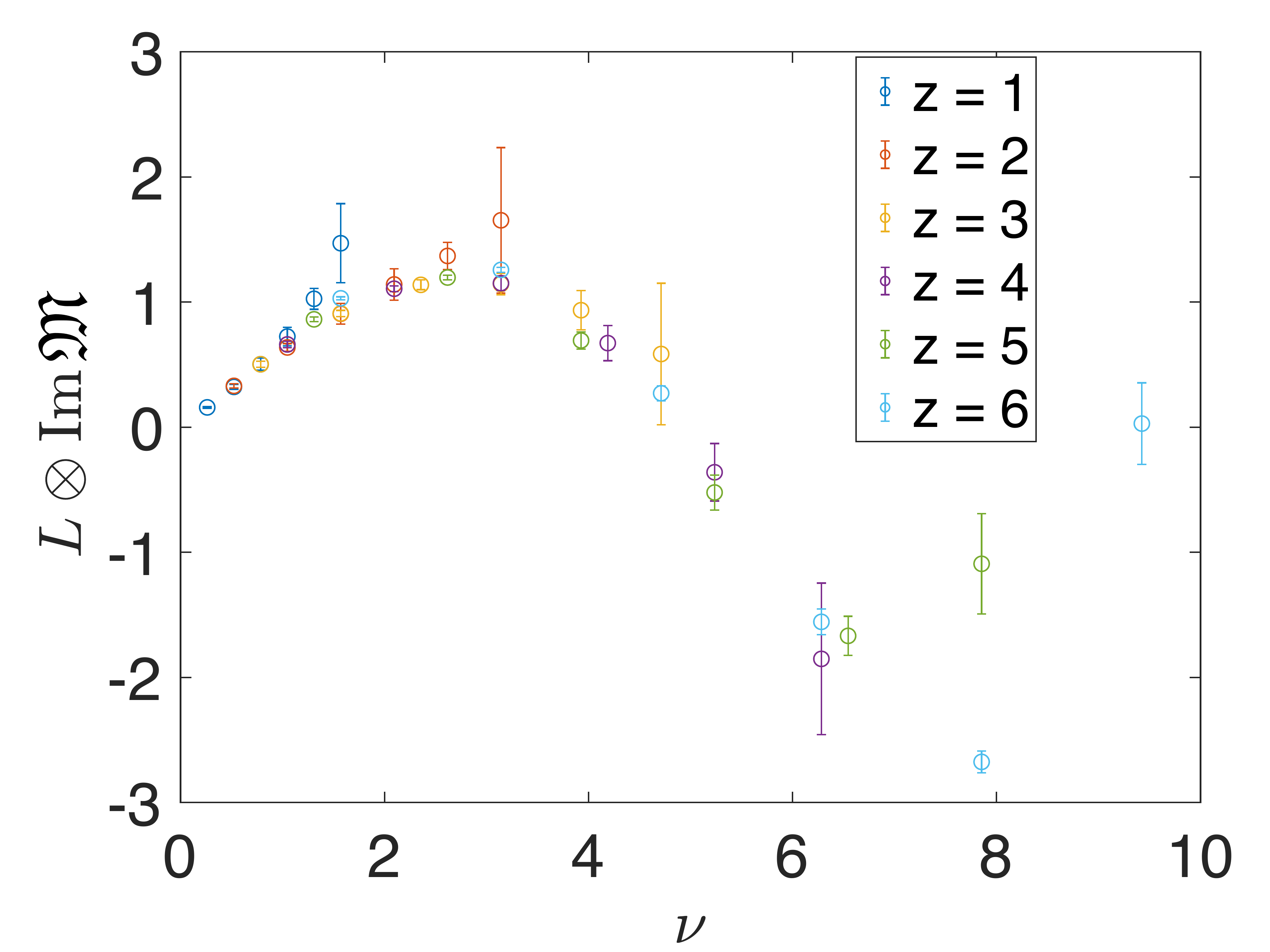}

\caption{The convolutions required for the evolution and matching of the reduced pseudo-ITD to the $\overline{\rm MS}$ ITD on the ensemble $a127m415$. The reduced pseudo-ITD was interpolated by fitting a polynomial.}

\label{fig:convos_coarse}
\end{figure}

\begin{figure}[ht]\centering
\includegraphics[width=0.495\textwidth]{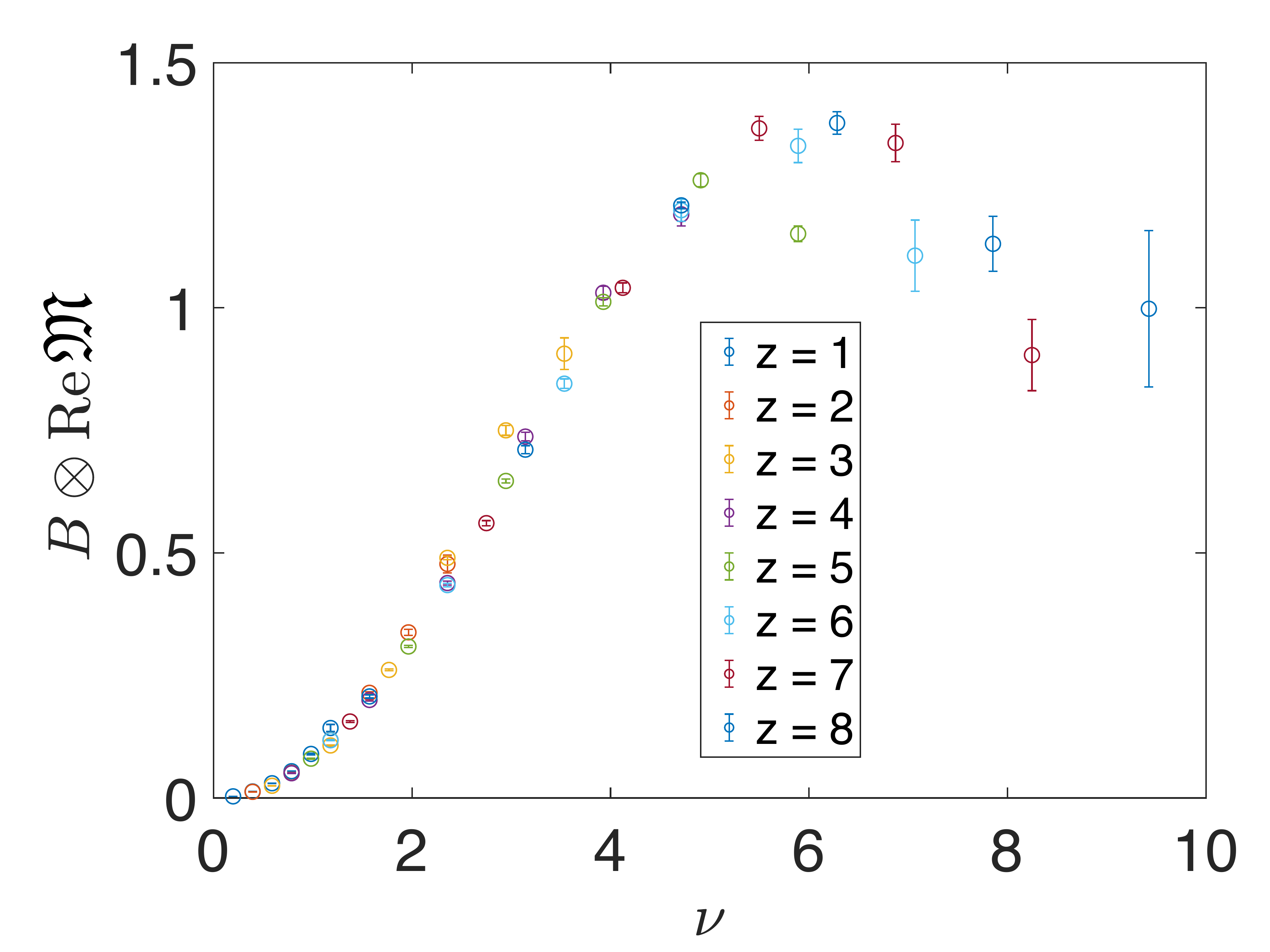}
\includegraphics[width=0.495\textwidth]{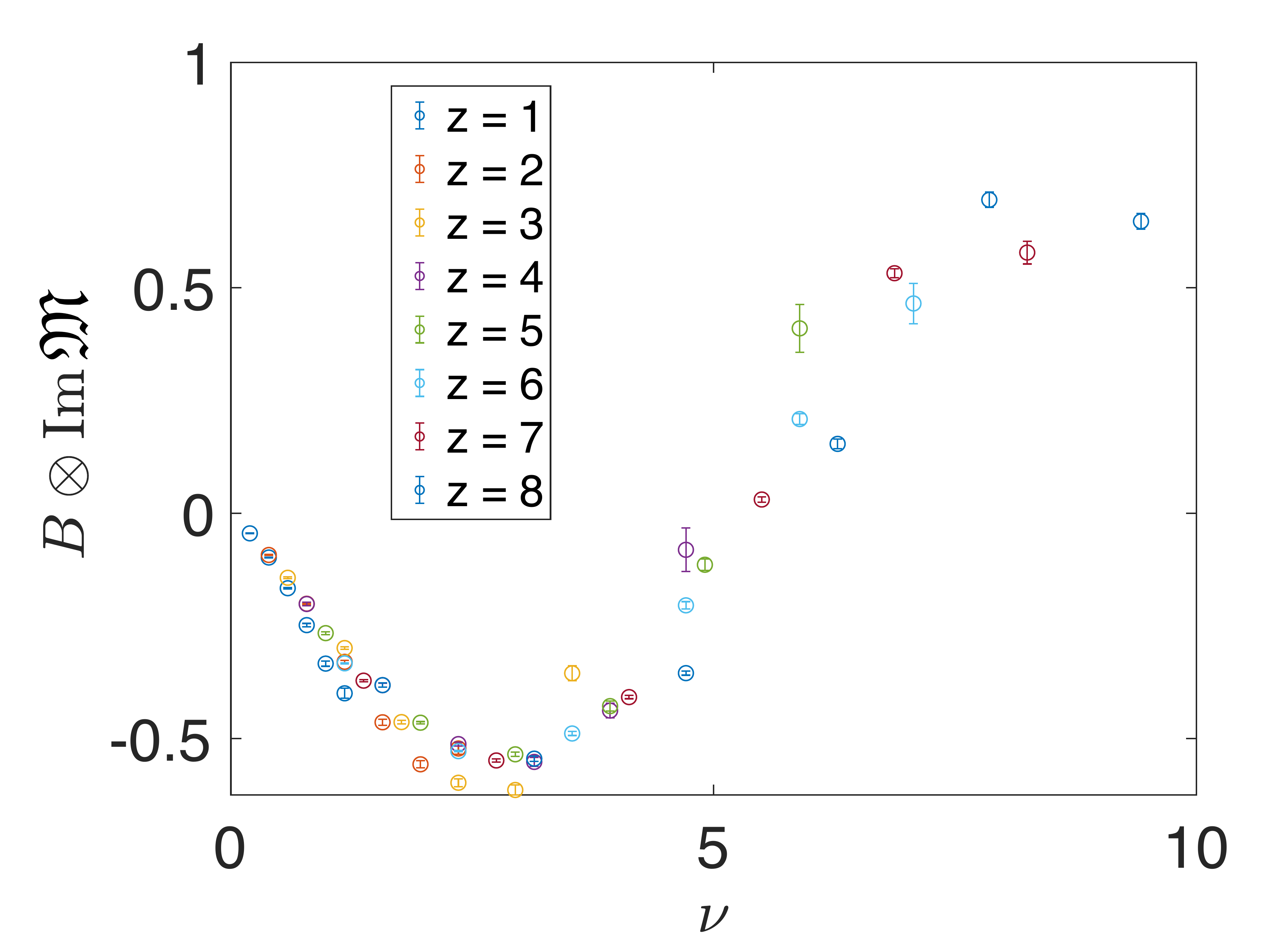}
\includegraphics[width=0.495\textwidth]{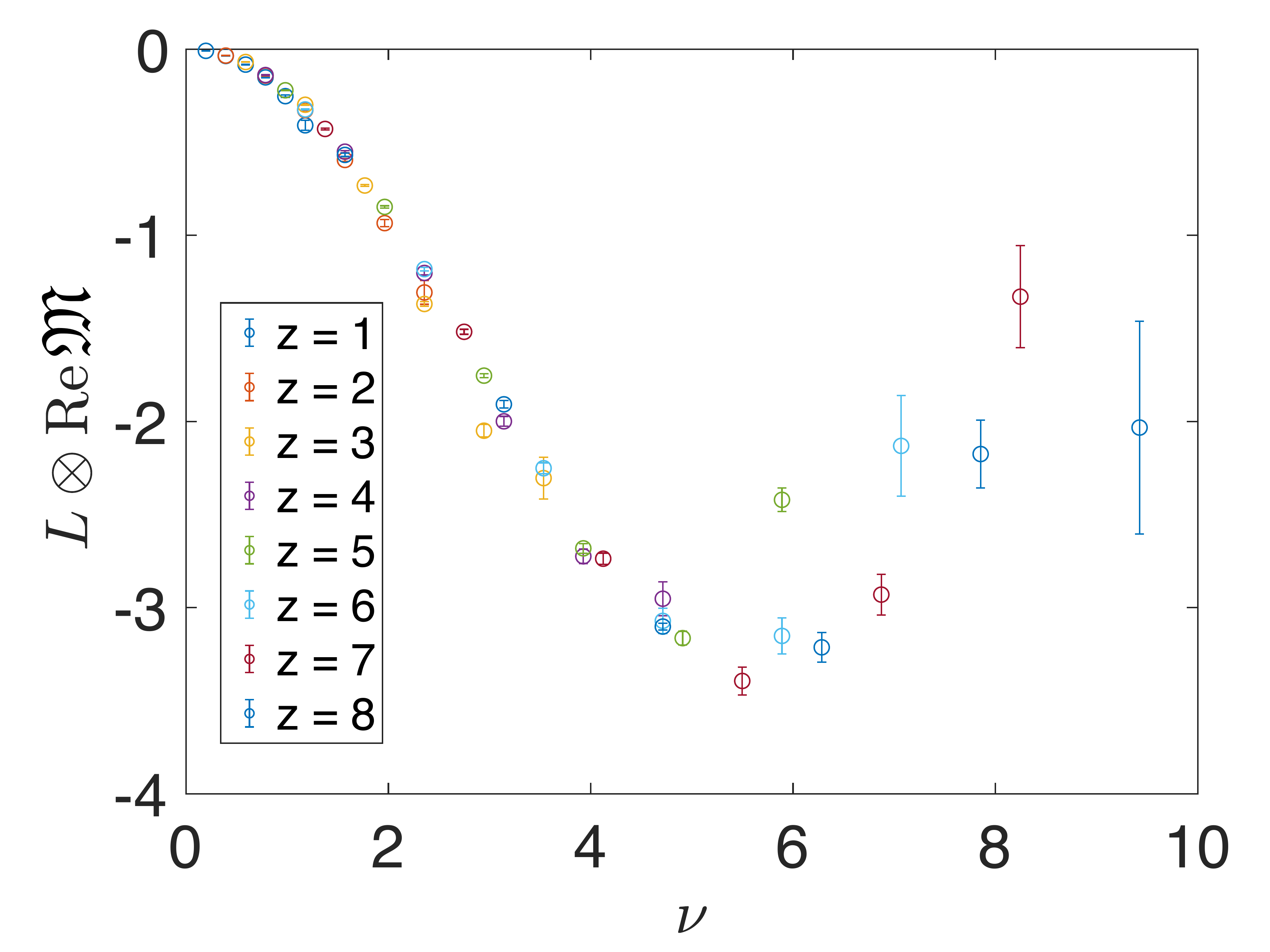}
\includegraphics[width=0.495\textwidth]{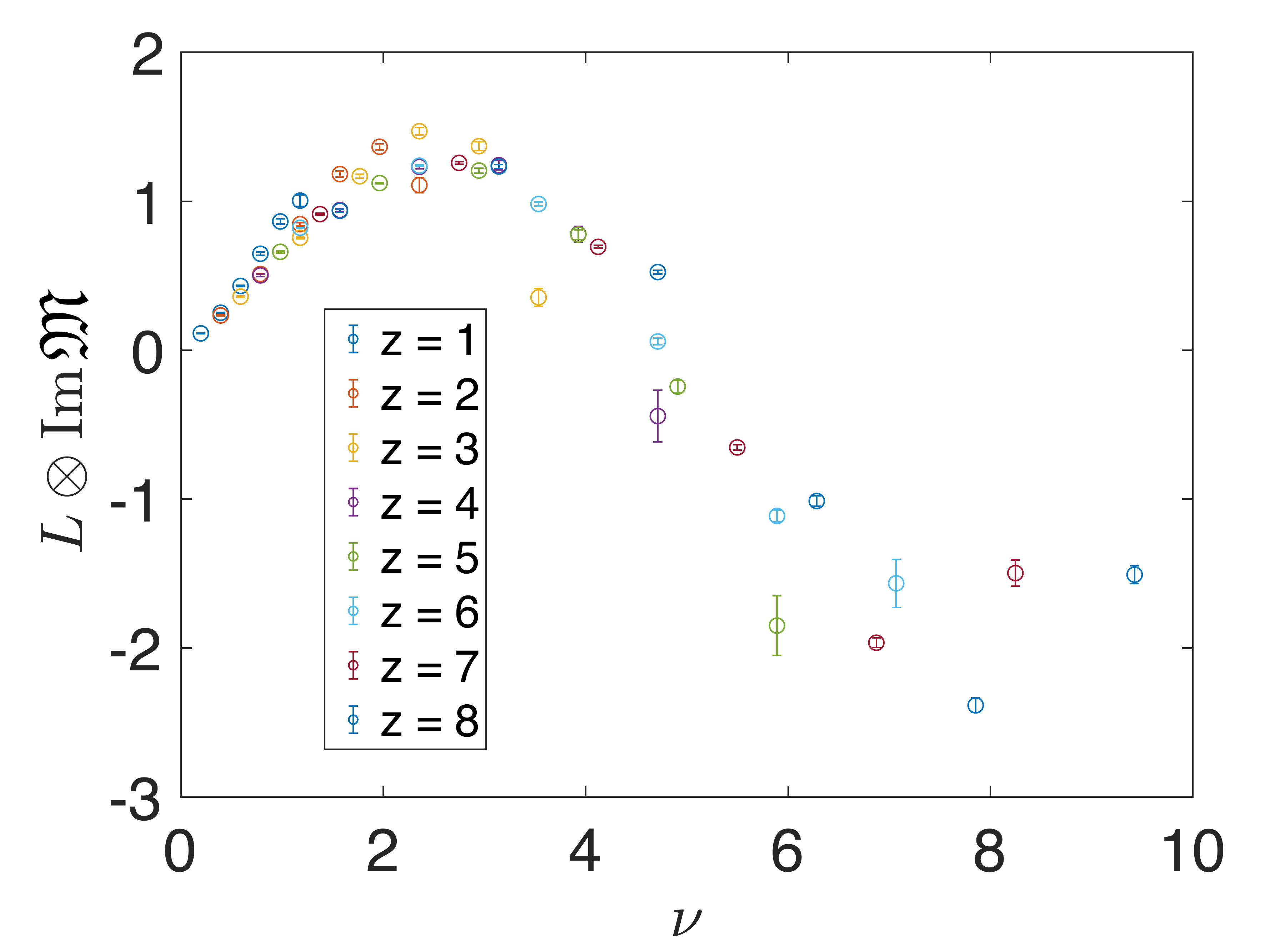}

\caption{The convolutions required for the evolution and matching of the reduced pseudo-ITD to the $\overline{\rm MS}$ ITD on the ensemble $a127m415L$. The reduced pseudo-ITD was interpolated by fitting a polynomial. }

\label{fig:convos_big}
\end{figure}

\begin{figure}[ht]\centering
\includegraphics[width=0.495\textwidth]{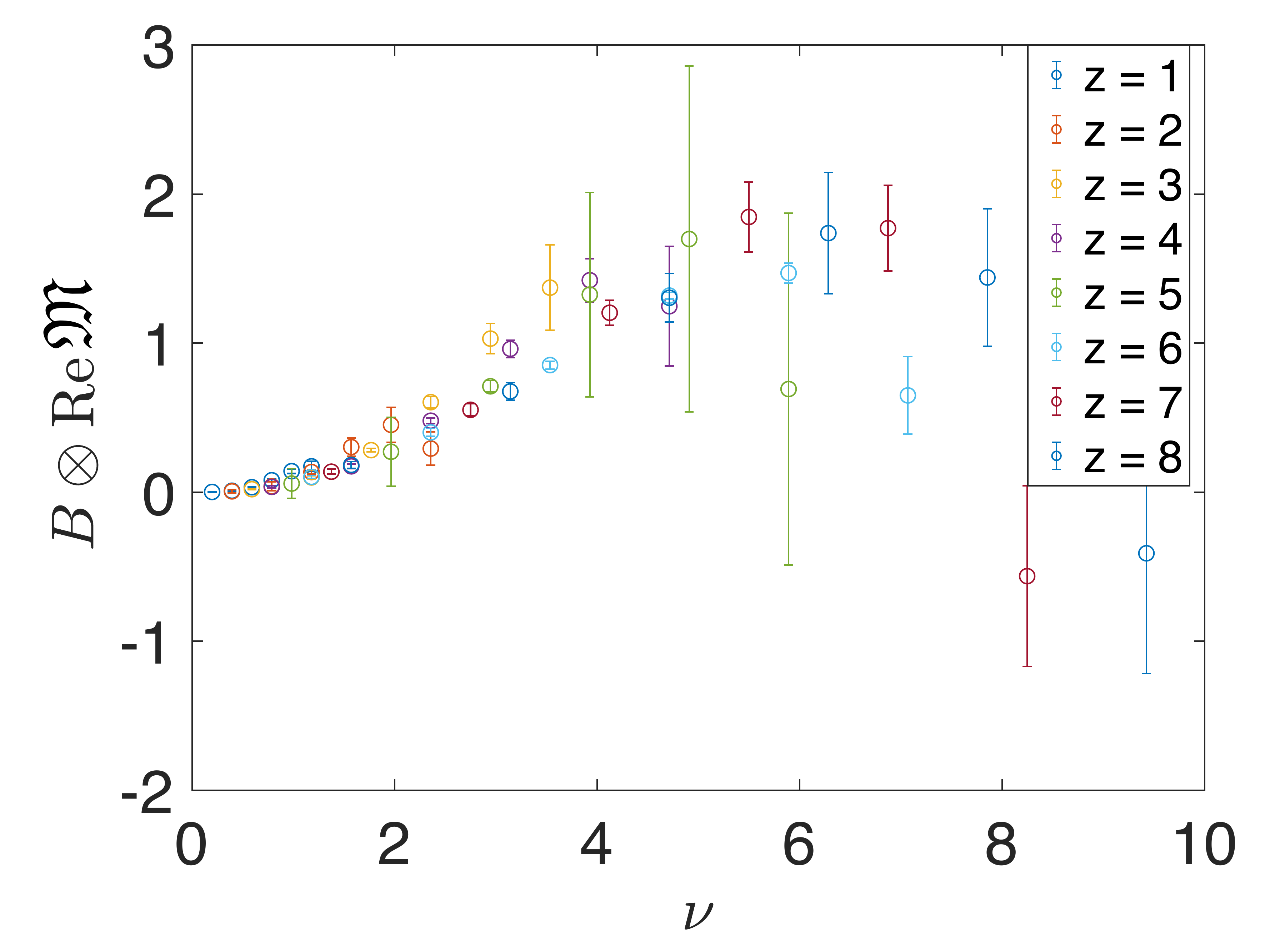}
\includegraphics[width=0.495\textwidth]{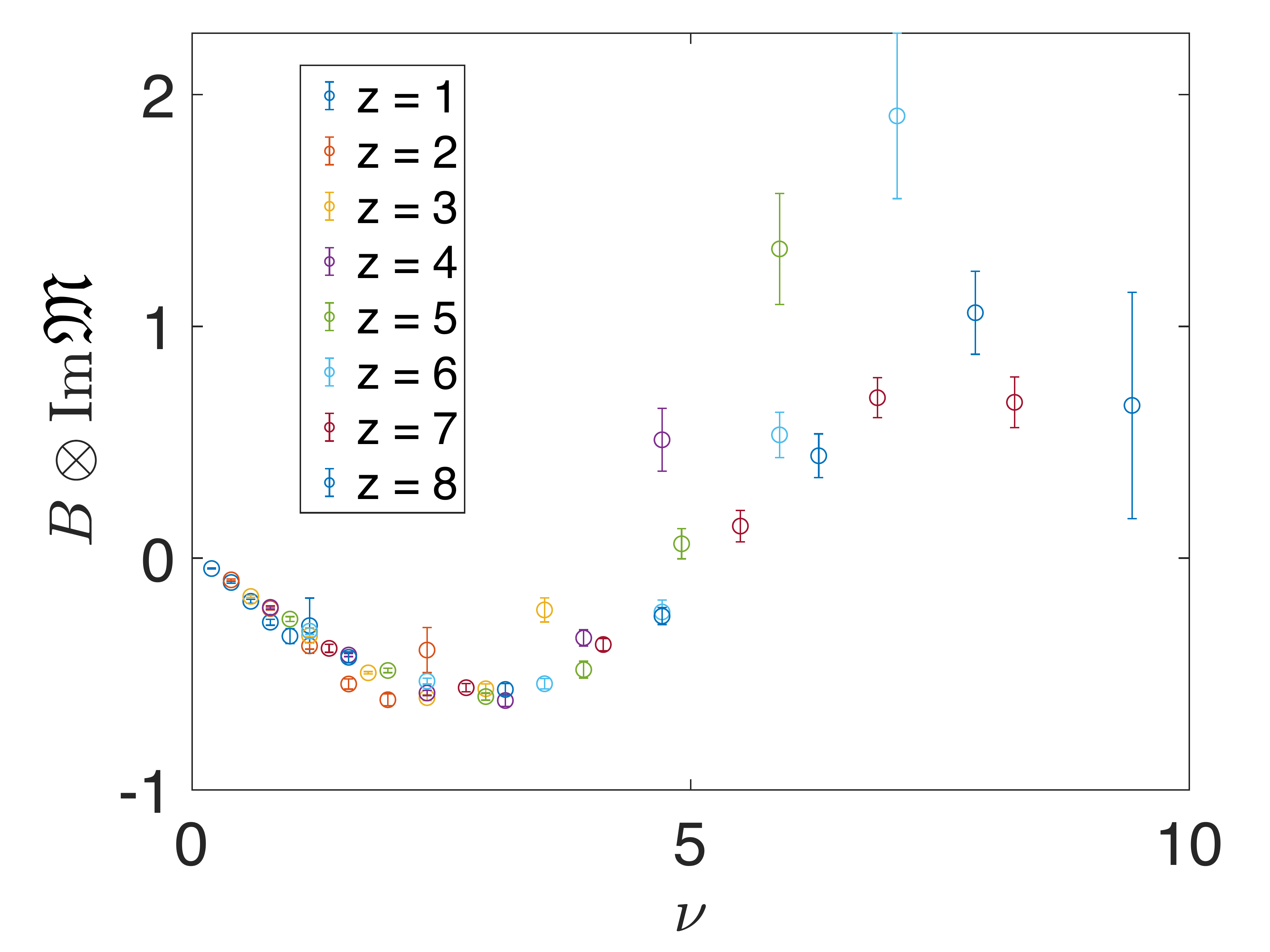}
\includegraphics[width=0.495\textwidth]{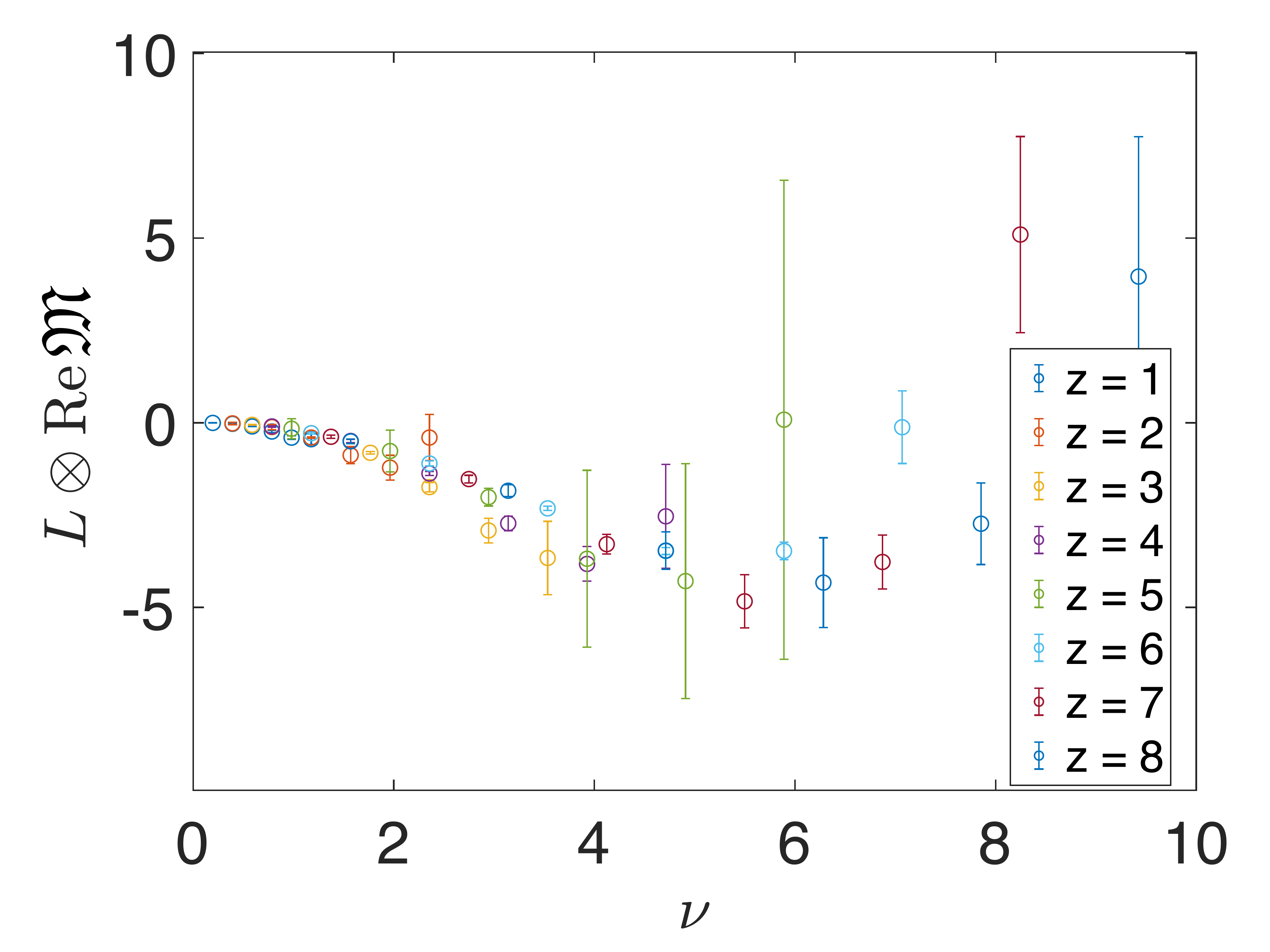}
\includegraphics[width=0.495\textwidth]{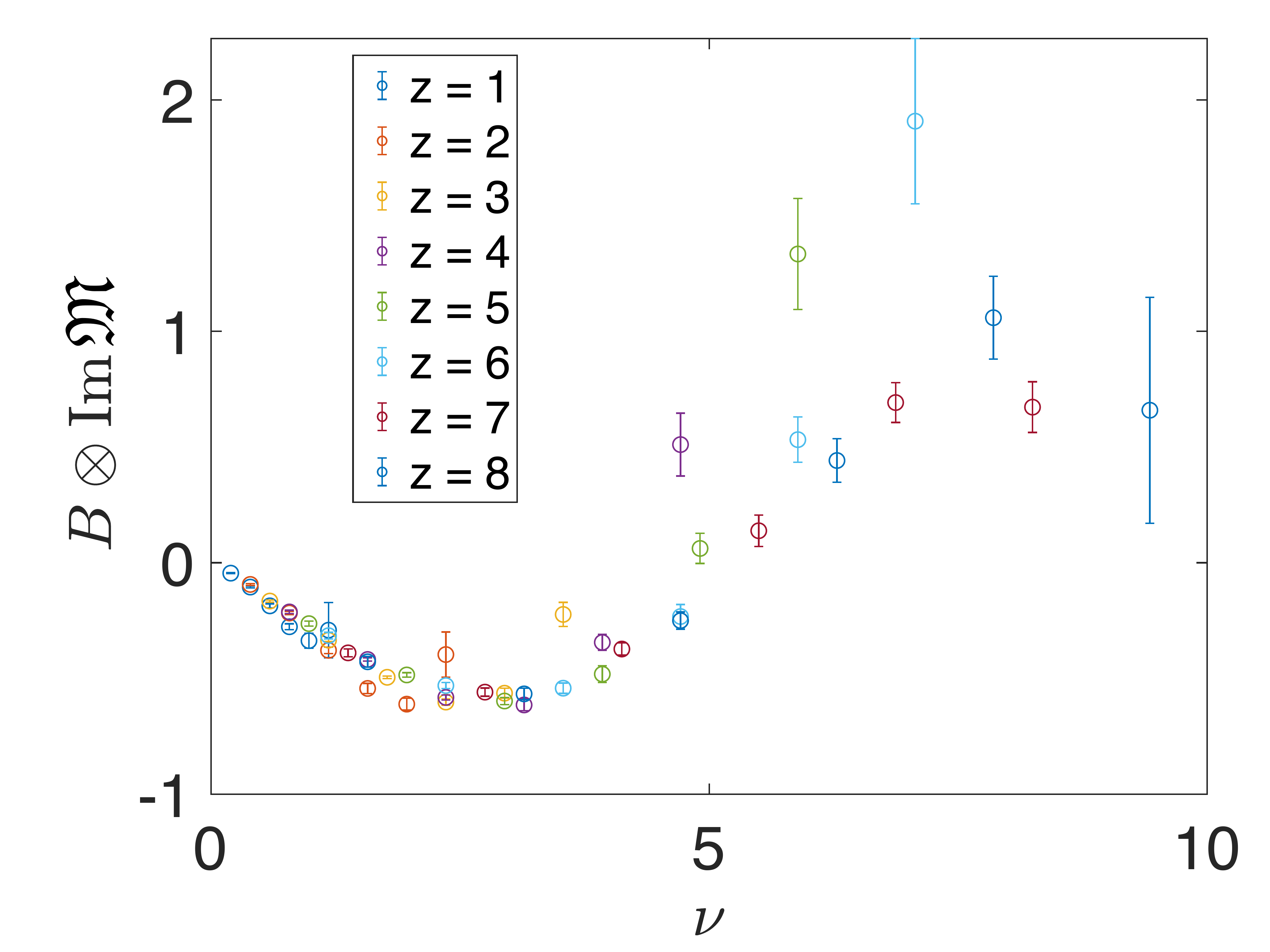}

\caption{The convolutions required for the evolution and matching of the reduced pseudo-ITD to the $\overline{\rm MS}$ ITD on the ensemble $a094m390$. The reduced pseudo-ITD was interpolated by fitting a polynomial.}

\label{fig:convos_fine}
\end{figure}

\begin{figure}[ht]\centering
\includegraphics[width=0.495\textwidth]{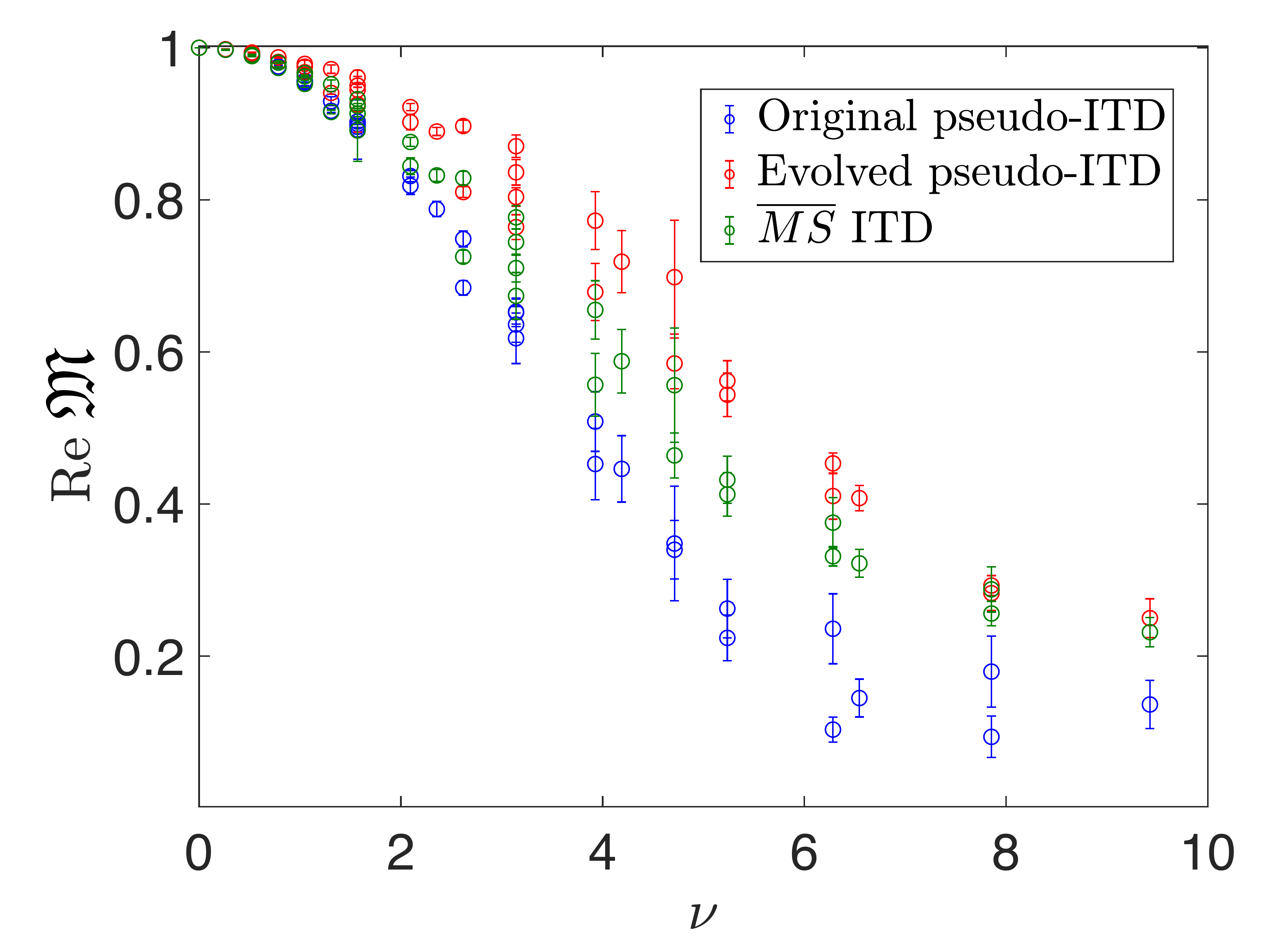}
\includegraphics[width=0.495\textwidth]{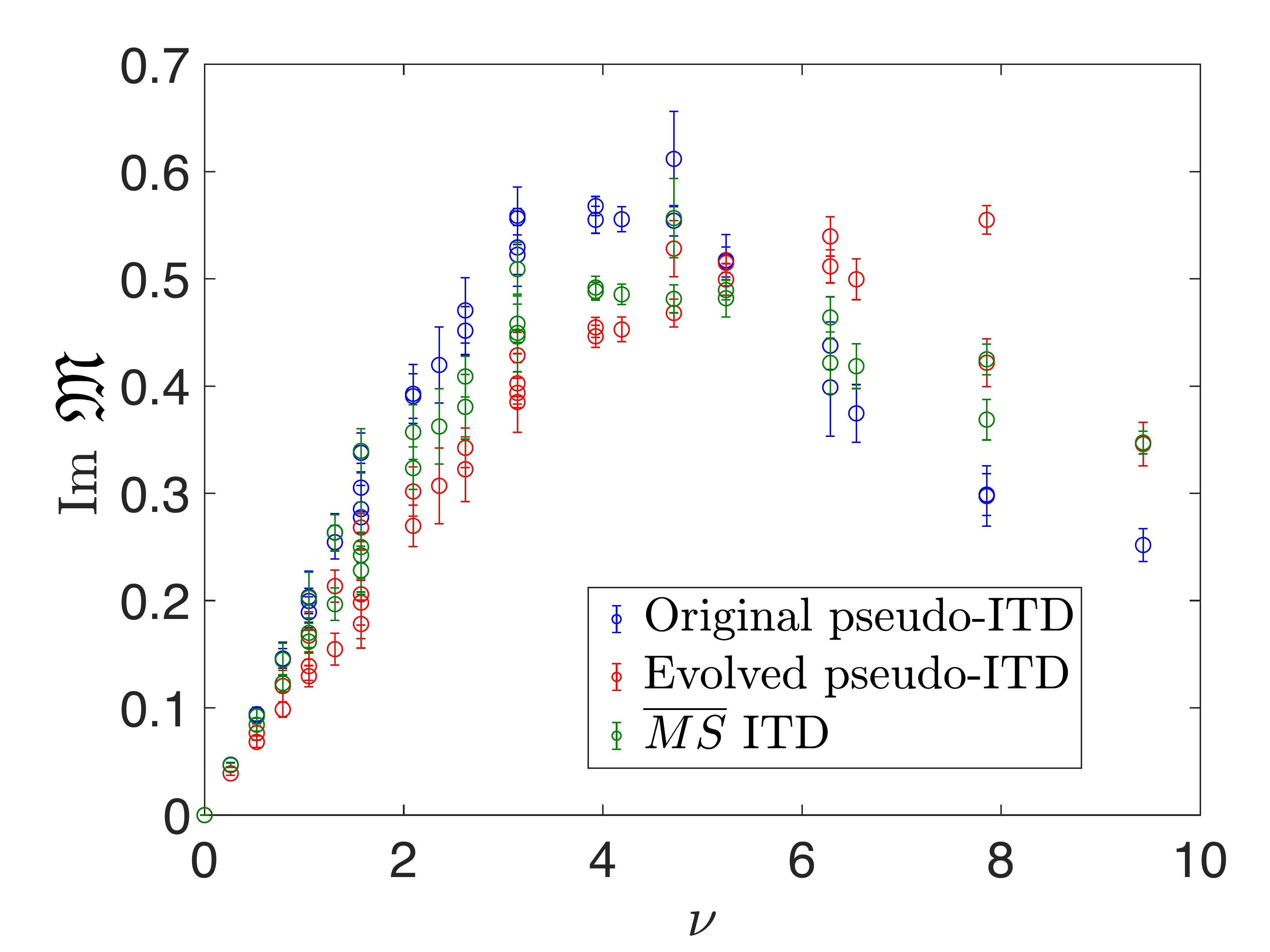}
\includegraphics[width=0.495\textwidth]{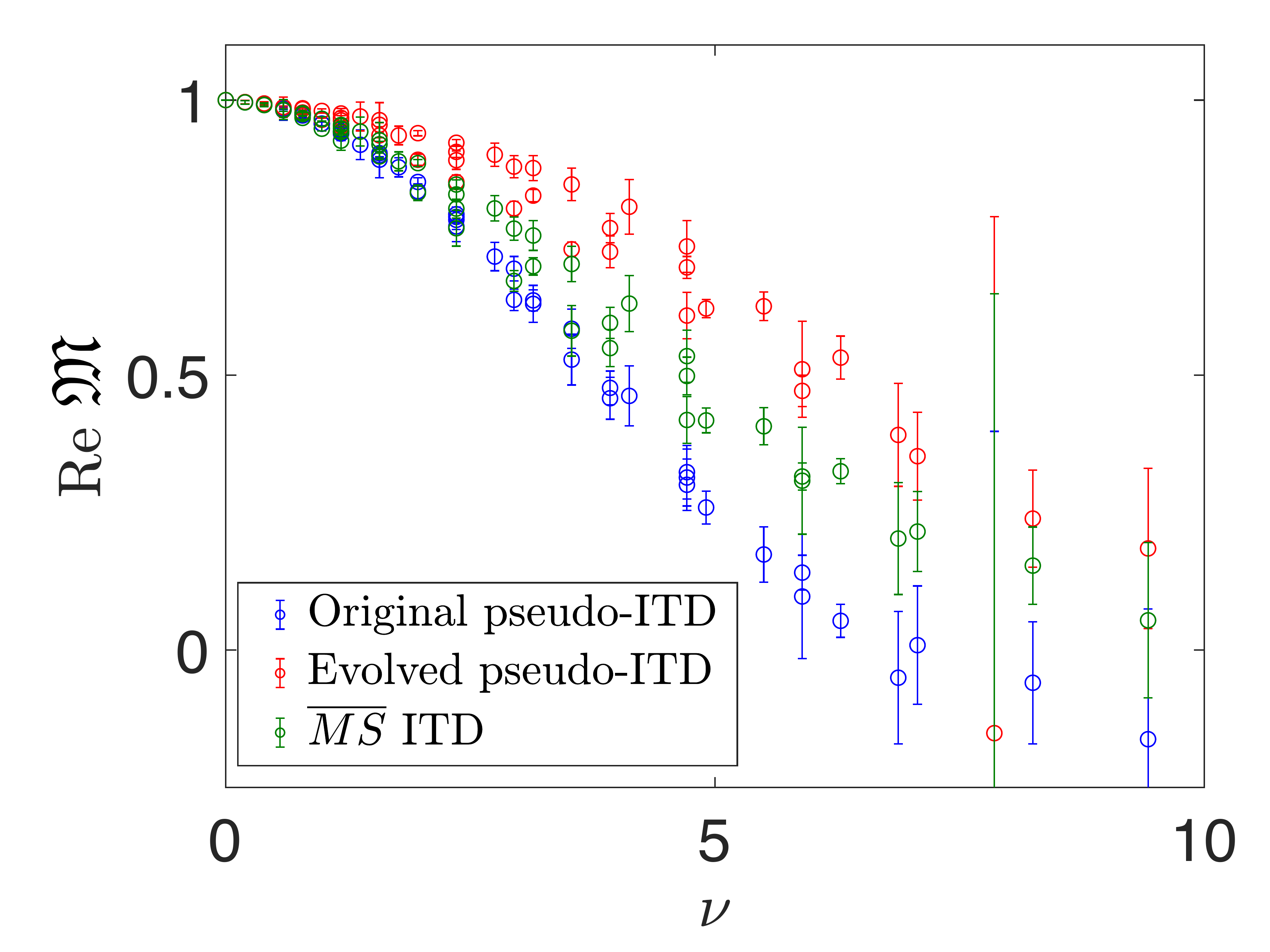}
\includegraphics[width=0.495\textwidth]{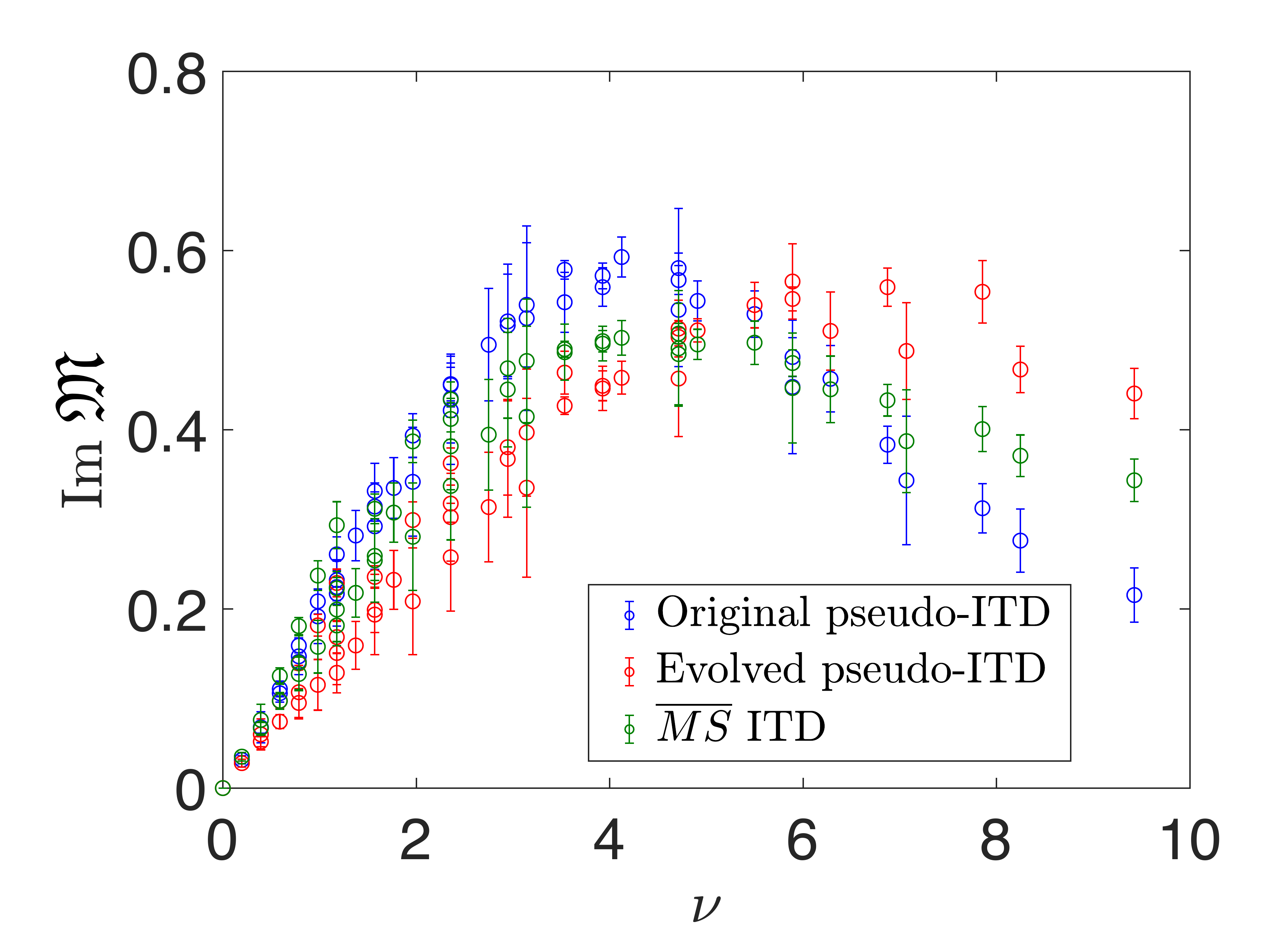}
\includegraphics[width=0.495\textwidth]{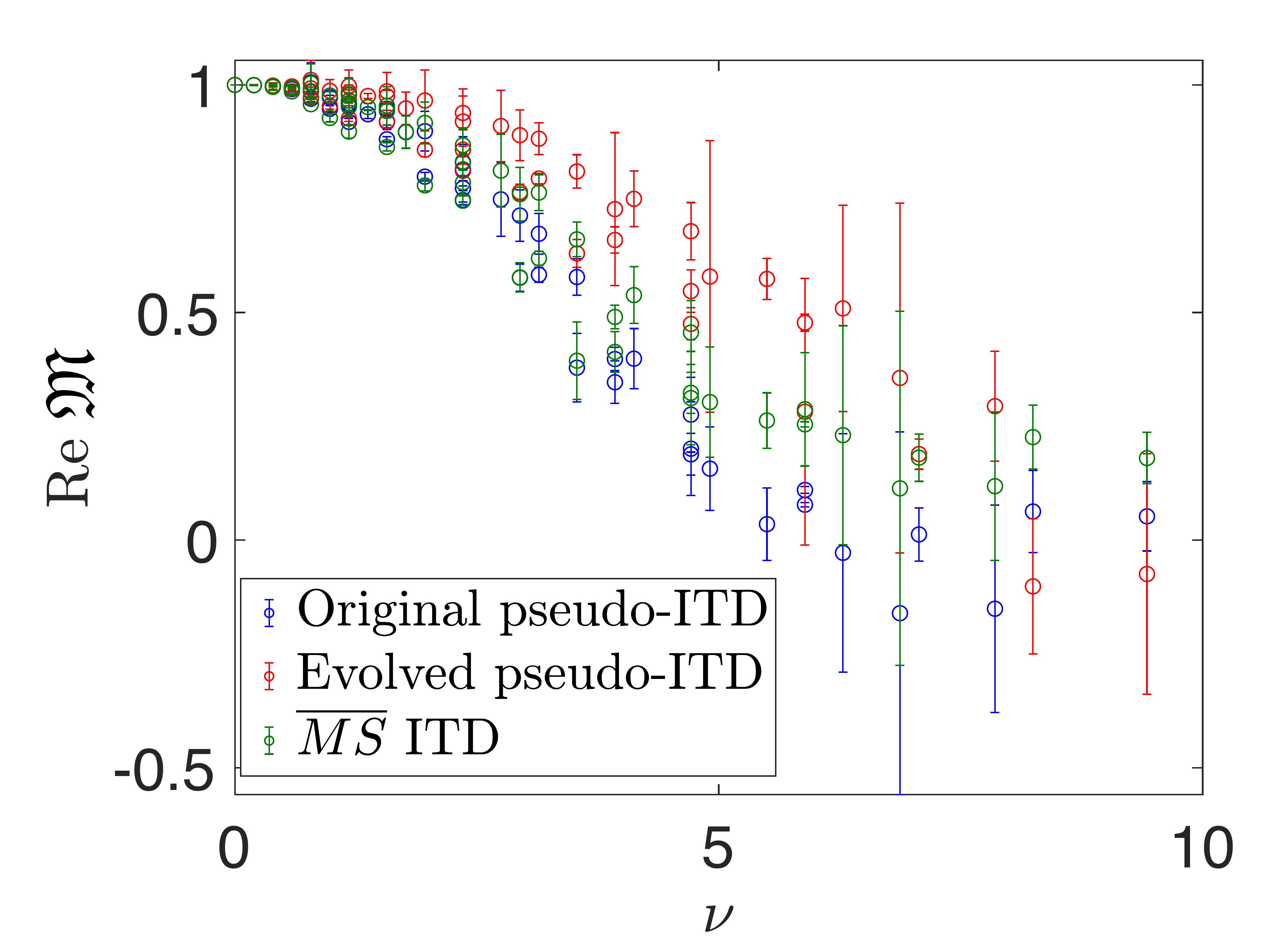}
\includegraphics[width=0.495\textwidth]{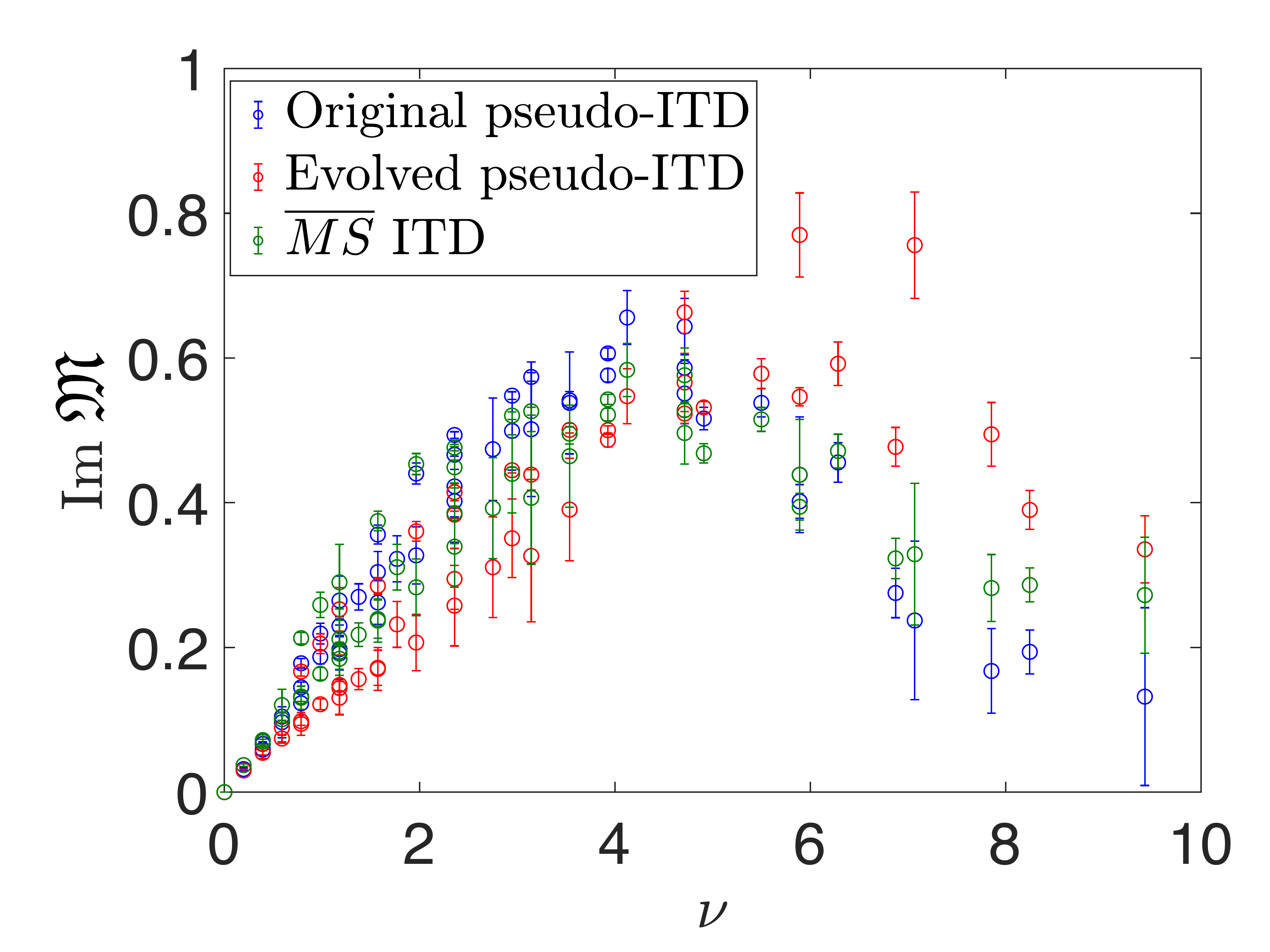}

\caption{The evolution and matching of the pseudo-ITD to the $\overline{\rm MS}$ ITD. The left and right plots show the real and imaginary components respectively. The top plots are from ensemble $a127m415$, the middle plots are from ensemble $a127m415L$, and the lower plots are from ensemble $a094m390$.}

\label{fig:orig_evo_match}
\end{figure}

In order to test the systematic effects of this choice, the real and imaginary components are also interpolated with a cubic spline. The results of these two integrations are consistent with each other. The convolutions calculated for the different ensembles are shown in Figs~\ref{fig:convos_coarse}-\ref{fig:convos_fine}. These convolutions follow the same trends as the model predictions shown in Fig.~\ref{fig:convo_theory}. Fig.~\ref{fig:orig_evo_match} shows the reduced pseudo-ITD evolved to $z^{-2} =4 e^{2\gamma_E +1}$ GeV$^2$. This particular scale was chosen, so that the reduced pseudo-ITD can be easily matched to the $\overline{\rm MS}$ ITD at $\mu=2$ GeV. The value of $\alpha_s(2~{\rm GeV})=  0.303$ was taken from the evolution used by LHAPDF~\cite{Buckley:2014ana} for the dataset {\tt cj15nlo} from the CTEQ-Jefferson Lab collaboration~\cite{CJ}.

There are two ways in which these convolutions can be used to evolve and match the reduced pseudo-ITD to the $\overline{\rm MS}$ ITD. The most straightforward is a direct inversion of Eq.~\eqref{eq:pITD_to_ITD} for data with different $z^2$ independently. An alternative, but equivalent, approach is to perform the $z^2$ evolution of the reduced pseudo-ITD, for each $z^2$ independently, using Eq.\eqref{eq:pseudo_evo} to the scale $z_0^2 = e^{-2\gamma_E -1} \mu^{-2}$. With all the data evolved to this common scale, the inverse of Eq.~\eqref{eq:pITD_to_ITD} can be applied to match the pseudo-ITD to the $\overline{\rm MS}$ ITD for the data originating with all $z^2$ simultaneously. The convolutions with $L$ are performed by fitting the evolved reduced pseudo-ITD to the same polynomials as before in Eq.~\eqref{eq:pseudo-poly}. The common scale was chosen such that the scale dependent logarithm in Eq.~\eqref{eq:pITD_to_ITD} vanishes when matching to the $\overline{\rm MS}$ ITD for a particular $\mu$. 

In this work, the scale $\mu=2$ GeV was chosen. The evolved reduced pseudo-ITD and the matched $\overline{\rm MS}$ ITD are shown in Fig~\ref{fig:orig_evo_match}. It has been tested that the evolution and matching procedure performed in a single step or being performed in two steps result in a consistent $\overline{\rm MS}$ ITD. For the remainder of this work, only the one step matching results will be used.

\section{PDF extraction}\label{pdfe}
Due to the restrictions in allowed quark-field separations and
momentum states on the lattice, the data lay discretized on an
interval of $\nu$ different than the full Brillouin zone. These issues
make the extraction of the PDF from Eqs.~\eqref{eq:pdf_fit_real} and
\eqref{eq:pdf_fit_imag} given lattice data an ill-posed inverse
problem. In order to reliably extract a PDF from the lattice data, one
will have to provide additional information. What information and how
it is applied constitute different solutions to the inverse problem, a
few of which were studied for use in PDF calculations
in~\cite{Karpie:2019eiq,Liang:2019frk,Cichy:2019ebf}.

\subsection{Moments of PDF and pseudo-PDF}
\begin{figure}
\centering
\includegraphics[width=0.495\textwidth]{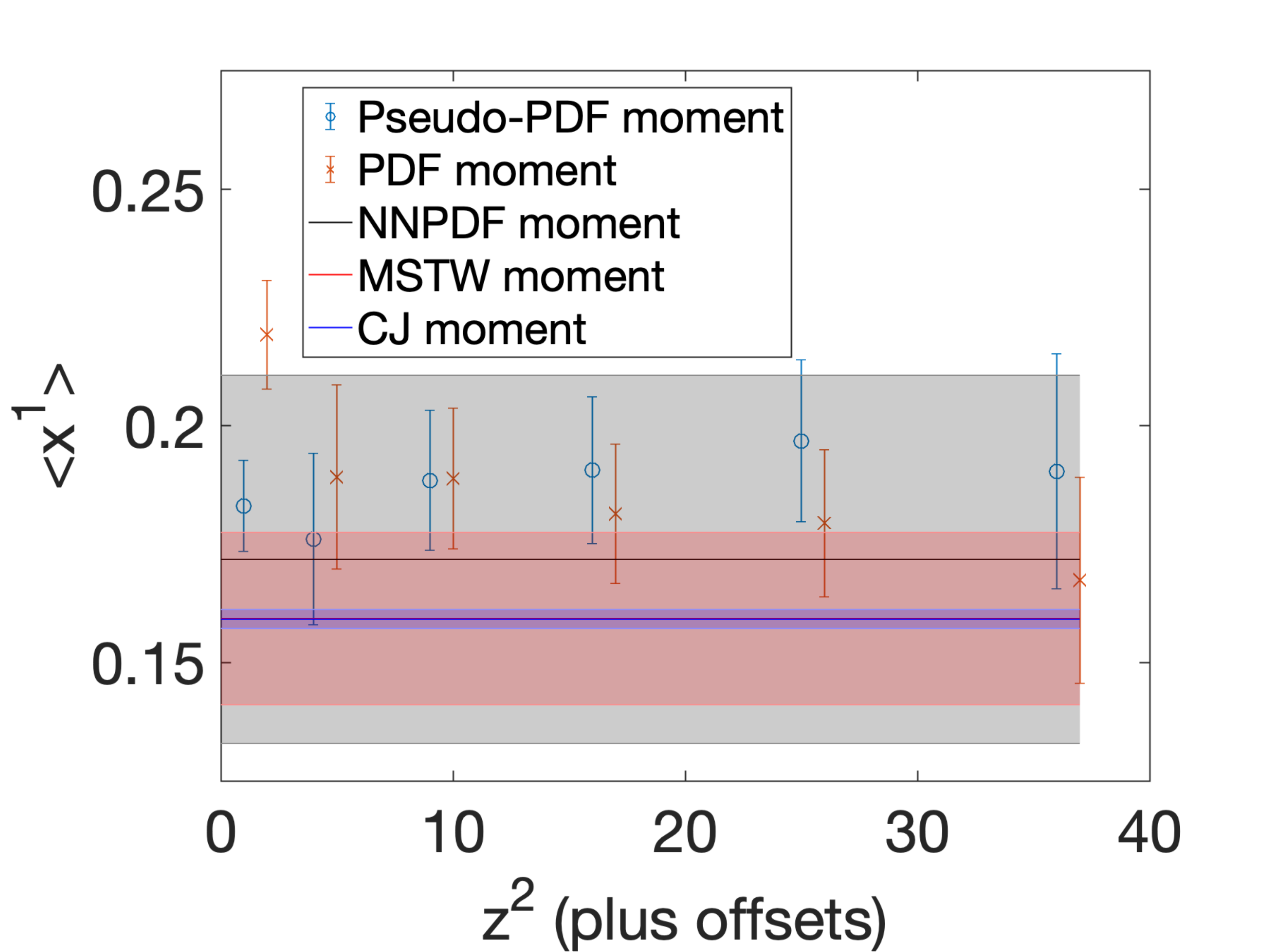}
\includegraphics[width=0.495\textwidth]{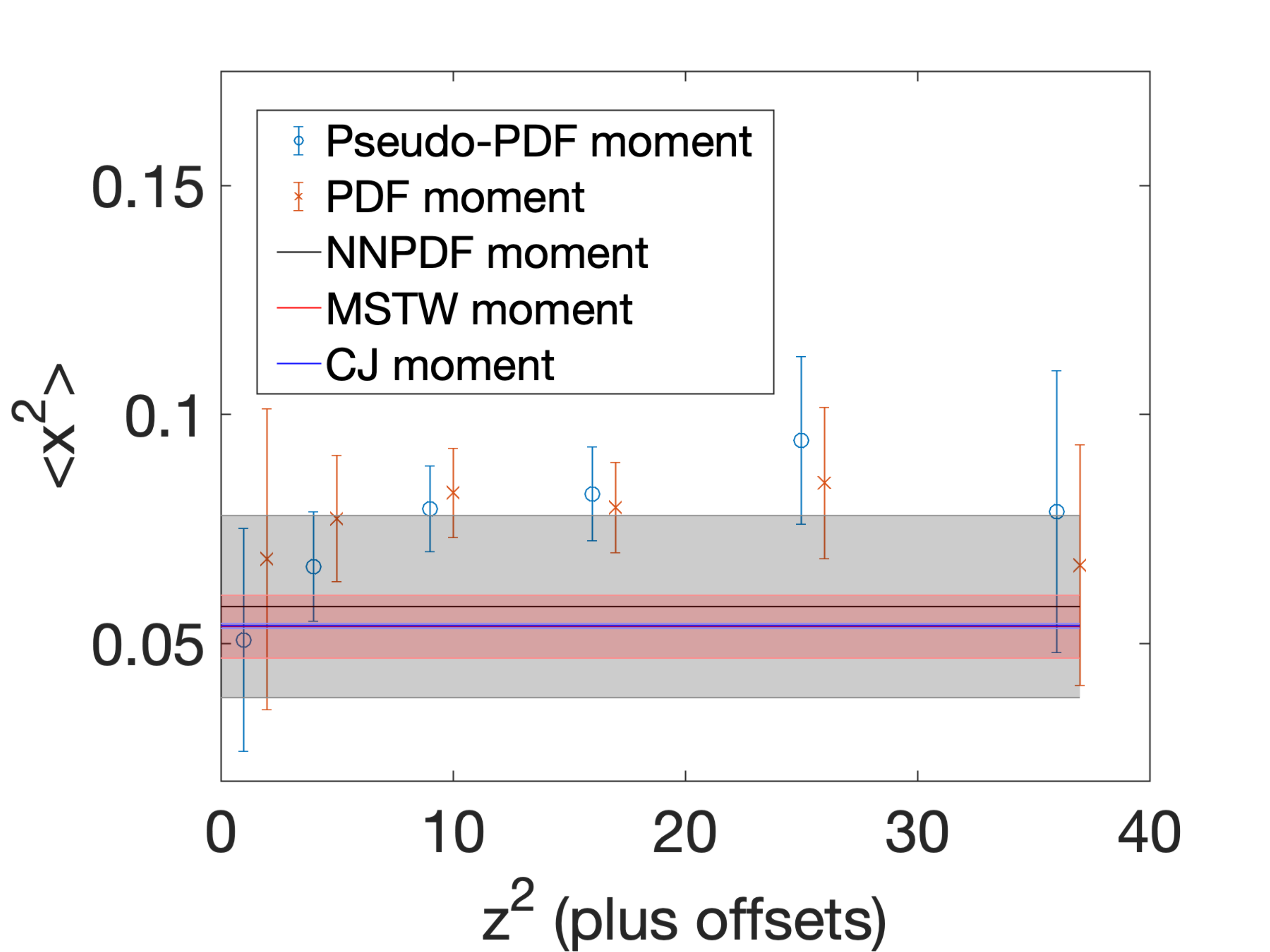}
\includegraphics[width=0.495\textwidth]{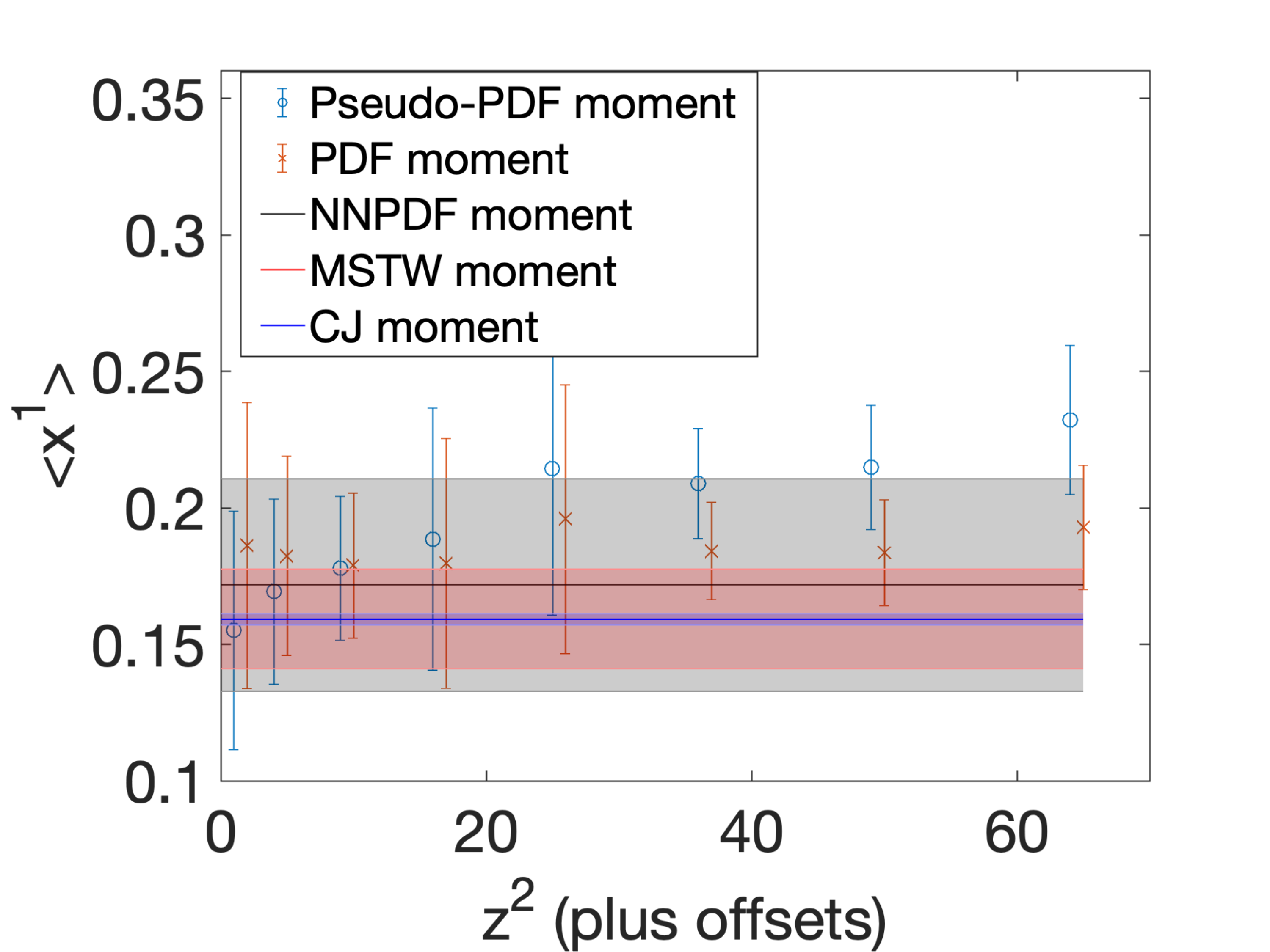}
\includegraphics[width=0.495\textwidth]{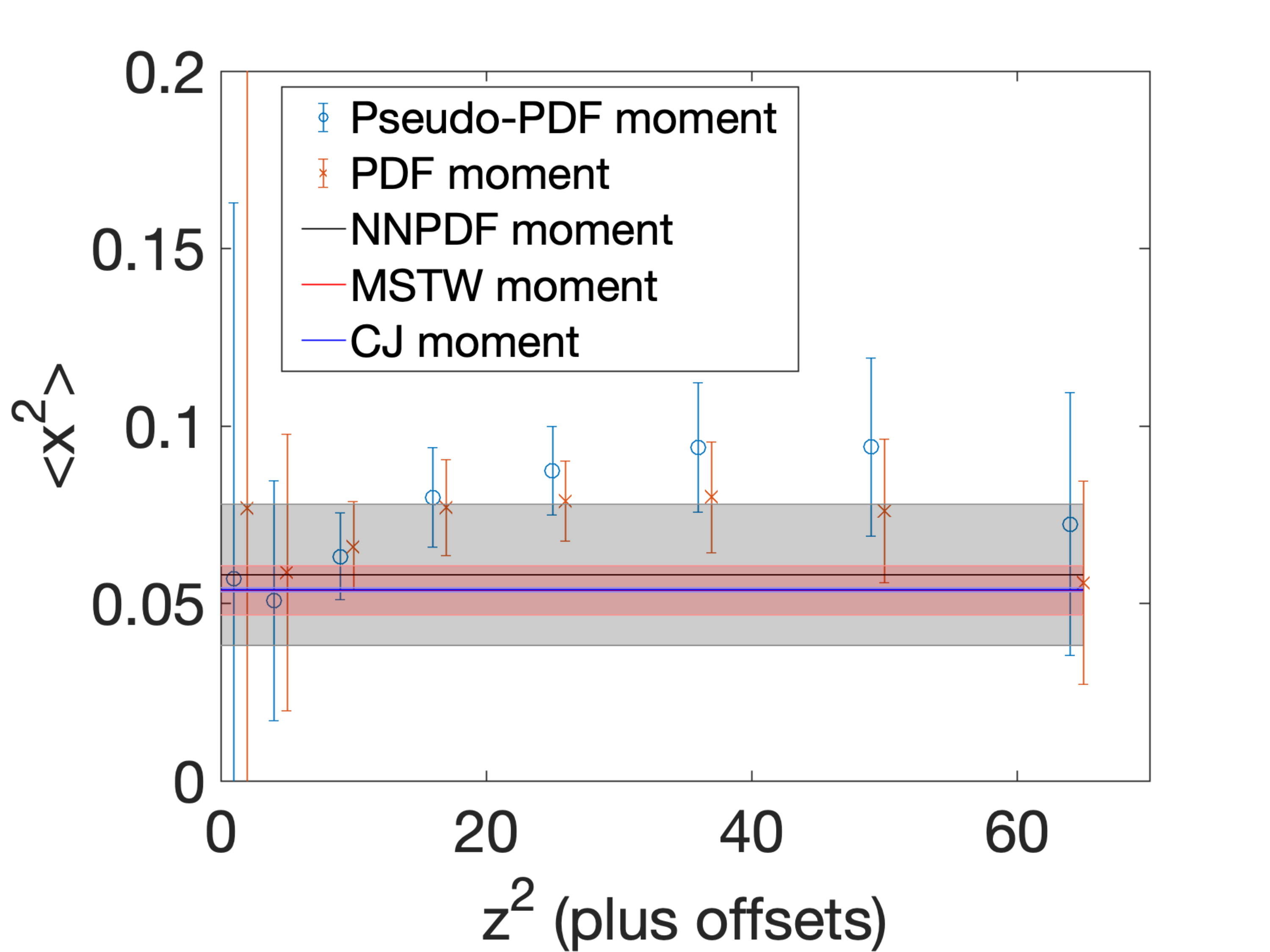}
\includegraphics[width=0.495\textwidth]{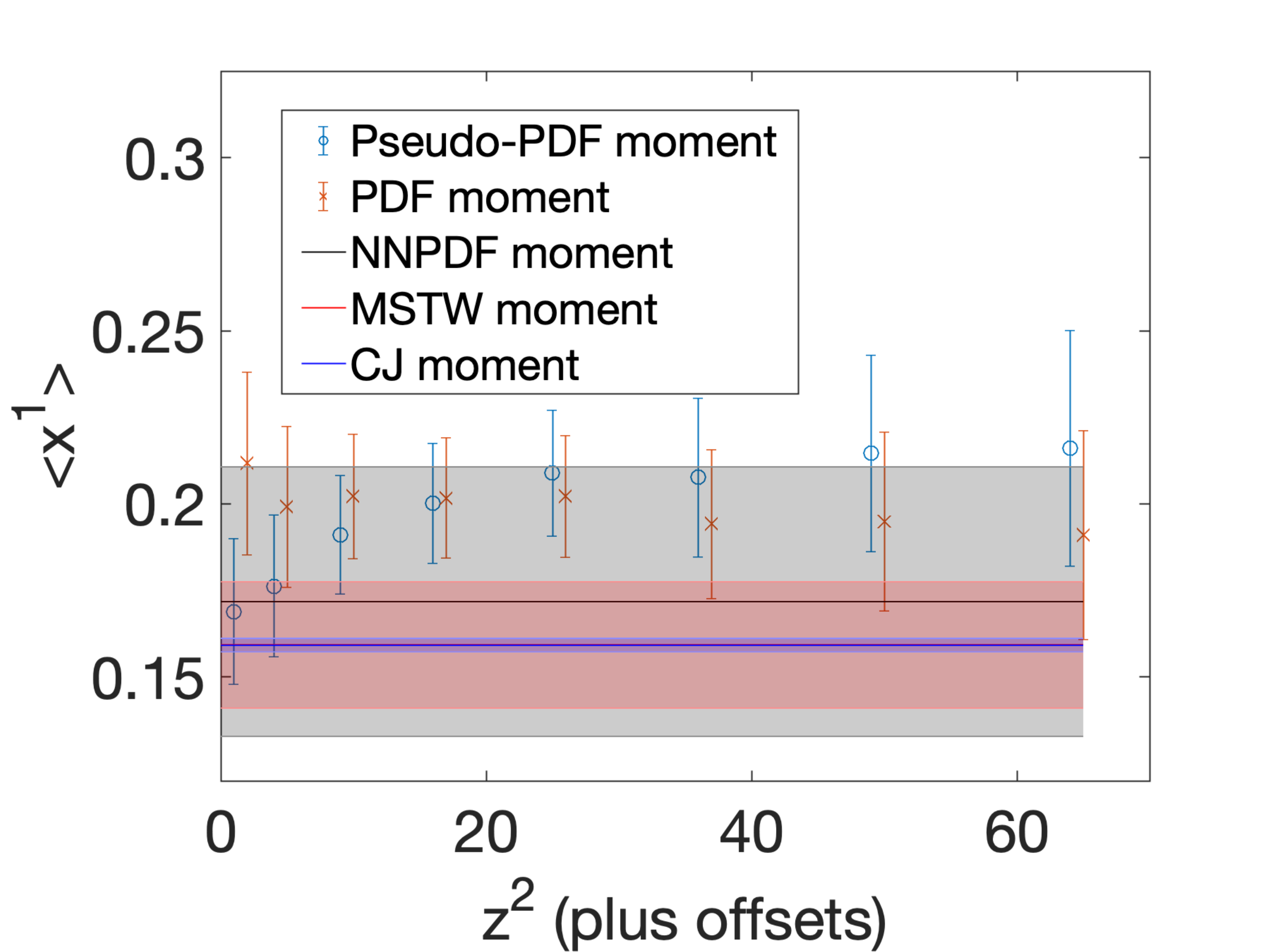}
\includegraphics[width=0.495\textwidth]{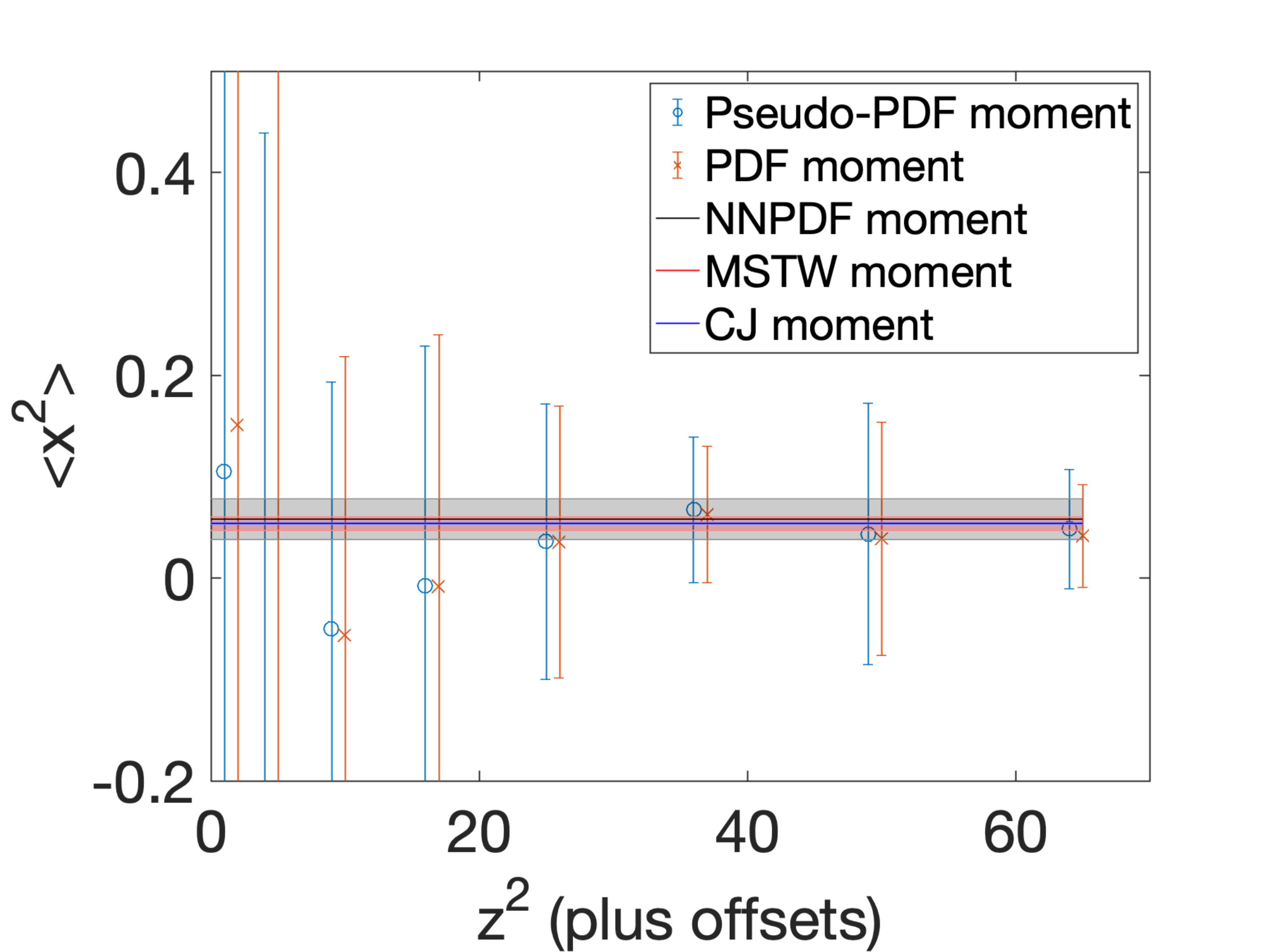}
\caption{The moments of the pseudo-PDF compared to phenomenologically determined PDF moments from the NLO global fit CJ15nlo~\cite{CJ}, and the NNLO global fits MSTW2008nnlo68cl\_nf4~\cite{Martin:2009iq} and NNPDF31\_nnlo\_pch\_as\_0118\_mc\_164~\cite{Ball:2017nwa} all evolved to 2 GeV. The top, middle, and bottom plots are from the ensembles $a127m415$, $a127m415L$ and $a094m390$ respectively. The left and right columns show the first and second moments respectively. Only the lowest two moments have signal for most $z$. The higher moments only have signal for the largest $z$ where the maximum Ioffe time used allows for identifying more than the leading behavior in the Taylor expansion.}
\label{fig:mom_big}
\end{figure}

Information about the PDF can still be determined from the reduced
pseudo-ITD without directly performing the Fourier transform. The
moments of the pseudo-PDF can be used to calculate PDF moments while
avoiding entirely the inverse problem~\cite{Karpie:2018zaz}. A
discretized version of the relationship in Eq.~\eqref{eq:pPDF_mom}
between the moments of the pseudo-PDF and the reduced pseudo-ITD data
can be written in the matrix form \be {\bf \mathfrak{M}} = {\bf C}
{\bf b}\,, \ee where $\bf \mathfrak{M}$ is a vector of $N$ data
points, ${\bf C}$ is known as the Vandermonde matrix in $\nu$, and ${\bf b}$ is a
vector of $M$ moments weighted by the factor of $i^n/n!$ mentioned in
Eq.~\eqref{eq:pPDF_mom}. The Vandermonde matrix is an $N\times M$
matrix of the form ${\bf C}_{in} = \nu_i^{n}$ where $\nu_i$ is the
Ioffe time for the $i$-th data point in ${\bf \mathfrak{M}}$. This
relationship can also be split into real and imaginary components
which only contain even and odd powers of $\nu$ and result in even and
odd moments of the pseudo-PDF respectively. This equation is inverted
for points with a fixed $z^2$. The results of the first and second
moments of the pseudo-PDF as well as the matched PDF moments are shown
in Fig~\ref{fig:mom_big}. As described in Sec.~\ref{sec:mom}, the
moments of the pseudo-PDF can then be matched to the $\overline{\rm MS}$
moments.

At small separations, the moments of the pseudo-PDF have small
dependence on $z^2$. After the application of the DGLAP evolution and
matching relationships from Eq.~\eqref{eq:pmom_match}, any residual
$z^2$ dependence of the moments of the PDF, which would be caused by
higher twist contaminations, appears negligible. Fig~\ref{fig:mom_big}
has a comparison of this calculation of the pseudo-PDF and
$\overline{\rm MS}$ PDF moments and those calculated from various global
fits. As is the case in the direct calculation of the local matrix
element~\cite{Alexandrou:2016hiy}, the PDF moments at heavy pion mass
are systematically higher than the phenomenologically determined
result.

In principle, one can also use this technique to extract the higher
moments. This procedure has been tested on the next two higher
moments. Only the results with the largest few $z^2$ had statistical
errors comparable to the lower moments. The range of Ioffe time for
those large $z$ data points had been sufficiently large to determine
the second variable in the Taylor expansion in
Eq.~\eqref{eq:pseudo-poly} with reasonable statistical precision.
Since these data potentially have significant higher twist
corrections, these results should be considered questionable. The
small $z$ data points appear almost entirely described by including
only the terms proportional to $c_1$ and $c_2$.  Further studies on finer
lattice spacings will be required to extend the range of Ioffe time
for low $z$ data in order to constrain the higher moments and confirm
the lack of higher twist contamination which was observed in the lower
moments.

Reconstructing the PDF from its moments in itself is an inverse problem. Instead of inverting a Fourier transform, like most of the procedures discussed in this chapter, this problem is the inversion of a Mellin transform. The PDF can be parameterized by some function such as the PDF Ansatz in Eq.~\eqref{eq:ansatz} or the moments can be parameterized by some function with a known inverse Mellin transform as was suggested in~\cite{Chambers:2017dov}. In this work, with at best two moments being constrained, there is not much hope for actually performing a reliable fit to any of these functional forms.

The determination of moments of the pseudo-PDF is a useful calculation while studying these Ioffe time distributions. The inversion of the Fourier transform is an ill-defined problem which comes along with many complications and the moments of the pseudo-PDF allow for a quick sanity check that the data contain reasonable information. The residual $z^2$ dependence of the $\overline{\rm MS}$ moments, derived from moments of the pseudo-PDF, is a non-trivial check of the size of higher twist effects. The lack of any statistically meaningful $z^2$ dependence in the $\overline{\rm MS}$ moments calculated on these ensembles can be used to justify the validity of using the data with all $z^2$ in the following PDF extractions.

\subsection{PDF fits}
All solutions to the ill posed inverse problem require adding additional
information to constrain a unique solution for the unknown
function. In the global PDF fitting community, the most common choice
for solving the inverse problem is to choose a physically motivated
functional form for the PDF. By choosing a PDF parameterization with
fewer parameters than existing data points, the inverse problem is
regulated. This model Ansatz can be designed to explicitly show some
limiting behaviors, physically motivated features, and satisfy some
possible constraints. The better motivated the information used to
create the Ansatz the more successful this technique will be, but any
particular choice will introduce a model-dependent bias into the final
result. The ill-posed inverse problem does not have a unique solution,
and ultimately some bias must be introduced into the PDF
determination. Ideally, several different parameterizations would be
checked and compared, and in effect this cross checking has occurred
amongst the several phenomenological PDF fits employed, each with
different choices of models.

It is known that the PDF can be 
reasonably well 
described simply by the following
expression that parameterizes its limiting behaviors,
\beq \label{eq:ansatz} f(x) = \frac{ x^a (1-x)^b } { B(a+1,b+1)}\,.
\eeq To add more generality, the
phenomenological PDFs are fit to a more flexible functional form,
\beq\label{eq:gen_pdf} f(x) = x^a (1-x)^b P(x) \,, \eeq where $P(x)$
is a yet to be specified interpolating function with more model
parameters. There exist well known features of PDFs, such as vanishing
as $x \to 1$, diverging as $x \to 0$, and the constraints of the PDF sum rules.  The limiting behaviors can be
seen through the signs of the model parameters $a$ and $b$ in
Eq.~\eqref{eq:gen_pdf} and the normalization can be fixed to satisfy the sum rules. By separating these features, $P(x)$ is
allowed to be a smoother and slower varying function which is easier
to determine. One choice of $P(x)$ for the valence quark PDF employed
by the CJ~\cite{CJ} and the MSTW collaborations~\cite{Martin:2009iq}
is given by

\beq\label{eq:jam_p}
P(x) = \frac{1 + c \,\sqrt x + d\, x}{B(a+1,b+1) + c\, B(a+1.5,b+1) + d\, B(a+2,b+1)} \,.
\eeq
 This functional form explicitly sets the PDF's sum rule and allows all moments to be directly calculated as a ratio of sums of Beta functions.

The statistical errors will be obtained with the jackknife resampling technique. In the dynamical quark ITDs, there does not appear to be a strong dependence on the initial separation $z$ with which the data point had been calculated. This $z^2$ independence indicates that large polynomial $z^2$ corrections do not appear to exist. Therefore we can fit all $z^2$ separations simultaneously.

\begin{figure}[h!]
  \centering
\includegraphics[width=0.495\textwidth]{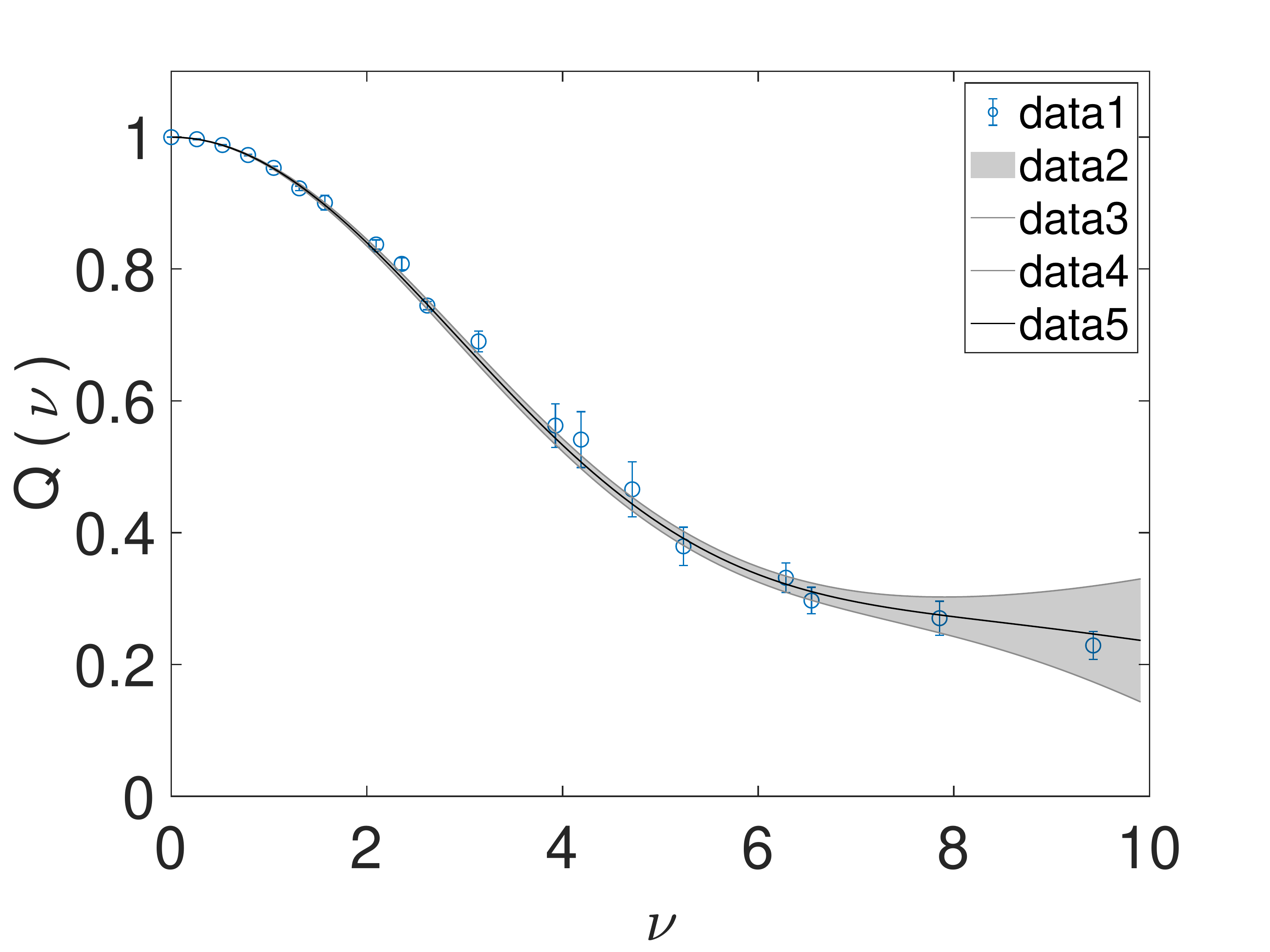}
\includegraphics[width=0.495\textwidth]{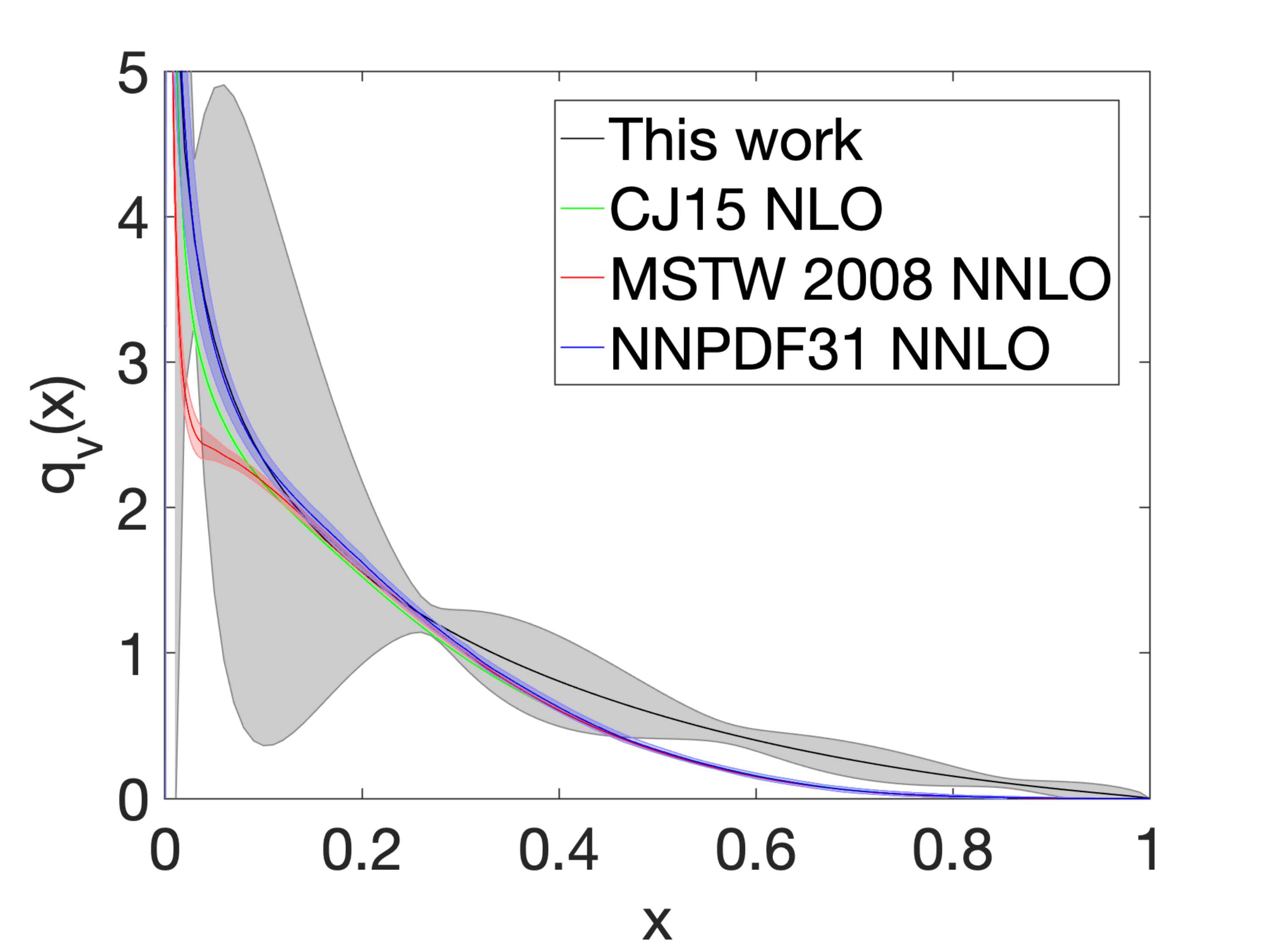}
  \caption{\label{fig:jam_coarse}
   The nucleon valence distribution obtained from the ensemble $a127m415$ fit to the form used by the JAM collaboration in Eq.~\eqref{eq:jam_p}. 
   The $\chi^2/{\rm d.o.f.}$ for the fit with all the data is 2.5(1.5). 
   The uncertainty band is obtained from the fits  to  the jackknife samples of the data. The resulting fits are compared to phenomenologically determined PDF moments from the NLO global fit CJ15nlo~\cite{CJ}, and the NNLO global fits MSTW2008nnlo68cl\_nf4~\cite{Martin:2009iq} and NNPDF31\_nnlo\_pch\_as\_0118\_mc\_164~\cite{Ball:2017nwa} all evolved to 2 GeV.}

\end{figure}

\begin{figure}[h!]
  \centering
\includegraphics[width=0.495\textwidth]{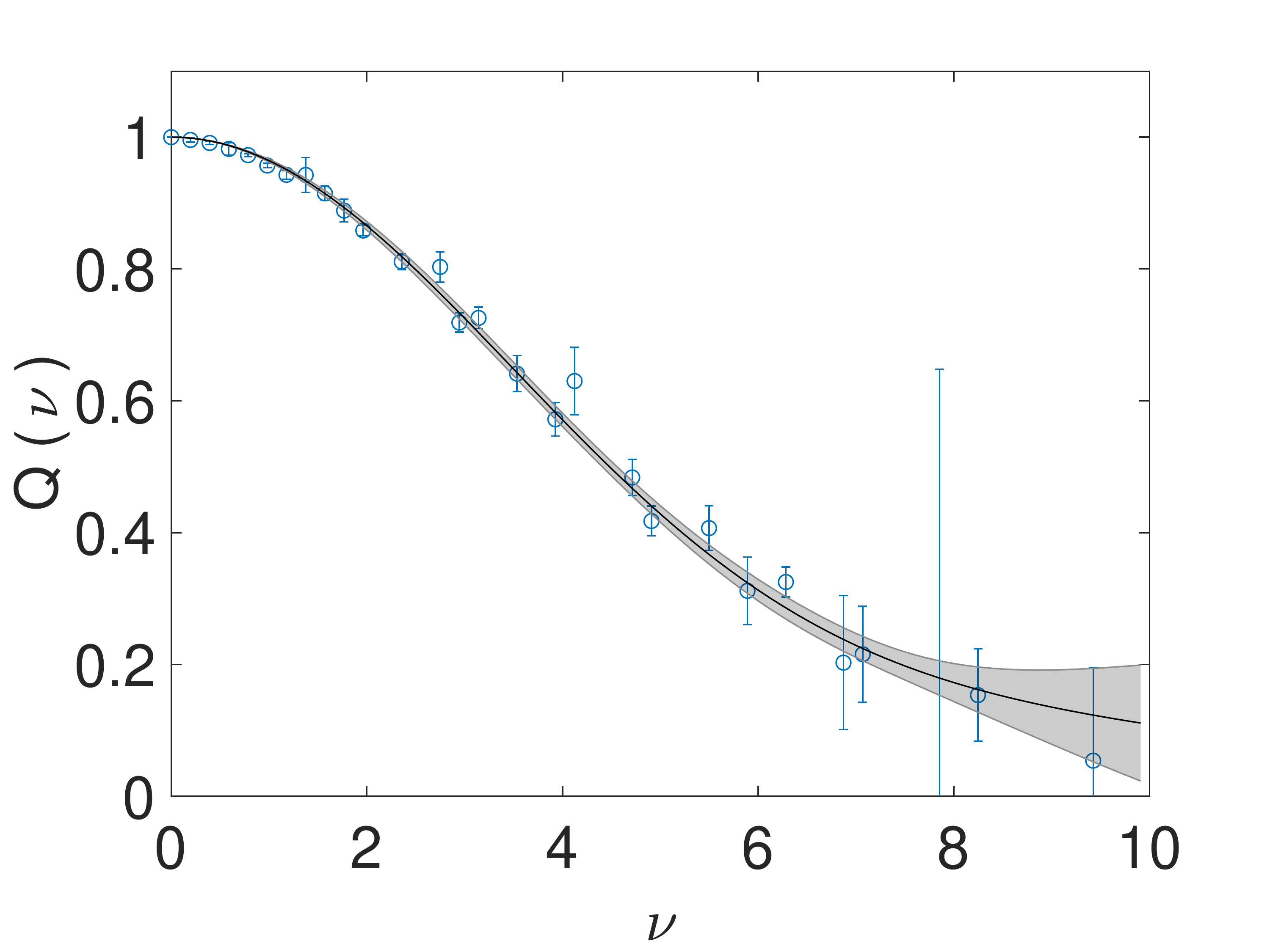}
\includegraphics[width=0.495\textwidth]{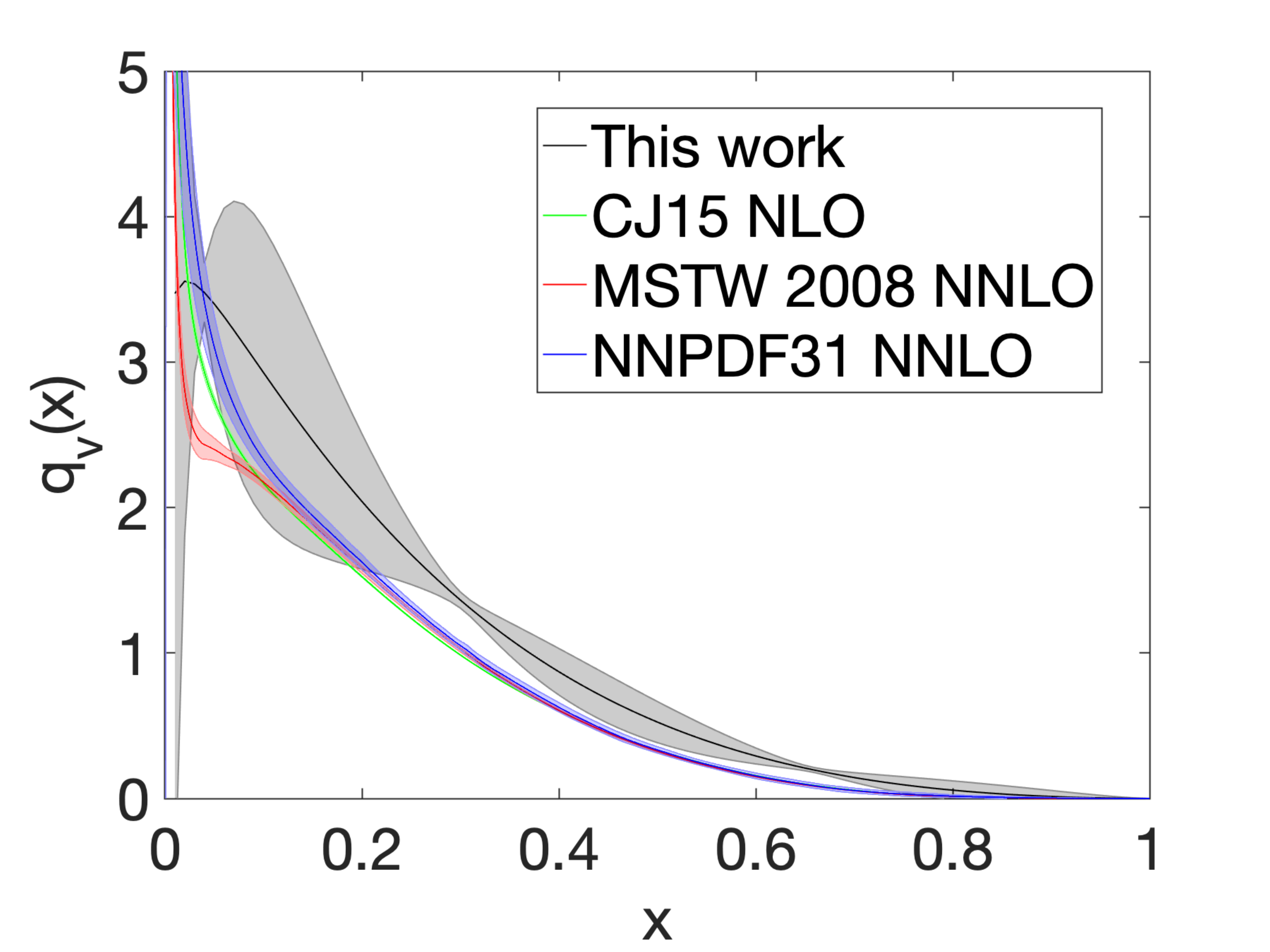}
  \caption{\label{fig:jam_big}
   The nucleon valence distribution obtained from the ensemble $a127m415L$  fit to the form used by the JAM collaboration in Eq.~\eqref{eq:jam_p}. 
   The $\chi^2/{\rm d.o.f.}$ for the fit with all the data is 2.1(6). 
   The uncertainty band is obtained from the fits  to  the jackknife samples of the data. The resulting fits are compared to phenomenologically determined PDF moments from the NLO global fit CJ15nlo~\cite{CJ}, and the NNLO global fits MSTW2008nnlo68cl\_nf4~\cite{Martin:2009iq} and NNPDF31\_nnlo\_pch\_as\_0118\_mc\_164~\cite{Ball:2017nwa} all evolved to 2 GeV.}
\end{figure}

\begin{figure}[h!]
  \centering
\includegraphics[width=0.495\textwidth]{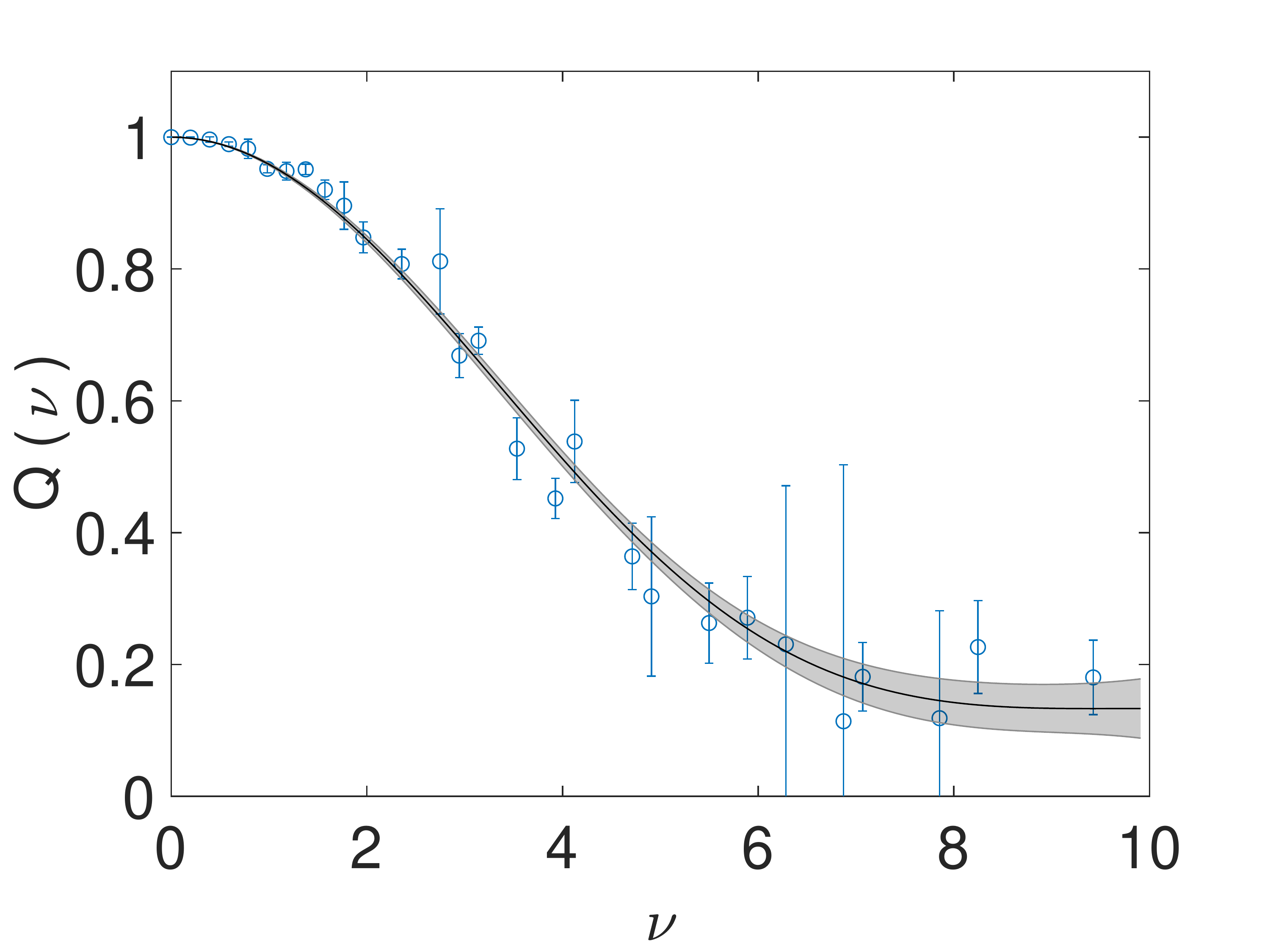}
\includegraphics[width=0.495\textwidth]{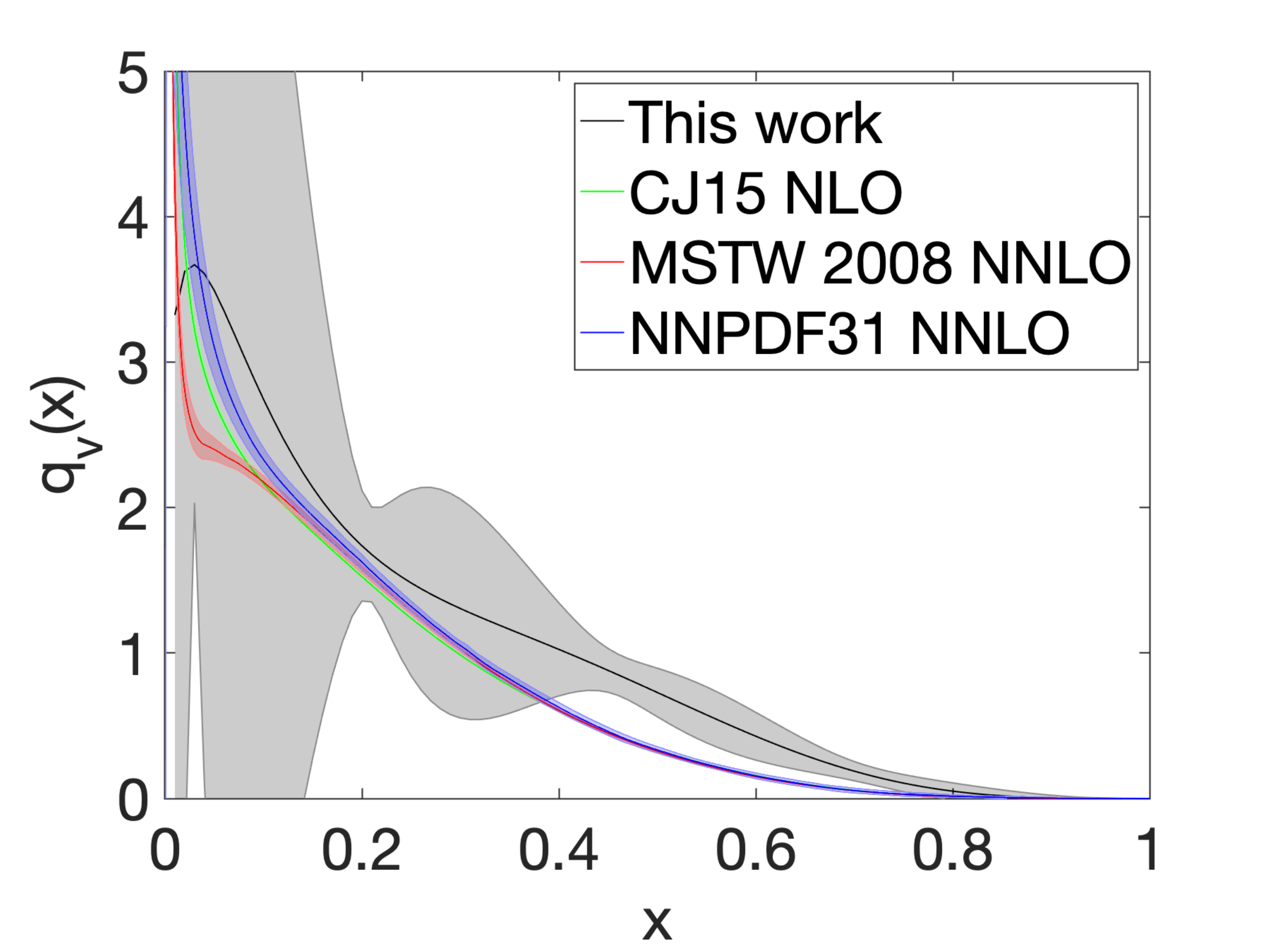}  \caption{\label{fig:jam_fine}
   The nucleon valence distribution obtained from the ensemble $a094m390$  fit to the form used by the JAM collaboration in Eq.~\eqref{eq:jam_p}. 
   The $\chi^2/{\rm d.o.f.}$ for the fit with all the data is 2.0(5). 
   The uncertainty band is obtained from the fits  to  the jackknife samples of the data. The resulting fits are compared to phenomenologically determined PDF moments from the NLO global fit CJ15nlo~\cite{CJ}, and the NNLO global fits MSTW2008nnlo68cl\_nf4~\cite{Martin:2009iq} and NNPDF31\_nnlo\_pch\_as\_0118\_mc\_164~\cite{Ball:2017nwa} all evolved to 2 GeV.}
\end{figure}

Fig.~\ref{fig:jam_coarse}-~\ref{fig:jam_fine} shows the results of fitting the ITD to the functional formed used by the CJ and MSTW collaborations. 
If the variance of the fit result at large values of the Ioffe time, where there is lack of data, the more the PDF result tends to have large oscillatory errors. It will be seen in other functional forms when the variance grows significantly at large Ioffe times, the PDF will contain oscillatory solutions which generate large errors. In these functional forms, the large Ioffe time behavior of the PDF is largely governed by the low $x$ parameter $a$. The large Ioffe time behavior of the ITD from the model in Eq.~\eqref{eq:ansatz} is given by
\beq
Q(\nu) \sim -\sin \left(\frac \pi 2 a \right) \frac{\Gamma(a+1)}{\nu^{a+1}} + b \cos \left(\frac \pi 2 a\right) \frac{\Gamma(a+2)}{\nu^{a+2}}  \,.
\eeq
The PDF must have a finite integral for the sum rules to be enforced. This feature restricts the power of the $x\to0$ divergence to $a > -1$ and this ITD must vanish in the limit $\nu \to \infty$. All of the polynomials tried above, will only add terms which force the ITD to converge to 0 more rapidly. Any of these fit solutions which does not eventually converge to 0 should be rejected, because it cannot have a finite integral and violates the sum rule.

\subsection{Continuum extrapolation}
Previous publications of quasi-PDFs and pseudo-PDFs from lattice QCD
have only been performed with one lattice spacing at a time. With results from these two  lattice spacings, we can get an estimate of the continuum extrapolation systematic. Though the
action is $O(a)$ improved, the quark bilinear operator $\bar \psi(0)
\gamma^\alpha W(0;z) \psi(z)$ is not; therefore, $O(a)$ effects are
still possible. With two lattice spacings, it is not actually possible
to extrapolate a quadratic form in $a$. If one is cavalier, it could be
supposed that $O(a)$ effects may have been significantly reduced or
even canceled in the ratios for the reduced matrix elements. This
feature is almost certainly true for the low $\nu$ region where 
the normalization explicitly fixes  the value to 1. This hope if is further supported by the fact that the results from both lattice spacings are statistically consistent with each other in this region. With the two available lattice spacings an extrapolation may not be reliable, however, the primary goal of this exercise is not the extrapolation in itself but to understand the regions of $p$ and $z$ which show signs of larger discretization errors. As will be argued, the large momentum discretization errors, proportional to $\mathcal{O}(ap)$, are more significant than the low separation discretization errors, proportional to $\mathcal{O}(a/z)$.

\begin{figure}[h!]
  \centering
\includegraphics[width=0.495\textwidth]{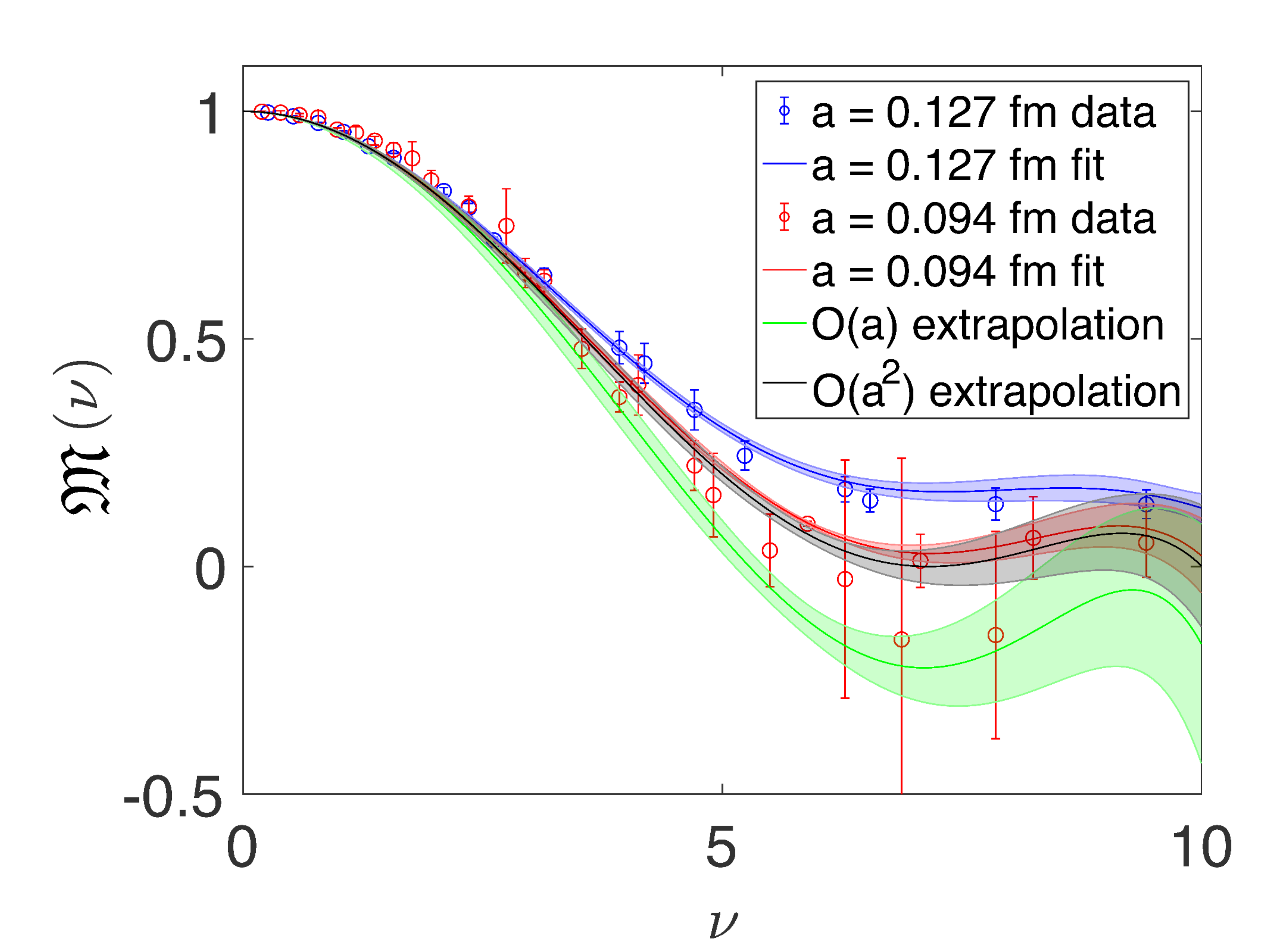}
\includegraphics[width=0.47\textwidth]{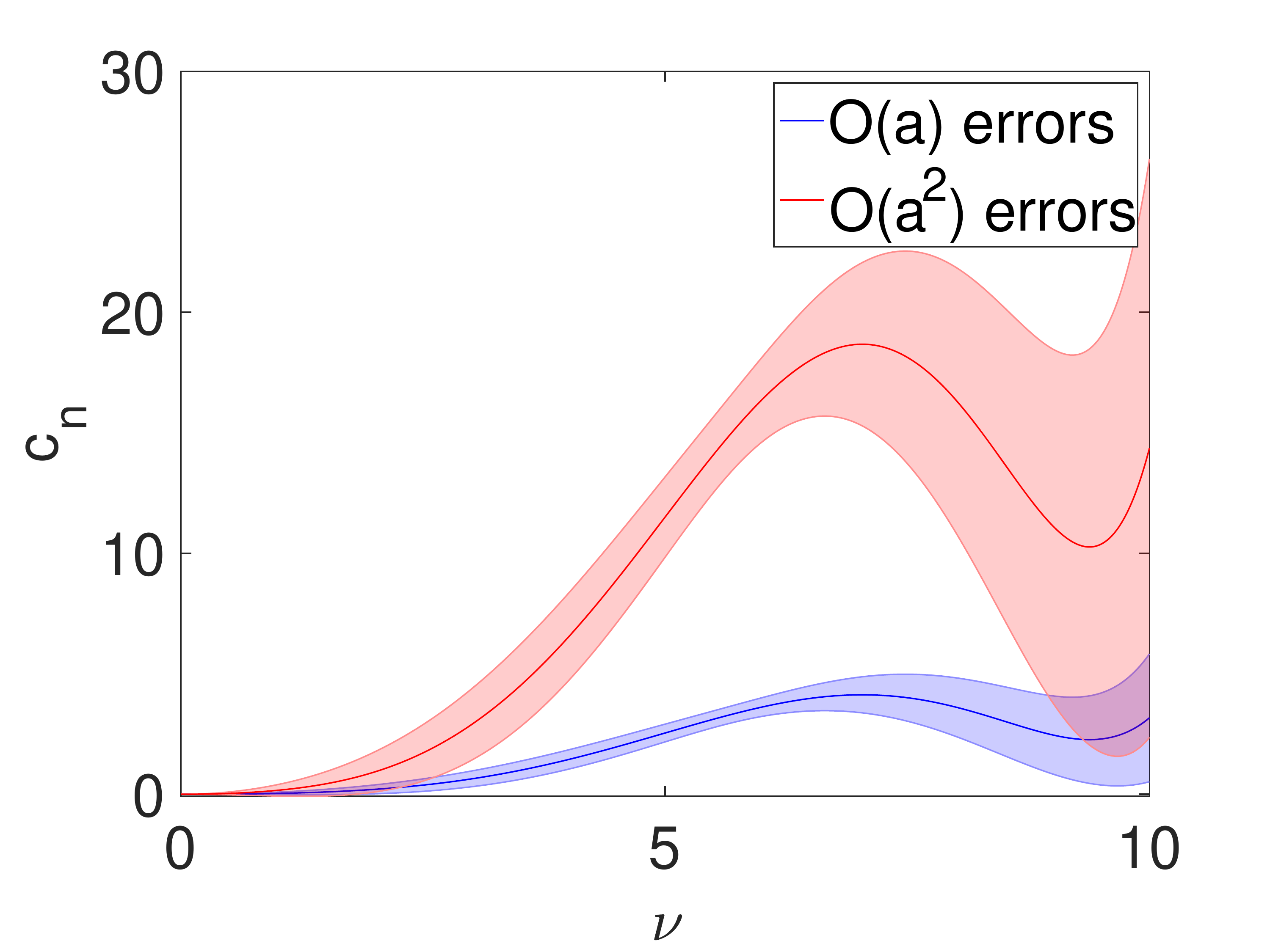}
  \caption{On the left is the reduced pseudo-ITD calculated from two lattice spacings and extrapolated to the continuum assuming either $O(a)$ or $O(a^2)$ errors. On the right is the coefficient of the discretization errors from Eq.~\eqref{eq:cont_error} shown for either $O(a)$ or $O(a^2)$ errors. The size of the discretization errors is small at low $\nu$ due to the normalization of the ITD.} 
  \label{fig:cont_lim}
\end{figure}

In order to study the discretization effects,
the real component of the reduced pseudo-ITD calculated on ensembles
$a094m400$ and $a127m440$, which are of approximately the same spatial
extent, are fit to a polynomial expansion
\beq\label{eq:itd_poly_expansion}
\mathfrak{M}(\nu) = 1 + a \nu^2 + b \nu^4 + c \nu^6 + d \nu^8\,. 
\eeq
For this fit, data with the same Ioffe time are averaged and the $z^2$ dependence is neglected. Due to the discretization of the allowed nucleon momentum $p$, the evolved reduced pseudo-ITD
$\mathfrak{M}(\nu)$ is calculated for a different set of $\nu$ on
configurations with different lattice lengths $L$, and therefore some
$\nu$ are in common between both ensembles but far from all of them.
The results of these fits as well as the
extrapolation to the continuum limit are shown in
Fig.~\ref{fig:cont_lim}. The discretization effects are assumed to
have the form \beq\label{eq:cont_error} \mathfrak{M}(\nu,a)_{\rm Latt} =
\mathfrak{M}(\nu)_{\rm Cont} + c_n(\nu) a^n\,, \eeq where $n=1,2$. Without more lattice spacings, particularly finer lattice spacings, these extrapolations should be considered with reservations. They are more of an attempt to quantify how significant the discretization effects can be especially for large values of the Ioffe time $\nu$ (originating from the region of large momenta). Consequently, no attempt to determine a continuum limit extrapolated PDF from these data will be made.

The discrepancy between the two lattice spacings is small at low $\nu$,
but becomes significant at large values of $\nu$. The low $\nu$ source
of discretization errors would come from effects proportional to
powers of $a/z$, but their size are restricted by the normalization of
the reduced pseudo-ITD. On the other hand, large $\nu$ discretization
errors are proportional to powers of $a p$ which are not constrained
in any way. The size of the coefficient $c_a(\nu)$ is shown in
Fig.~\ref{fig:cont_lim} for errors proportional to $O(a)$ and
$O(a^2)$.

The discretization errors appear to raise the lattice results at these fairly coarse lattice spacing. For obtaining results in the critical large Ioffe time region, finer lattice spacings are  required, in order to  both reduce  discretization effects and also to produce more data in this region due to finer resolution in momentum. 
It should be noted that lattice spacing errors at large Ioffe time would be dominated by $O(a p)$ effects rather than short distance effects that scale as $O(\frac az)$. Therefore, careful study of the continuum extrapolation of this large Ioffe time region should be performed.

\subsection{Finite Volume effects}
Another potential pitfall in the study of PDFs in numerical Lattice QCD arises from the non local operators used in these studies. Numerical Lattice QCD requires a finite volume to be used, whose effect on local matrix elements typically is exponentially suppressed as
\beq
M_{\rm Inf} = M_{\rm Latt} + C_L e^{-mL}\,,
\eeq
where $L$ is the length of the lattice and $m$ is the mass of the lightest particle of the theory with the appropriate quantum numbers. This effect can be thought of as coming from a particle traveling from the operator across the boundary and back to the operator. Generally, the lightest such particle, in QCD simulations with dynamical quarks, is a pion and lattices are designed to make $m_\pi L$ to be sufficiently large so that these effects are small. This picture for local matrix elements is modified by the finite size of the operator. In the case of a non local operator of size $z$, the lightest particle does not have to travel the full distance $L$ to return to the operator, but instead it must travel a distance $L-z$,
\beq
\mathfrak{M}_{\rm Inf} = \mathfrak{M}_{\rm Latt} + C_L(\nu) e^{-m(L-z)}\,.
\eeq 
The case of a two current operator has been studied for a model of scalar ``pions'' and ``nucleons'' in~\cite{Briceno:2018lfj}. This operator, $O(z) = J(z) J(0)$ has a periodic behavior under shifts of the lattice size
\beq
O(z) = O(z+L)\,.
\eeq
This periodicity drives the significant finite volume effects observed in~\cite{Briceno:2018lfj}, particularly for distances such as $z\gtrsim L/2$. It is also possible that $C_L$ should be augmented by powers of $(L-z)$ as was found in~\cite{Briceno:2018lfj}. For simplicity, in the following, these unknown powers will be neglected. In a proper study of the finite volume effects, the functional form would need to be determined for the Wilson line matrix element.

\begin{figure}[ht]
\includegraphics[width=3in]{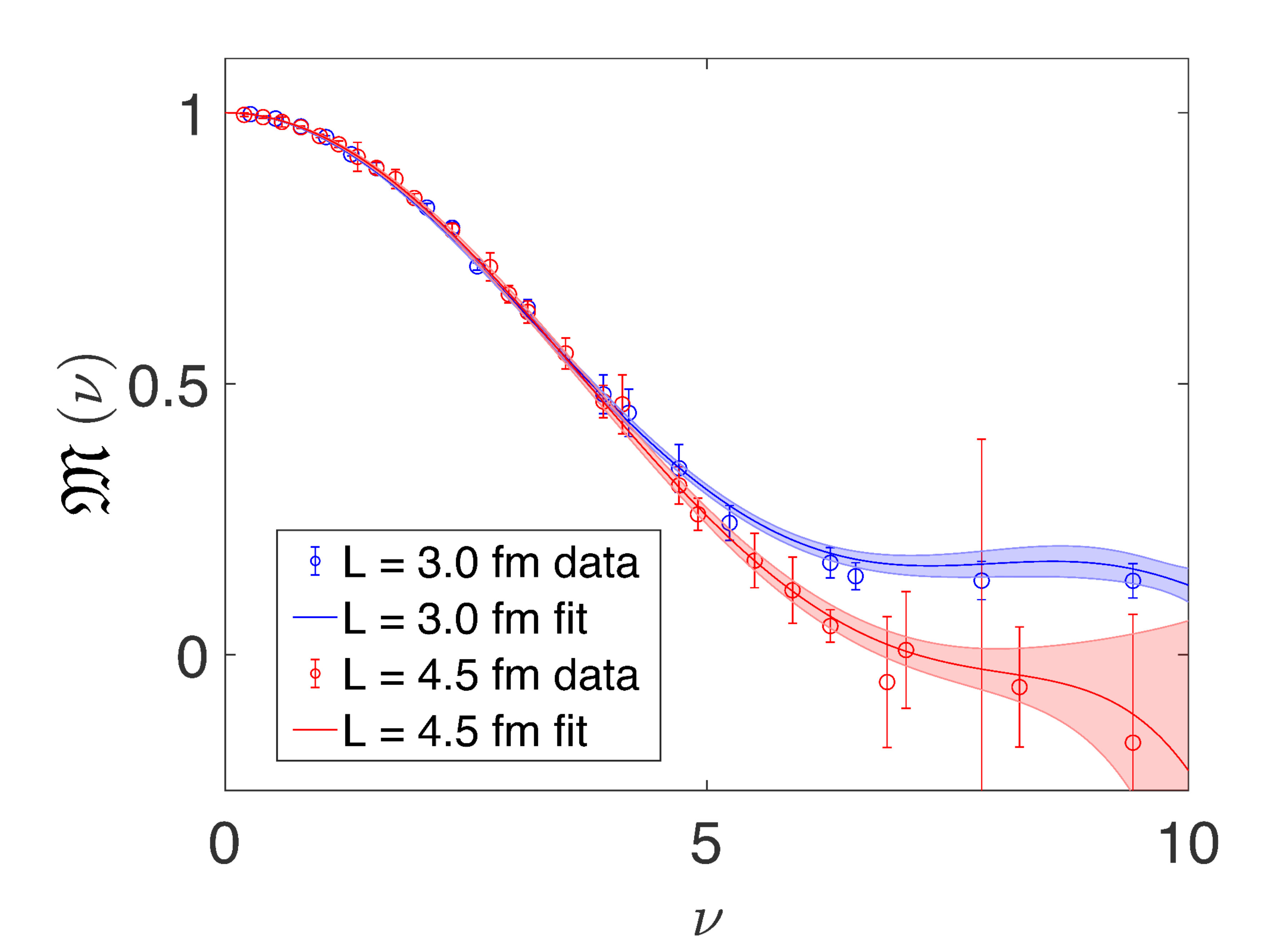}
\includegraphics[width=3in]{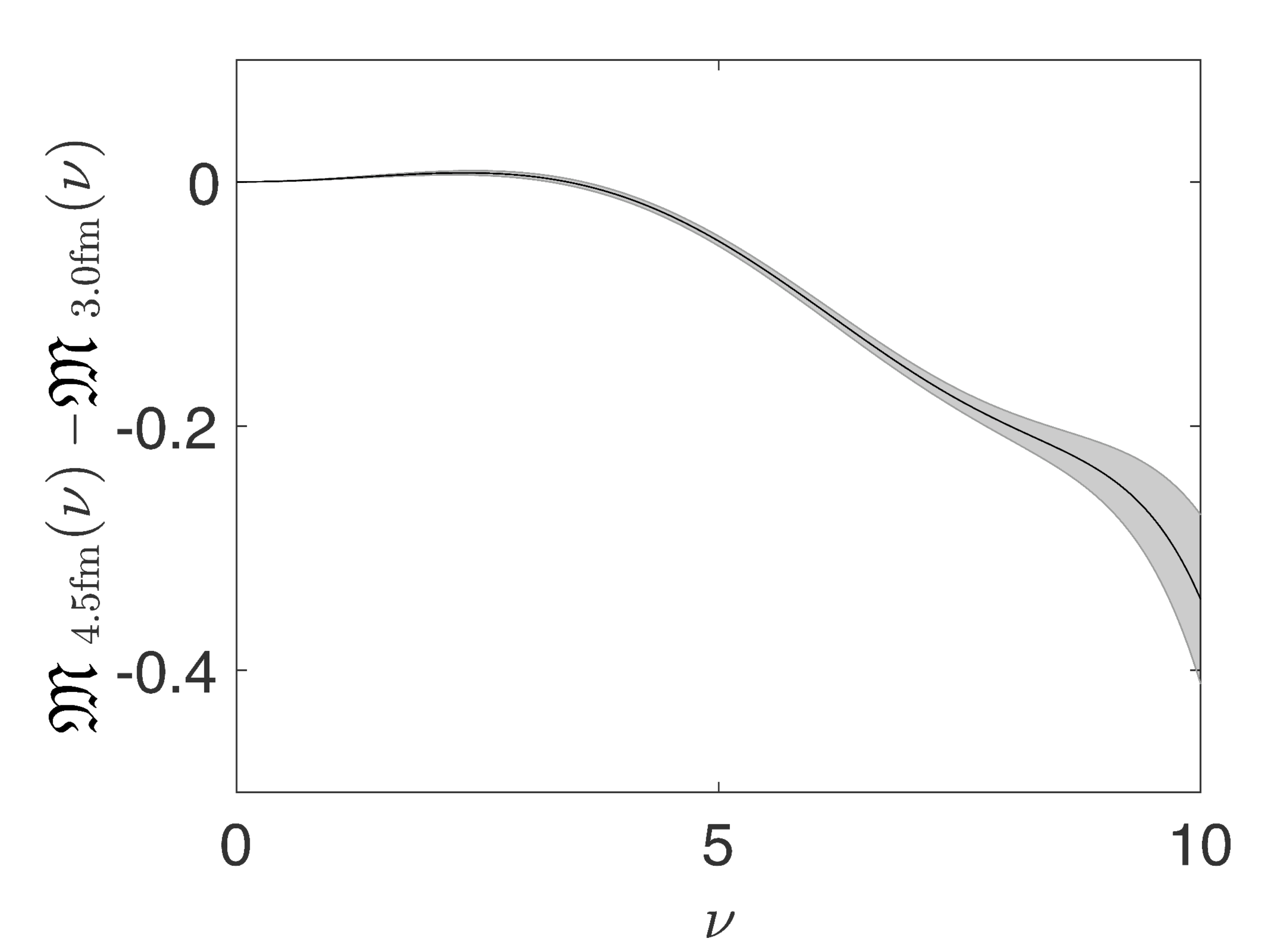}
\caption{The real components of the reduced pseudo-ITD from the ensembles $a127m415$ and $a127m415L$ with volumes of approximately 3 fm and 4.5 fm respectively. There appear to be slight finite volume effects whose difference is plotted on the right.}

\label{fig:fin_vol}
\end{figure}

The Wilson line operator defining the pseudo-ITD does not have this same periodic feature, but finite volume effects can still be significant. The two ensembles with lattice spacing $a=0.127$ fm, $a127m415$ and $a127m415L$ have volumes of approximately 3 fm and 4.5 fm respectively. Just as was done when studying lattice spacing effects, the reduced pseudo-ITD is fit to the form in Eq.~\eqref{eq:itd_poly_expansion}. In Fig.~\ref{fig:fin_vol}, the reduced pseudo-ITDs calculated on both ensembles are compared. There appears only a slight sign of deviation for the results on these two volumes from the data, but the fits to the polynomial expression show clear deviations. The finite volume effects do appear to be small particularly at small Ioffe time and the difference between the fit results are shown in Fig~\ref{fig:fin_vol}. These differences are largest in the large Ioffe time region, $\nu \gtrsim 5$, where the data originated from large $z$ and the matrix elements are the least precise. The largest $z$ used on either of these lattices was $L/4$. With these heavy pion masses, the product $m_\pi L$ is fairly large for a lattice calculation. One needs to keep in mind when performing a lighter pion mass calculation that the significance of finite volume effects should be checked.

\section{Conclusions}\label{sec:con}
In this work we presented a detailed and systematic analysis of the
extraction of nucleon PDFs based on the formalism of pseudo-PDFs. We
have employed lattice ensembles with $N_{\rm f}=2+1$ Wilson-clover
fermions with stout smearing and tree-level tadpole improved Symanzik
gauge action with two values of the lattice spacing namely $a$=0.127
fm and $a$=0.094fm. While two values of the lattice spacing are not
sufficient for a stringent control of the continuum extrapolation,
they do provide us with a lot of information regarding the size and source of
discretization errors for our formalism. In future studies, more
lattice spacings should be used for a more robust extrapolation to the
continuum limit as well as studying the functional forms used to
interpolate the $\nu$ dependence.
Moreover, since toy model
calculations of PDFs employing formalisms that are based on spatially
nonlocal operators could potentially suffer from enhanced finite
volume effects~\cite{Briceno:2018lfj} we have also addressed two
different physical volumes for the case of the coarser lattice
spacings. Our studies did reveal the presence of slight finite volume
effects for the values of the parameters that we investigated. The volume dependence does appear to be larger for matrix elements that come from large separations as expected from the toy model calculation. For a proper study of the finite volume effects, the functional form for the Wilson line matrix element will need to be determined. However, we believe that the functional form used in our studies is well motivated if one assumes that finite volume effects arise from pion exchanges that wrap around the periodic volume.  Also just 
as with the lattice spacing analysis, the dependence of this
discrepancy on the functional form used in the interpolation of $\nu$
should be studied in future work as well as including more
volumes.

The authors of~\cite{Alexandrou:2019lfo} stressed the
necessity of controlling the contamination from excited states which
becomes an increasing concern as one approaches the limit of the
physical pion mass. In this respect we discussed in detail how our
method for the extraction of the matrix element possesses a number of
advantages compared to the commonly utilized sequential source
technique. As was shown in ~\cite{Chang:2018uxx}, the matrix element
extraction based on the Feynman-Hellman theorem can begin at much
earlier times and this is very advantageous for simulations with
physical pion mass. Additionally, we have also employed the method of
momentum smearing which has proven to substantially improve the
overlap of the interpolating fields with the boosted hadron ground
state.

Beyond the extraction of the $x$-dependence of the PDF we also
perform the extraction of the lowest two moments of the PDF and we
compare to the pertinent phenomenological determinations from CJ 15,
NNPDF and MSTW collaborations. Our results, lie above the results of
CJ 15 and MSTW, as expected, due to the relatively heavy masses of our
pions but agree within errors with NNPDF due to the larger error bars
of the latter. As was analyzed in detail in~\cite{Lin:2017snn} the
complementary synergies between the communities of global fits and
Lattice QCD would be very fruitful in the forthcoming years. In the
latter article different scenarios of lattice data included in the
global fits analyses were presented. In all cases, the conclusion was
that a close collaboration of the two communities is necessary in
order to achieve the best possible PDF extraction.

In this work despite the relatively heavy pions we have addressed many
of the pertinent systematics of the extraction of light cone PDFs with
the method of pseudo-PDFs. Our studies have shown that the lattice
community as the time goes by can have under better control all
systematics of these calculations and steady progress is being
made. In our forthcoming studies we plan to employ lattice ensembles
which have pion masses at the physical pion mass and consider also the
pion PDF besides that of the nucleon.

\section{Acknowledgements}
AR thanks V. Braun,  X. Ji,  J. Qiu and Y. Zhao for discussions and comments. SZ thanks
K. Cichy for useful discussions.  This work is supported by Jefferson
Science Associates, LLC under U.S. DOE Contract \#DE-AC05-06OR23177.
KO was supported in part by U.S.  DOE grant \mbox{
  \#DE-FG02-04ER41302}, and in part  by STFC consolidated grant ST/P000681/1, and  acknowledges the hospitality from DAMTP and Clare Hall at Cambridge University.
    AR was supported in part by U.S. DOE Grant
\mbox{\#DE-FG02-97ER41028. } SZ acknowledges support by the DFG
Collaborative Research Centre SFB 1225 (ISOQUANT).  J.K. was supported
in part by the U.S. Department of Energy under contract
DE-FG02-04ER41302 and Department of Energy Office of Science Graduate
Student Research fellowships, through the U.S. Department of Energy,
Office of Science, Office of Workforce Development for Teachers and
Scientists, Office of Science Graduate Student Research (SCGSR)
program.  The authors gratefully acknowledge the computing time
granted by the John von Neumann Institute for Computing (NIC) and
provided on the supercomputer JURECA at J\"ulich Supercomputing Centre
(JSC)~\cite{jureca}.  This work was performed in part using computing
facilities at the College of William and Mary which were provided by
contributions from the National Science Foundation (MRI grant
PHY-1626177), the Commonwealth of Virginia Equipment Trust Fund and
the Office of Naval Research. In addition, this work used resources at
NERSC, a DOE Office of Science User Facility supported by the Office
of Science of the U.S. Department of Energy under Contract
\#DE-AC02-05CH11231, as well as resources of the Oak Ridge Leadership Computing Facility at the Oak Ridge National Laboratory, which is supported by the Office of Science of the U.S. Department of Energy under Contract No. \mbox{\#DE-AC05-00OR22725}.

\clearpage

\bibliography{dynppdfs.bib}

\providecommand{\href}[2]{#2}\begingroup\raggedright\begin{thebibliography}{10}

\bibitem{Feynman:1973xc}
R.~P. Feynman, \emph{{\it Photon-hadron interactions}}.
\newblock Reading, 1972.

\bibitem{Bali:2013gya}
G.~Bali, S.~Collins, B.~Glässle, M.~Göckeler, N.~Javadi-Motaghi, J.~Najjar
  et~al., \emph{{Pion structure from lattice QCD}},
  \href{https://doi.org/10.22323/1.187.0447}{\emph{PoS} {\bfseries LATTICE2013}
  (2014) 447}, [\href{https://arxiv.org/abs/1311.7639}{{\ttfamily 1311.7639}}].

\bibitem{Abdel-Rehim:2015owa}
A.~Abdel-Rehim et~al., \emph{{Nucleon and pion structure with lattice QCD
  simulations at physical value of the pion mass}},
  \href{https://doi.org/10.1103/PhysRevD.92.114513,
  10.1103/PhysRevD.93.039904}{\emph{Phys. Rev.} {\bfseries D92} (2015) 114513},
  [\href{https://arxiv.org/abs/1507.04936}{{\ttfamily 1507.04936}}].

\bibitem{Alexandrou:2017oeh}
C.~Alexandrou, M.~Constantinou, K.~Hadjiyiannakou, K.~Jansen, C.~Kallidonis,
  G.~Koutsou et~al., \emph{{Nucleon Spin and Momentum Decomposition Using
  Lattice QCD Simulations}},
  \href{https://doi.org/10.1103/PhysRevLett.119.142002}{\emph{Phys. Rev. Lett.}
  {\bfseries 119} (2017) 142002},
  [\href{https://arxiv.org/abs/1706.02973}{{\ttfamily 1706.02973}}].

\bibitem{Oehm:2018jvm}
M.~Oehm, C.~Alexandrou, M.~Constantinou, K.~Jansen, G.~Koutsou, B.~Kostrzewa
  et~al., \emph{{$\langle x\rangle$ and $\langle x^2\rangle$ of the pion PDF
  from lattice QCD with $N_f=2+1+1$ dynamical quark flavors}},
  \href{https://doi.org/10.1103/PhysRevD.99.014508}{\emph{Phys. Rev.}
  {\bfseries D99} (2019) 014508},
  [\href{https://arxiv.org/abs/1810.09743}{{\ttfamily 1810.09743}}].

\bibitem{Bali:2019ecy}
G.~S. Bali et~al., \emph{{Baryon distribution amplitudes in QCD}},
  \href{https://arxiv.org/abs/1903.12590}{{\ttfamily 1903.12590}}.

\bibitem{Bali:2019dqc}
G.~S. Bali, V.~M. Braun, S.~Bürger, M.~Göckeler, M.~Gruber, F.~Hutzler
  et~al., \emph{{Light-cone distribution amplitudes of pseudoscalar mesons from
  lattice QCD}}, \href{https://doi.org/10.1007/JHEP08(2019)065}{\emph{JHEP}
  {\bfseries 08} (2019) 065},
  [\href{https://arxiv.org/abs/1903.08038}{{\ttfamily 1903.08038}}].

\bibitem{Detmold:2001dv}
W.~Detmold, W.~Melnitchouk and A.~W. Thomas, \emph{{Parton distributions from
  lattice QCD}}, \href{https://doi.org/10.1007/s1010501c0013}{\emph{Eur. Phys.
  J.direct} {\bfseries 3} (2001) 13},
  [\href{https://arxiv.org/abs/hep-lat/0108002}{{\ttfamily hep-lat/0108002}}].

\bibitem{Ji:2013dva}
X.~Ji, \emph{Parton physics on a euclidean lattice},
  \href{https://doi.org/10.1103/PhysRevLett.110.262002}{\emph{Phys. Rev. Lett.}
  {\bfseries 110} (Jun, 2013) 262002}.

\bibitem{Lin:2014zya}
H.-W. Lin, J.-W. Chen, S.~D. Cohen and X.~Ji, \emph{{Flavor Structure of the
  Nucleon Sea from Lattice QCD}},
  \href{https://doi.org/10.1103/PhysRevD.91.054510}{\emph{Phys. Rev.}
  {\bfseries D91} (2015) 054510},
  [\href{https://arxiv.org/abs/1402.1462}{{\ttfamily 1402.1462}}].

\bibitem{Chen:2016utp}
J.-W. Chen, S.~D. Cohen, X.~Ji, H.-W. Lin and J.-H. Zhang, \emph{{Nucleon
  Helicity and Transversity Parton Distributions from Lattice QCD}},
  \href{https://doi.org/10.1016/j.nuclphysb.2016.07.033}{\emph{Nucl. Phys.}
  {\bfseries B911} (2016) 246--273},
  [\href{https://arxiv.org/abs/1603.06664}{{\ttfamily 1603.06664}}].

\bibitem{Alexandrou:2015rja}
C.~Alexandrou, K.~Cichy, V.~Drach, E.~Garcia-Ramos, K.~Hadjiyiannakou,
  K.~Jansen et~al., \emph{{Lattice calculation of parton distributions}},
  \href{https://doi.org/10.1103/PhysRevD.92.014502}{\emph{Phys. Rev.}
  {\bfseries D92} (2015) 014502},
  [\href{https://arxiv.org/abs/1504.07455}{{\ttfamily 1504.07455}}].

\bibitem{Alexandrou:2016jqi}
C.~Alexandrou, K.~Cichy, M.~Constantinou, K.~Hadjiyiannakou, K.~Jansen,
  F.~Steffens et~al., \emph{{Updated Lattice Results for Parton
  Distributions}},
  \href{https://doi.org/10.1103/PhysRevD.96.014513}{\emph{Phys. Rev.}
  {\bfseries D96} (2017) 014513},
  [\href{https://arxiv.org/abs/1610.03689}{{\ttfamily 1610.03689}}].

\bibitem{Monahan:2016bvm}
C.~Monahan and K.~Orginos, \emph{{Quasi parton distributions and the gradient
  flow}}, \href{https://doi.org/10.1007/JHEP03(2017)116}{\emph{JHEP} {\bfseries
  03} (2017) 116}, [\href{https://arxiv.org/abs/1612.01584}{{\ttfamily
  1612.01584}}].

\bibitem{Zhang:2017bzy}
J.-H. Zhang, J.-W. Chen, X.~Ji, L.~Jin and H.-W. Lin, \emph{{Pion Distribution
  Amplitude from Lattice QCD}},
  \href{https://doi.org/10.1103/PhysRevD.95.094514}{\emph{Phys. Rev.}
  {\bfseries D95} (2017) 094514},
  [\href{https://arxiv.org/abs/1702.00008}{{\ttfamily 1702.00008}}].

\bibitem{Alexandrou:2017huk}
C.~Alexandrou, K.~Cichy, M.~Constantinou, K.~Hadjiyiannakou, K.~Jansen,
  H.~Panagopoulos et~al., \emph{{A complete non-perturbative renormalization
  prescription for quasi-PDFs}},
  \href{https://doi.org/10.1016/j.nuclphysb.2017.08.012}{\emph{Nucl. Phys.}
  {\bfseries B923} (2017) 394--415},
  [\href{https://arxiv.org/abs/1706.00265}{{\ttfamily 1706.00265}}].

\bibitem{Green:2017xeu}
J.~Green, K.~Jansen and F.~Steffens, \emph{{Nonperturbative Renormalization of
  Nonlocal Quark Bilinears for Parton Quasidistribution Functions on the
  Lattice Using an Auxiliary Field}},
  \href{https://doi.org/10.1103/PhysRevLett.121.022004}{\emph{Phys. Rev. Lett.}
  {\bfseries 121} (2018) 022004},
  [\href{https://arxiv.org/abs/1707.07152}{{\ttfamily 1707.07152}}].

\bibitem{Stewart:2017tvs}
I.~W. Stewart and Y.~Zhao, \emph{{Matching the quasiparton distribution in a
  momentum subtraction scheme}},
  \href{https://doi.org/10.1103/PhysRevD.97.054512}{\emph{Phys. Rev.}
  {\bfseries D97} (2018) 054512},
  [\href{https://arxiv.org/abs/1709.04933}{{\ttfamily 1709.04933}}].

\bibitem{Monahan:2017hpu}
C.~Monahan, \emph{{Smeared quasidistributions in perturbation theory}},
  \href{https://doi.org/10.1103/PhysRevD.97.054507}{\emph{Phys. Rev.}
  {\bfseries D97} (2018) 054507},
  [\href{https://arxiv.org/abs/1710.04607}{{\ttfamily 1710.04607}}].

\bibitem{Broniowski:2017gfp}
W.~Broniowski and E.~Ruiz~Arriola, \emph{{Partonic quasidistributions of the
  proton and pion from transverse-momentum distributions}},
  \href{https://doi.org/10.1103/PhysRevD.97.034031}{\emph{Phys. Rev.}
  {\bfseries D97} (2018) 034031},
  [\href{https://arxiv.org/abs/1711.03377}{{\ttfamily 1711.03377}}].

\bibitem{Alexandrou:2018pbm}
C.~Alexandrou, K.~Cichy, M.~Constantinou, K.~Jansen, A.~Scapellato and
  F.~Steffens, \emph{{Light-Cone Parton Distribution Functions from Lattice
  QCD}}, \href{https://doi.org/10.1103/PhysRevLett.121.112001}{\emph{Phys. Rev.
  Lett.} {\bfseries 121} (2018) 112001},
  [\href{https://arxiv.org/abs/1803.02685}{{\ttfamily 1803.02685}}].

\bibitem{Alexandrou:2018eet}
C.~Alexandrou, K.~Cichy, M.~Constantinou, K.~Jansen, A.~Scapellato and
  F.~Steffens, \emph{{Transversity parton distribution functions from lattice
  QCD}}, \href{https://doi.org/10.1103/PhysRevD.98.091503}{\emph{Phys. Rev.}
  {\bfseries D98} (2018) 091503},
  [\href{https://arxiv.org/abs/1807.00232}{{\ttfamily 1807.00232}}].

\bibitem{Alexandrou:2019lfo}
C.~Alexandrou, K.~Cichy, M.~Constantinou, K.~Hadjiyiannakou, K.~Jansen,
  A.~Scapellato et~al., \emph{{Systematic uncertainties in parton distribution
  functions from lattice QCD simulations at the physical point}},
  \href{https://doi.org/10.1103/PhysRevD.99.114504}{\emph{Phys. Rev.}
  {\bfseries D99} (2019) 114504},
  [\href{https://arxiv.org/abs/1902.00587}{{\ttfamily 1902.00587}}].

\bibitem{Izubuchi:2019lyk}
T.~Izubuchi, L.~Jin, C.~Kallidonis, N.~Karthik, S.~Mukherjee, P.~Petreczky
  et~al., \emph{{Valence parton distribution function of pion from fine
  lattice}},  \href{https://arxiv.org/abs/1905.06349}{{\ttfamily 1905.06349}}.

\bibitem{Detmold:2005gg}
W.~Detmold and C.~J.~D. Lin, \emph{{Deep-inelastic scattering and the operator
  product expansion in lattice QCD}},
  \href{https://doi.org/10.1103/PhysRevD.73.014501}{\emph{Phys. Rev.}
  {\bfseries D73} (2006) 014501},
  [\href{https://arxiv.org/abs/hep-lat/0507007}{{\ttfamily hep-lat/0507007}}].

\bibitem{Braun:2007wv}
V.~Braun and D.~Müller, \emph{{Exclusive processes in position space and the
  pion distribution amplitude}},
  \href{https://doi.org/10.1140/epjc/s10052-008-0608-4}{\emph{Eur. Phys. J.}
  {\bfseries C55} (2008) 349--361},
  [\href{https://arxiv.org/abs/0709.1348}{{\ttfamily 0709.1348}}].

\bibitem{Chambers:2017dov}
A.~J. Chambers, R.~Horsley, Y.~Nakamura, H.~Perlt, P.~E.~L. Rakow,
  G.~Schierholz et~al., \emph{{Nucleon Structure Functions from Operator
  Product Expansion on the Lattice}},
  \href{https://doi.org/10.1103/PhysRevLett.118.242001}{\emph{Phys. Rev. Lett.}
  {\bfseries 118} (2017) 242001},
  [\href{https://arxiv.org/abs/1703.01153}{{\ttfamily 1703.01153}}].

\bibitem{Liang:2019frk}
J.~Liang, T.~Draper, K.-F. Liu, A.~Rothkopf and Y.-B. Yang, \emph{{Towards the
  nucleon hadronic tensor from lattice QCD}},
  \href{https://arxiv.org/abs/1906.05312}{{\ttfamily 1906.05312}}.

\bibitem{Radyushkin:2016hsy}
A.~Radyushkin, \emph{{Nonperturbative Evolution of Parton
  Quasi-Distributions}},
  \href{https://doi.org/10.1016/j.physletb.2017.02.019}{\emph{Phys. Lett.}
  {\bfseries B767} (2017) 314--320},
  [\href{https://arxiv.org/abs/1612.05170}{{\ttfamily 1612.05170}}].

\bibitem{Radyushkin:2017gjd}
A.~V. Radyushkin, \emph{{Pion Distribution Amplitude and Quasi-Distributions}},
  \href{https://doi.org/10.1103/PhysRevD.95.056020}{\emph{Phys. Rev.}
  {\bfseries D95} (2017) 056020},
  [\href{https://arxiv.org/abs/1701.02688}{{\ttfamily 1701.02688}}].

\bibitem{Musch:2010ka}
B.~U. Musch, P.~Hagler, J.~W. Negele and A.~Schafer, \emph{{Exploring quark
  transverse momentum distributions with lattice QCD}},
  \href{https://doi.org/10.1103/PhysRevD.83.094507}{\emph{Phys. Rev.}
  {\bfseries D83} (2011) 094507},
  [\href{https://arxiv.org/abs/1011.1213}{{\ttfamily 1011.1213}}].

\bibitem{Radyushkin:2017cyf}
A.~V. Radyushkin, \emph{{Quasi-parton distribution functions, momentum
  distributions, and pseudo-parton distribution functions}},
  \href{https://doi.org/10.1103/PhysRevD.96.034025}{\emph{Phys. Rev.}
  {\bfseries D96} (2017) 034025},
  [\href{https://arxiv.org/abs/1705.01488}{{\ttfamily 1705.01488}}].

\bibitem{Ma:2017pxb}
Y.-Q. Ma and J.-W. Qiu, \emph{{Exploring Partonic Structure of Hadrons Using ab
  initio Lattice QCD Calculations}},
  \href{https://doi.org/10.1103/PhysRevLett.120.022003}{\emph{Phys. Rev. Lett.}
  {\bfseries 120} (2018) 022003},
  [\href{https://arxiv.org/abs/1709.03018}{{\ttfamily 1709.03018}}].

\bibitem{Orginos:2017wcl}
K.~Orginos, A.~Radyushkin, J.~Karpie and S.~Zafeiropoulos, \emph{{Lattice QCD
  exploration of parton pseudo-distribution functions}},
  \href{https://doi.org/10.1103/PhysRevD.96.094503}{\emph{Phys. Rev.}
  {\bfseries D96} (2017) 094503},
  [\href{https://arxiv.org/abs/1706.05373}{{\ttfamily 1706.05373}}].

\bibitem{Karpie:2017bzm}
J.~Karpie, K.~Orginos, A.~Radyushkin and S.~Zafeiropoulos, \emph{{Parton
  distribution functions on the lattice and in the continuum}},
  \href{https://doi.org/10.1051/epjconf/201817506032}{\emph{EPJ Web Conf.}
  {\bfseries 175} (2018) 06032},
  [\href{https://arxiv.org/abs/1710.08288}{{\ttfamily 1710.08288}}].

\bibitem{Bali:2017gfr}
G.~S. Bali et~al., \emph{{Pion distribution amplitude from Euclidean
  correlation functions}},
  \href{https://doi.org/10.1140/epjc/s10052-018-5700-9}{\emph{Eur. Phys. J.}
  {\bfseries C78} (2018) 217},
  [\href{https://arxiv.org/abs/1709.04325}{{\ttfamily 1709.04325}}].

\bibitem{Bali:2018spj}
G.~S. Bali, V.~M. Braun, B.~Gläßle, M.~Göckeler, M.~Gruber, F.~Hutzler
  et~al., \emph{{Pion distribution amplitude from Euclidean correlation
  functions: Exploring universality and higher-twist effects}},
  \href{https://doi.org/10.1103/PhysRevD.98.094507}{\emph{Phys. Rev.}
  {\bfseries D98} (2018) 094507},
  [\href{https://arxiv.org/abs/1807.06671}{{\ttfamily 1807.06671}}].

\bibitem{Sufian:2019bol}
R.~S. Sufian, J.~Karpie, C.~Egerer, K.~Orginos, J.-W. Qiu and D.~G. Richards,
  \emph{{Pion Valence Quark Distribution from Matrix Element Calculated in
  Lattice QCD}}, \href{https://doi.org/10.1103/PhysRevD.99.074507}{\emph{Phys.
  Rev.} {\bfseries D99} (2019) 074507},
  [\href{https://arxiv.org/abs/1901.03921}{{\ttfamily 1901.03921}}].

\bibitem{Lin:2017snn}
H.-W. Lin et~al., \emph{{Parton distributions and lattice QCD calculations: a
  community white paper}},
  \href{https://doi.org/10.1016/j.ppnp.2018.01.007}{\emph{Prog. Part. Nucl.
  Phys.} {\bfseries 100} (2018) 107--160},
  [\href{https://arxiv.org/abs/1711.07916}{{\ttfamily 1711.07916}}].

\bibitem{Cichy:2018mum}
K.~Cichy and M.~Constantinou, \emph{{A guide to light-cone PDFs from Lattice
  QCD: an overview of approaches, techniques and results}},
  \href{https://doi.org/10.1155/2019/3036904}{\emph{Adv. High Energy Phys.}
  {\bfseries 2019} (2019) 3036904},
  [\href{https://arxiv.org/abs/1811.07248}{{\ttfamily 1811.07248}}].

\bibitem{Monahan:2018euv}
C.~Monahan, \emph{{Recent Developments in $x$-dependent Structure
  Calculations}}, {\emph{PoS} {\bfseries LATTICE2018} (2018) 018},
  [\href{https://arxiv.org/abs/1811.00678}{{\ttfamily 1811.00678}}].

\bibitem{Qiu:2019kyy}
J.-W. Qiu, \emph{{Nucleon Structure from Lattice QCD Calculations}},  in
  \emph{{8th International Conference on Quarks and Nuclear Physics (QNP2018)
  Tsukuba, Japan, November 13-17, 2018}}, 2019,
  \href{https://arxiv.org/abs/1903.11902}{{\ttfamily 1903.11902}}.

\bibitem{Rossi:2017muf}
G.~C. Rossi and M.~Testa, \emph{{Note on lattice regularization and equal-time
  correlators for parton distribution functions}},
  \href{https://doi.org/10.1103/PhysRevD.96.014507}{\emph{Phys. Rev.}
  {\bfseries D96} (2017) 014507},
  [\href{https://arxiv.org/abs/1706.04428}{{\ttfamily 1706.04428}}].

\bibitem{Rossi:2018zkn}
G.~Rossi and M.~Testa, \emph{{Euclidean versus Minkowski short distance}},
  \href{https://doi.org/10.1103/PhysRevD.98.054028}{\emph{Phys. Rev.}
  {\bfseries D98} (2018) 054028},
  [\href{https://arxiv.org/abs/1806.00808}{{\ttfamily 1806.00808}}].

\bibitem{Ji:2017rah}
X.~Ji, J.-H. Zhang and Y.~Zhao, \emph{{More On Large-Momentum Effective Theory
  Approach to Parton Physics}},
  \href{https://doi.org/10.1016/j.nuclphysb.2017.09.001}{\emph{Nucl. Phys.}
  {\bfseries B924} (2017) 366--376},
  [\href{https://arxiv.org/abs/1706.07416}{{\ttfamily 1706.07416}}].

\bibitem{Radyushkin:2018nbf}
A.~V. Radyushkin, \emph{{Structure of parton quasi-distributions and their
  moments}}, \href{https://doi.org/10.1016/j.physletb.2018.11.047}{\emph{Phys.
  Lett.} {\bfseries B788} (2019) 380--387},
  [\href{https://arxiv.org/abs/1807.07509}{{\ttfamily 1807.07509}}].

\bibitem{Karpie:2018zaz}
J.~Karpie, K.~Orginos and S.~Zafeiropoulos, \emph{{Moments of Ioffe time parton
  distribution functions from non-local matrix elements}},
  \href{https://doi.org/10.1007/JHEP11(2018)178}{\emph{JHEP} {\bfseries 11}
  (2018) 178}, [\href{https://arxiv.org/abs/1807.10933}{{\ttfamily
  1807.10933}}].

\bibitem{Dawson:1997ic}
C.~Dawson, G.~Martinelli, G.~C. Rossi, C.~T. Sachrajda, S.~R. Sharpe, M.~Talevi
  et~al., \emph{{New lattice approaches to the delta I = 1/2 rule}},
  \href{https://doi.org/10.1016/S0550-3213(97)00756-6}{\emph{Nucl. Phys.}
  {\bfseries B514} (1998) 313--335},
  [\href{https://arxiv.org/abs/hep-lat/9707009}{{\ttfamily hep-lat/9707009}}].

\bibitem{Martinelli:1998hz}
G.~Martinelli, \emph{{Hadronic weak interactions of light quarks}},
  \href{https://doi.org/10.1016/S0920-5632(99)85007-5}{\emph{Nucl. Phys. Proc.
  Suppl.} {\bfseries 73} (1999) 58--71},
  [\href{https://arxiv.org/abs/hep-lat/9810013}{{\ttfamily hep-lat/9810013}}].

\bibitem{Bali:2016lva}
G.~S. Bali, B.~Lang, B.~U. Musch and A.~Schäfer, \emph{{Novel quark smearing
  for hadrons with high momenta in lattice QCD}},
  \href{https://doi.org/10.1103/PhysRevD.93.094515}{\emph{Phys. Rev.}
  {\bfseries D93} (2016) 094515},
  [\href{https://arxiv.org/abs/1602.05525}{{\ttfamily 1602.05525}}].

\bibitem{Radyushkin:2017lvu}
A.~V. Radyushkin, \emph{{Quark pseudodistributions at short distances}},
  \href{https://doi.org/10.1016/j.physletb.2018.04.023}{\emph{Phys. Lett.}
  {\bfseries B781} (2018) 433--442},
  [\href{https://arxiv.org/abs/1710.08813}{{\ttfamily 1710.08813}}].

\bibitem{Radyushkin:2017sfi}
A.~Radyushkin, \emph{{Quasi-PDFs and pseudo-PDFs}},
  \href{https://doi.org/10.22323/1.308.0021}{\emph{PoS} {\bfseries QCDEV2017}
  (2017) 021}, [\href{https://arxiv.org/abs/1711.06031}{{\ttfamily
  1711.06031}}].

\bibitem{Karpie:2019eiq}
J.~Karpie, K.~Orginos, A.~Rothkopf and S.~Zafeiropoulos, \emph{{Reconstructing
  parton distribution functions from Ioffe time data: from Bayesian methods to
  Neural Networks}}, \href{https://doi.org/10.1007/JHEP04(2019)057}{\emph{JHEP}
  {\bfseries 04} (2019) 057},
  [\href{https://arxiv.org/abs/1901.05408}{{\ttfamily 1901.05408}}].

\bibitem{Orginos:2017kos}
K.~Orginos, A.~Radyushkin, J.~Karpie and S.~Zafeiropoulos, \emph{{Lattice QCD
  exploration of parton pseudo-distribution functions}},
  \href{https://doi.org/10.1103/PhysRevD.96.094503}{\emph{Phys. Rev.}
  {\bfseries D96} (2017) 094503},
  [\href{https://arxiv.org/abs/1706.05373}{{\ttfamily 1706.05373}}].

\bibitem{Braun:2018brg}
V.~M. Braun, A.~Vladimirov and J.-H. Zhang, \emph{{Power corrections and
  renormalons in parton quasidistributions}},
  \href{https://doi.org/10.1103/PhysRevD.99.014013}{\emph{Phys. Rev.}
  {\bfseries D99} (2019) 014013},
  [\href{https://arxiv.org/abs/1810.00048}{{\ttfamily 1810.00048}}].

\bibitem{Dotsenko:1979wb}
V.~S. Dotsenko and S.~N. Vergeles, \emph{{Renormalizability of Phase Factors in
  the Nonabelian Gauge Theory}},
  \href{https://doi.org/10.1016/0550-3213(80)90103-0}{\emph{Nucl. Phys.}
  {\bfseries B169} (1980) 527--546}.

\bibitem{Brandt:1981kf}
R.~A. Brandt, F.~Neri and M.-a. Sato, \emph{{Renormalization of Loop Functions
  for All Loops}}, \href{https://doi.org/10.1103/PhysRevD.24.879}{\emph{Phys.
  Rev.} {\bfseries D24} (1981) 879}.

\bibitem{Ishikawa:2017faj}
T.~Ishikawa, Y.-Q. Ma, J.-W. Qiu and S.~Yoshida, \emph{{Renormalizability of
  quasiparton distribution functions}},
  \href{https://doi.org/10.1103/PhysRevD.96.094019}{\emph{Phys. Rev.}
  {\bfseries D96} (2017) 094019},
  [\href{https://arxiv.org/abs/1707.03107}{{\ttfamily 1707.03107}}].

\bibitem{Radyushkin:2018cvn}
A.~Radyushkin, \emph{{One-loop evolution of parton pseudo-distribution
  functions on the lattice}},
  \href{https://doi.org/10.1103/PhysRevD.98.014019}{\emph{Phys. Rev.}
  {\bfseries D98} (2018) 014019},
  [\href{https://arxiv.org/abs/1801.02427}{{\ttfamily 1801.02427}}].

\bibitem{Zhang:2018ggy}
J.-H. Zhang, J.-W. Chen and C.~Monahan, \emph{{Parton distribution functions
  from reduced Ioffe-time distributions}},
  \href{https://doi.org/10.1103/PhysRevD.97.074508}{\emph{Phys. Rev.}
  {\bfseries D97} (2018) 074508},
  [\href{https://arxiv.org/abs/1801.03023}{{\ttfamily 1801.03023}}].

\bibitem{Izubuchi:2018srq}
T.~Izubuchi, X.~Ji, L.~Jin, I.~W. Stewart and Y.~Zhao, \emph{{Factorization
  Theorem Relating Euclidean and Light-Cone Parton Distributions}},
  \href{https://doi.org/10.1103/PhysRevD.98.056004}{\emph{Phys. Rev.}
  {\bfseries D98} (2018) 056004},
  [\href{https://arxiv.org/abs/1801.03917}{{\ttfamily 1801.03917}}].

\bibitem{Vogt:2004mw}
A.~Vogt, S.~Moch and J.~A.~M. Vermaseren, \emph{{The Three-loop splitting
  functions in QCD: The Singlet case}},
  \href{https://doi.org/10.1016/j.nuclphysb.2004.04.024}{\emph{Nucl. Phys.}
  {\bfseries B691} (2004) 129--181},
  [\href{https://arxiv.org/abs/hep-ph/0404111}{{\ttfamily hep-ph/0404111}}].

\bibitem{Moch:2004pa}
S.~Moch, J.~A.~M. Vermaseren and A.~Vogt, \emph{{The Three loop splitting
  functions in QCD: The Nonsinglet case}},
  \href{https://doi.org/10.1016/j.nuclphysb.2004.03.030}{\emph{Nucl. Phys.}
  {\bfseries B688} (2004) 101--134},
  [\href{https://arxiv.org/abs/hep-ph/0403192}{{\ttfamily hep-ph/0403192}}].

\bibitem{lattices}
R.~Edwards, B.~Jo\'o, K.~Orginos, D.~Richards and F.~Winter{\emph{U.S. 2+1
  flavor clover lattice generation program} (2016) },
  [\href{https://arxiv.org/abs/unpublished}{{\ttfamily unpublished}}].

\bibitem{w0}
S.~Borsanyi et~al., \emph{{High-precision scale setting in lattice QCD}},
  \href{https://doi.org/10.1007/JHEP09(2012)010}{\emph{JHEP} {\bfseries 09}
  (2012) 010}, [\href{https://arxiv.org/abs/1203.4469}{{\ttfamily 1203.4469}}].

\bibitem{Bouchard:2016heu}
C.~Bouchard, C.~C. Chang, T.~Kurth, K.~Orginos and A.~Walker-Loud, \emph{{On
  the Feynman-Hellmann Theorem in Quantum Field Theory and the Calculation of
  Matrix Elements}},
  \href{https://doi.org/10.1103/PhysRevD.96.014504}{\emph{Phys. Rev.}
  {\bfseries D96} (2017) 014504},
  [\href{https://arxiv.org/abs/1612.06963}{{\ttfamily 1612.06963}}].

\bibitem{Chang:2018uxx}
C.~C. Chang et~al., \emph{{A per-cent-level determination of the nucleon axial
  coupling from quantum chromodynamics}},
  \href{https://doi.org/10.1038/s41586-018-0161-8}{\emph{Nature} {\bfseries
  558} (2018) 91--94}, [\href{https://arxiv.org/abs/1805.12130}{{\ttfamily
  1805.12130}}].

\bibitem{Buckley:2014ana}
A.~Buckley, J.~Ferrando, S.~Lloyd, K.~Nordström, B.~Page, M.~Rüfenacht
  et~al., \emph{{LHAPDF6: parton density access in the LHC precision era}},
  \href{https://doi.org/10.1140/epjc/s10052-015-3318-8}{\emph{Eur. Phys. J.}
  {\bfseries C75} (2015) 132},
  [\href{https://arxiv.org/abs/1412.7420}{{\ttfamily 1412.7420}}].

\bibitem{CJ}
A.~Accardi, L.~T. Brady, W.~Melnitchouk, J.~F. Owens and N.~Sato,
  \emph{{Constraints on large-$x$ parton distributions from new weak boson
  production and deep-inelastic scattering data}},
  \href{https://doi.org/10.1103/PhysRevD.93.114017}{\emph{Phys. Rev.}
  {\bfseries D93} (2016) 114017},
  [\href{https://arxiv.org/abs/1602.03154}{{\ttfamily 1602.03154}}].

\bibitem{Cichy:2019ebf}
K.~Cichy, L.~Del~Debbio and T.~Giani, \emph{{Parton distributions from lattice
  data: the nonsinglet case}},
  \href{https://arxiv.org/abs/1907.06037}{{\ttfamily 1907.06037}}.

\bibitem{Alexandrou:2016hiy}
C.~Alexandrou, \emph{{Novel applications of Lattice QCD: Parton Distributions,
  proton charge radius and neutron electric dipole moment}},
  \href{https://doi.org/10.1051/epjconf/201713701004}{\emph{EPJ Web Conf.}
  {\bfseries 137} (2017) 01004},
  [\href{https://arxiv.org/abs/1612.04644}{{\ttfamily 1612.04644}}].

\bibitem{Martin:2009iq}
A.~D. Martin, W.~J. Stirling, R.~S. Thorne and G.~Watt, \emph{{Parton
  distributions for the LHC}},
  \href{https://doi.org/10.1140/epjc/s10052-009-1072-5}{\emph{Eur. Phys. J.}
  {\bfseries C63} (2009) 189--285},
  [\href{https://arxiv.org/abs/0901.0002}{{\ttfamily 0901.0002}}].

\bibitem{Ball:2017nwa}
{\scshape NNPDF} collaboration, R.~D. Ball et~al., \emph{{Parton distributions
  from high-precision collider data}},
  \href{https://doi.org/10.1140/epjc/s10052-017-5199-5}{\emph{Eur. Phys. J.}
  {\bfseries C77} (2017) 663},
  [\href{https://arxiv.org/abs/1706.00428}{{\ttfamily 1706.00428}}].

\bibitem{Briceno:2018lfj}
R.~A. Briceño, J.~V. Guerrero, M.~T. Hansen and C.~J. Monahan,
  \emph{{Finite-volume effects due to spatially nonlocal operators}},
  \href{https://doi.org/10.1103/PhysRevD.98.014511}{\emph{Phys. Rev.}
  {\bfseries D98} (2018) 014511},
  [\href{https://arxiv.org/abs/1805.01034}{{\ttfamily 1805.01034}}].

\bibitem{jureca}
{J\"{u}lich Supercomputing Centre}, \emph{{JURECA: Modular supercomputer at
  J\"{u}lich Supercomputing Centre}},
  \href{https://doi.org/10.17815/jlsrf-4-121-1}{\emph{Journal of large-scale
  research facilities} {\bfseries 4} (2018) }.

\end{thebibliography}\endgroup
\bibliographystyle{jhep}

\end{document}